\documentclass[manuscript]{emulateapj}

\slugcomment{Accepted 2013 June 28 for publication in the Astrophysical Journal}
\shorttitle{ {\it Kepler}-field RR Lyrae Stars}
\shortauthors{Nemec, Cohen, Ripepi {\it et al.}}

\begin{document}

\title{Metal Abundances, Radial Velocities and other Physical \\ Characteristics for the RR~Lyrae Stars in the {\it Kepler} Field \altaffilmark{$\star$}}

\author{James M. Nemec\altaffilmark{1,2} 
Judith G. Cohen\altaffilmark{3},
Vincenzo Ripepi\altaffilmark{4}, 
Aliz Derekas\altaffilmark{5,6}, \\   
Pawel Moskalik\altaffilmark{7}, 
Branimir Sesar\altaffilmark{3}, 
Merieme Chadid\altaffilmark{8} \& 
Hans Bruntt\altaffilmark{9} }

\altaffiltext{$\star$}{Based in part on observations made at the W.M. Keck Observatory, which is operated as a 
scientific partnership among the California Institute of Technology, the University of California and the National 
Aeronautics and Space Administration. The Keck Observatory was made possible by the generous financial support of the 
W.M. Keck Foundation.  Also, based in part on observations obtained at the Canada-France-Hawaii Telescope (CFHT) which is operated by
the National Research Council of Canada, the Institut National Des Sciences de l'Univers of the Centre National de la
Recherche Scientifique of France, and the University of Hawaii.}
\altaffiltext{1}{Department of Physics \& Astronomy, Camosun College, Victoria, British Columbia, V8P~5J2, Canada, {\it nemec@camosun.ca}}
\altaffiltext{2}{International Statistics \& Research Corp., Brentwood Bay, BC, V8M~1R3, Canada, {\it jmn@isr.bc.ca}}
\altaffiltext{3}{Department of Physics and Astronomy, Caltech, USA, {\it jlc@astro.caltech.edu}, {\it bsesar@astro.caltech.edu} }
\altaffiltext{4}{INAF-Osservatorio Astronomico di Capodimonte, Salita Moiariello 16, I-80131, Napoli, Italy, {\it ripepi@na.astro.it}}
\altaffiltext{5}{Konkoly Observatory, Research Centre for Astronomy and Earth Sciences, Hungarian Academy of Sciences, H-1121 Budapest, Hungary, {\it derekas@konkoly.hu} }
\altaffiltext{6}{Sydney Institute for Astronomy, School of Physics, University of Sydney, NSW 2006, Australia}
\altaffiltext{7}{Copernicus Astronomical Centre, ul.Bartycka 18, 00-716, Warsaw, Poland {\it pam@camk.edu.pl} }
\altaffiltext{8}{Observatoire de la C\^ote d'Azur, Univ.\,Nice, Sophia-Antipolis, UMR 6525, Parc Valrose, 06108, Nice Cedex 02, France, {\it chadid@marseille.fr}}
\altaffiltext{9}{Department of Physics \& Astronomy, Aarhus Univ., DK-8000 Aarhus C, Denmark {\it bruntt@phys.au.dk}}

\begin{abstract}

Spectroscopic iron-to-hydrogen ratios, radial velocities, atmospheric parameters, and 
new  photometric analyses are presented for 41 RR~Lyrae stars (and one probable high-amplitude $\delta$~Sct star) located in the field-of-view of the  
{\it Kepler} space telescope.  Thirty-seven of the RR~Lyrae stars are fundamental-mode pulsators ({\it i.e.}, RRab stars) of which 16 exhibit the Blazhko effect.
Four of the stars are multiperiodic RRc pulsators oscillating primarily in the first-overtone mode.  
Spectroscopic [Fe/H] values for the 34 stars for which we were able to derive estimates range from
$-2.54\pm0.13$ (NR~Lyr) to $-0.05\pm0.13$\,dex (V784~Cyg), and for the 19 {\it Kepler}-field non-Blazhko stars studied by 
Nemec {\it et al.} (2011) the abundances agree will with their photometric [Fe/H] values. 
Four non-Blazhko RR~Lyrae stars that they identified as metal-rich
(KIC\,6100702, V2470~Cyg, V782~Cyg and V784~Cyg) are confirmed as such, and four additional stars  
(V839~Cyg, KIC\,5520878, KIC\,8832417, KIC\,3868420) are also shown here to be metal-rich.  
Five of the non-Blazhko RRab stars are found to be more metal-rich than
[Fe/H]$\sim -0.9$\,dex while all of the 16 Blazhko stars are more metal-poor than this value.  
New $P$-$\phi_{\rm 31}^s$-[Fe/H] relationships are derived based on  $\sim$970 days of
quasi-continuous high-precison Q0-Q11 long- and short-cadence {\it Kepler} photometry. 
With the exception of some Blazhko stars, the spectroscopic and photometric [Fe/H] values are in good agreement.  
Several stars with unique photometric characteristics are
identified, including a Blazhko variable with the smallest known amplitude and frequency modulations
(V838~Cyg).

\end{abstract}

\keywords{variable stars: RR~Lyrae stars, fundamental parameters --- stars: abundances --- telescopes: {\it Kepler}, Keck, CFHT}

\section{INTRODUCTION}
\label{ln:intro}

A knowledge of the chemical compositions and kinematics of individual RR~Lyrae stars is important
for studying problems concerning stellar and galactic evolution and for understanding the evolution
of stellar populations in our Galaxy and other nearby galaxies.  With iron-to-hydrogen abundances,
[Fe/H], ranging from smaller than $-2.5$ dex to larger than $0.0$ dex,  RR~Lyrae stars help to
define the halo and old-disk stellar populations.  They are ubiquitous, having been been found in
most globular clusters (see Clement {\it et al.} 2001), throughout the halo of our Galaxy (Le Borgne {\it et al.}
2012), in the Galactic bulge (Udalski {\it et al.} 1997; Alcock {\it et al.} 1998;  Moskalik \&
Poretti 2003; Collinge {\it et al.} 2006;   Kunder \& Chaboyer 2008, 2009; 
Soszynski {\it et al.} 2011; Pietrukowicz {\it et al.} 2012), 
and in a growing number
of nearby galaxies.  Of particular interest are the large numbers found in the 
Large Magellanic Cloud (Alcock {\it et al.} 1998, 2000; Gratton {\it et
al.} 2004; Nemec, Walker \& Jeon 2009; Soszynski {\it et al.} 2009),  in the Small Magellanic
Cloud (Soszynski {\it et al.} 2010), and in nearby dwarf
spheroidal galaxies (see Garofalo {\it et al.} 2013, and references therein).

RR~Lyrae stars are particularly valuable for stellar evolution studies because they are in an
advanced (post-RGB) evolutionary stage and therefore serve to improve our understanding of mass
loss, convection, rotation and magnetic fields.  One of their primary uses has been for distance
estimation using an absolute magnitude, metallicity relation (see Sandage 1981).  The linear form of
this relation for field and cluster RR~Lyr stars is usually written as $M_V(RR) = \alpha {\rm
[Fe/H]}+ \beta$, where $\beta\sim0.6$ and $\alpha$ ranges from $\sim$0.18-0.30;
however, there is strong evidence that the $M_V$-[Fe/H] relationship is nonlinear over the range
[Fe/H]$\sim-2.5$ to $\sim$0 dex (Cassisi {\it et al.} 1999; Caputo {\it et al.} 2000; 
Rey {\it et al.} 2000;  Bono {\it et al.} 2007).  In either case it now seems clear 
that the more metal-rich RR~Lyr stars have lower luminosities than their more metal-poor counterparts, at least for those
stars near the zero-age horizontal branch (ZAHB) that are not evolved away from the ZAHB.

Another astrophysical problem of great interest is the unsolved Blazhko effect (see Kov\'acs 2009, Buchler \& Koll\'{a}th 2011).
Whatever the eventual explanation, the correct model will undoubtedly be subject to observational
constraints such as are currently being provided by the Konkoly and Vienna Blazhko star surveys, and
the  high-precision, long-timeline photometry from the MOST, CoRoT and {\it Kepler} space
telescopes, as well as from ground-based follow-up surveys.  These surveys suggest that a significant
fraction of all RR~Lyr stars, maybe as many as half, exhibit Blazhko characteristics;  according to
the Konkoly survey the fraction is at least 40$\%$ (Jurcsik {\it et al.} 2009c; Szeidl {\it et al.} 2012). 
For RR~Lyr stars in the LMC the Blazhko effect is seen to be more frequent for 
stars with [Fe/H]$<-1.4$\,dex than for more metal rich RR Lyrae stars (Smolec 2005).    
To test for a similar trend in our Galaxy,
and for the derivation of distances and evolutionary states,   
requires accurate photometry and metal abundances.       

In the 115 deg$^2$ field of view of NASA's {\it Kepler} space telescope $\sim$45 RR Lyrae stars have
been identified among stars down to 17.5 mag.  Although this number is small compared with the
sample sizes for the MACHO (Alcock et al. 2003, 2004), OGLE (Soszynski {\it et al.} 2009, 2010), 
SDSS (Sesar {\it et al.} 2010), ASAS and other modern large surveys, {\it Kepler} photometry
provides the most precise and extensive set of photometry ever assembled for RR~Lyr stars.  

An early discovery made possible with {\it Kepler} data was the phenomenon of `period doubling', which manifests
itself as alternating high and low amplitude pulsations (Sz\'abo
{\it et al.} 2010; Kolenberg {\it et al.} 2010; Buchler \& Koll\'{a}th 2011). The effect
most probably is associated with the 9:2 resonance of the fundamental and overtone modes
that destablizes the fundamental-mode full amplitude pulsation (see Koll\'ath, Moln\'ar \& Szab\'o 2011).
Such period doubling usually
is not detectable from the ground because day-night bias prevents observations of successive
pulsation cycles.   Period doubling is well-known in RV~Tau and W~Vir stars, and from models 
has been predicted in classical Cepheids (Moskalik \& Buchler 1991) 
and in BL Her stars (Buchler \& Moskalik 1992), the latter having been seen recently by Smolec {\it et al.} (2012).

The RR~Lyrae stars in the {\it Kepler} field have been identified and studied previously by Kolenberg
{\it et al.} (2010, 2011; hereafter K10, K11), Benk\H o {\it et al.} (2010, hereafter B10), Nemec
{\it et al.} (2011, hereafter N11) and  Guggenberger {\it et al.} (2012, hereafter G12).     K10
published the first results  based on early-release (Q0,Q1) {\it Kepler} data, and carried out a
detailed analysis of  the Blazhko stars RR~Lyrae and V783~Cyg.    B10 presented a detailed analysis
of 29 {\it Kepler}-field RR~Lyrae stars based on the long cadence photometry from the first 138 days
(Q0-Q2) of the {\it Kepler} observations, and found that almost half of the stars (14/29) show
amplitude modulations.  K11 studied RR~Lyrae itself in great detail using Q1-Q2 long cadence data.
The N11 study consisted of an extensive Fourier decomposition analysis (using Q0-Q5 {\it Kepler} photometry) of 19
non-Blazhko {\it Kepler}-field RRab stars.  And most recently, G12 studied in detail the most
extreme Blazhko star, V445~Lyr. The RRc stars, all of which are multiperiodic, are being studied 
in detail by Moskalik {\it et al.} (2012).

The primary goal of the present investigation was to derive spectroscopic [Fe/H]$_{\rm spec}$
abundances (and the corresponding atmospheric parameters) for the {\it Kepler}-field RR~Lyrae stars.
Prior to the current investigation only photometric [Fe/H]$_{\rm phot}$ values were known  for the
19 non-Blazhko stars studied by N11;  and neither spectroscopic nor photometric abundances had been
measured for the RRc and Blazhko stars (a notable exception being RR~Lyrae itself).

The reliability of the N11 photometric metallicities, especially for the four non-Blazhko stars
identified as probable metal-rich old-disk stars, was of particular interest.  Thus, a secondary
goal was to confirm the {\it Kepler}-based [Fe/H]$_{\rm phot}$ values for the non-Blazhko stars, and
to use the longer time frame of the Q0-Q11 {\it Kepler} photometry (approaching 1000 days of
quasi-continuous measurements) and the improved {\it Kepler} pipeline to derive [Fe/H]$_{\rm phot}$
abundances for the Blazhko and RRc stars. 

With [Fe/H]$_{\rm spec}$ and [Fe/H]$_{\rm phot}$ measurements in hand, a comparison of the metal
abundances was possible.   The good agreement found between the spectroscopic and photometric
abundances  suggests that high-precision photometry alone can be used to derive iron-to-hydrogen
ratios, thereby providing a reliable means for the derivation of metallicities for stars too faint
for high-dispersion spectroscopy.

In $\S2$ the program stars are identified, and in $\S3$ the {\it Kepler} photometry is
described and used to derive photometric characteristics for the stars.  In $\S4$ the CFHT and Keck 
high-dispersion spectra are described, from which radial velocities ({\it RVs}), spectroscopic metallicities,
and correlations of the resulting atmospheric parameters are derived and discussed.   The
{\it Kepler} photometry and the new spectroscopic results are used in $\S5$ to better establish the 
relationship between metal abundance and light curve morphology, in particular the RRab and 
RRc $P$-$\phi_{\rm 31}$-[Fe/H] relations.  Our results are summarized in $\S6$.

\begin{deluxetable*}{lrlllllcc}
\tabletypesize{\scriptsize}
\tablewidth{0pt}
\tablecaption{Some Basic Photometric Characteristics for the {\it Kepler}-field RR~Lyrae Stars  
\label{tab:Table1}}
\tablehead{
\colhead{Star    }&\colhead{KIC }&\colhead{$\langle Kp \rangle$}&\colhead{$P_{\rm puls}$}&\colhead{$t_0$\thinspace(BJD)}& \colhead{$A_{\rm tot}$} & \colhead{$A_1$}  &\colhead{$ \phi_{31}^s $ }
&  {[Fe/H]$_{\rm phot}$ }     \\ 
\colhead{     }   &\colhead{ }   &\colhead{[mag]}               &\colhead{ [day]  }       &  \colhead{2400000+} & \colhead{[mag] }   & \colhead{ [mag] }  & \colhead{ [rad] }    
&   \colhead{  }  \\
\colhead{(1)} &\colhead{(2)} &\colhead{(3)} & \colhead{(4)} &\colhead{(5)} & \colhead{(6)} & \colhead{(7)} & \colhead{(8)} & \colhead{(9)}   }  
\startdata
\multicolumn{9}{c}{(a) 21 Non-Blazhko RRab-type stars} \\[0.1cm]
NR~Lyr        &  3733346 & 12.684 & 0.6820264     & 54964.7403 & 0.767&0.266 &5.120 &  $-2.51\pm0.06$   \\ 
V715~Cyg      &  3866709 & 16.265 & 0.47070609    & 54964.6037 & 0.988&0.338 &4.901 &  $-1.18\pm0.04$  \\ 
V782~Cyg      &  5299596 & 15.392 & 0.5236377     & 54964.5059 & 0.523&0.190 &5.808 &  $-0.30\pm0.04$  \\  
V784~Cyg      &  6070714 & 15.37  & 0.5340941     & 54964.8067 & 0.634&0.234 &6.084 &  $-0.14\pm0.07$  \\  
KIC\,6100702  &  6100702 & 13.458 & 0.4881457     & 54953.8399 & 0.575&0.209 &5.747 &  $-0.21\pm0.04$ \\ 
NQ~Lyr        &  6763132 & 13.075 & 0.5877887     & 54954.0702 & 0.811&0.280 &5.096 &  $-1.81\pm0.03$ \\ 
FN~Lyr        &  6936115 & 12.876 & 0.52739847    & 54953.2656 & 1.081&0.380 &4.818 &  $-1.90\pm0.04$ \\ 
KIC\,7021124  &  7021124 & 13.550 & 0.6224925     & 54965.6471 & 0.831&0.283 &5.060 &  $-2.18\pm0.04$   \\ 
KIC\,7030715  &  7030715 & 13.452 & 0.68361247    & 54953.8427 & 0.647&0.231 &5.616 &  $-1.28\pm0.05$  \\ 
V349~Lyr      &  7176080 & 17.433 & 0.5070740     & 54964.9588 & 0.988&0.346 &4.850 &  $-1.63\pm0.04$    \\ 
V368~Lyr      &  7742534 & 16.002 & 0.4564851     & 54964.7860 & 1.133&0.405 &4.772 &  $-1.29\pm0.05$  \\ 
V1510~Cyg     &  7988343 & 14.494 & 0.5811436     & 54964.6700 & 0.980&0.345 &5.075 &  $-1.80\pm0.03$  \\ 
V346~Lyr      &  8344381 & 16.421 & 0.5768288     & 54964.9231 & 0.964&0.330 &5.052 &  $-1.82\pm0.03$  \\ 
V350~Lyr      &  9508655 & 15.696 & 0.5942369     & 54964.7820 & 0.973&0.340 &5.119 &  $-1.81\pm0.03$ \\ 
V894~Cyg      &  9591503 & 13.293 & 0.5713866     & 54953.5624 & 1.105&0.377 &5.067 &  $-1.74\pm0.03$  \\ 
KIC\,9658012  &  9658012 & 16.001 & 0.533206      & 55779.9450 & 0.924&0.312 &5.144 &  $-1.28\pm0.02$  \\ 
KIC\,9717032  &  9717032 & 17.194 & 0.5569092     & 55779.8956 & 0.844&0.293 &5.183 &  $-1.38\pm0.02$  \\ 
V2470~Cyg     &  9947026 & 13.300 & 0.5485905     & 54953.7832 & 0.599&0.220 &5.737 &  $-0.47\pm0.03$  \\ 
V1107~Cyg     & 10136240 & 15.648 & 0.5657781     & 54964.7551 & 0.818&0.280 &5.196 &  $-1.42\pm0.02$   \\ 
V839~Cyg      & 10136603 & 14.066 & 0.4337747     & 55778.7060 & 0.793&0.273 &5.600 &  $-0.06\pm0.05$  \\ 
AW~Dra        & 11802860 & 13.053 & 0.6872160     & 54954.2160 & 0.892&0.307 &5.558 &  $-1.42\pm0.05$   \\ [0.1cm]  
\multicolumn{9}{c}{(b) 16 Blazhko RRab-type stars} \\[0.1cm]                                                        
V2178~Cyg     &  3864443 & 15.593 & 0.4869538     & 54976.3672 & 0.749&0.305 &4.895 &  $-1.35\pm0.03$   \\ 
V808~Cyg      &  4484128 & 15.363 & 0.5478642     & 54970.2834 & 0.879&0.298 &5.254 &  $-1.18\pm0.03$  \\ 
V783~Cyg      &  5559631 & 14.643 & 0.62070001    & 54975.5439 & 0.799&0.271 &5.517 &  $-1.15\pm0.03$   \\
V354~Lyr      &  6183128 & 16.260 & 0.561691      & 55245.1590 & 0.815&0.303 &5.191 &  $-1.40\pm0.02$   \\ 
V445~Lyr      &  6186029 & 17.401 & 0.5131158     & 55160.5957 & 0.591&0.255 &5.198 &  $-1.02\pm0.03$   \\ 
RR~Lyrae      &  7198959 &\phn7.862 & 0.566788    & 55278.2263 & 0.717&0.255 &5.307 &  $-1.21\pm0.03$  \\ 
KIC\,7257008  &  7257008 & 16.542 & 0.51177516    & 55758.5859 & 0.818&0.290 &5.191 &  $-1.02\pm0.03$   \\ 
V355~Lyr      &  7505345 & 14.080 & 0.4737027     & 55124.7072 & 0.955&0.373 &4.928 &  $-1.16\pm0.04$  \\ 
V450~Lyr      &  7671081 & 16.653 & 0.5046123     & 54996.3226 & 0.850&0.335 &4.976 &  $-1.35\pm0.03$  \\ 
V353~Lyr      & \phn9001926 &16.914&0.5568016     & 55082.6820 & 0.834&0.293 &5.123 &  $-1.50\pm0.02$ \\ 
V366~Lyr      &  9578833 & 16.537 & 0.5270283     & 55326.1915 & 0.856&0.308 &5.167 &  $-1.18\pm0.02$ \\ 
V360~Lyr      &  9697825 & 16.001 & 0.5575765     & 54988.9332 & 0.647&0.257 &5.065 &  $-1.63\pm0.02$  \\ 
KIC\,9973633  &  9973633 & 16.999 & 0.51075       & 55780.3655 & 0.848&0.293 &4.122 &  $-1.17\pm0.03$  \\ 
V838~Cyg      & 10789273 & 13.770 & 0.48027971    & 55807.9302 & 1.100&0.390 &4.857 &  $-1.36\pm0.04$ \\ 
KIC\,11125706 & 11125706 & 11.367 & 0.6132200     & 54981.0658 & 0.471&0.180 &5.824 &  $-0.61\pm0.05$  \\ 
V1104~Cyg     & 12155928 & 15.033 & 0.43638507    & 55120.8363 & 1.103&0.394 &4.861 &  $-0.93\pm0.05$ \\ [0.1cm] 
\multicolumn{9}{c}{(c) 4 RRc-type stars   } \\[0.1cm]
KIC\,4064484  &  4064484 & 14.641 & 0.33700953    & 55552.1567 & 0.377&0.193 &0.494 &  $-1.59\pm0.02$   \\ 
KIC\,5520878  &  5520878 & 14.214 & 0.26917082    & 55800.0883 & 0.325&0.164 &0.908 &  $-0.36\pm0.06$  \\ 
KIC\,8832417  &  8832417 & 13.051 & 0.2485492     & 54964.6391 & 0.281&0.141 &0.727 &  $-0.20\pm0.07$   \\ 
KIC\,9453114  &  9453114 & 13.419 & 0.3660236     & 55740.6420 & 0.414&0.209 &0.865 &  $-1.70\pm0.02$   \\ [0.2cm] 
\multicolumn{9}{c}{(d) High-amplitude Delta Scuti (HADS) star?   } \\[0.1cm]
KIC\,3868420  &  3868420 & 10.110 & 0.2082275     & 54960.5192 & 0.171&0.083 &4.109 &  {\hfil \nodata \hfil}  \\ [0.1cm]  
\enddata
\tablecomments{The columns contain: 
(1) star name (GCVS or other);  
(2) number in the Kepler Input Catalog (KIC);  
(3) mean {\it Kepler} magnitude (given in the KIC catalog);
(4) pulsation period, with the number of digits giving a measure of the accuracy of the value; 
(5) time of maximum light (BJD $-$ 2400000);
(6) total amplitude ({\it Kp} system) - for the Blazhko and RRc variables the mean values are given;
(7) Fourier $A_1$ coefficient ($Kp$ system);
(8) average value of the Fourier $\phi_{\rm 31}^s$({\it Kp}) parameter;
(9) photometric metal abundance, [Fe/H]$_{\rm phot}$, on the high dispersion spectroscopy scale (c9),
and derived using the nonlinear regression analyses discussed in $\S5$;  the uncertainties are standard errors of the regression fit. } 
\end{deluxetable*}

\section{PROGRAM STARS}

The {\it Kepler}-field RR~Lyrae stars studied here can be sorted into four distinct categories: (a) non-Blazhko RRab stars;  (b) Blazhko RRab stars; 
(c) RRc stars; and (d) RRc/HADS? star.  
{\bf Table~1} contains some basic photometric characteristics for the stars. The first three columns
contain the common star name (GCVS or other), the star number in the {\it Kepler} Input Catalog
(KIC, see Brown {\it et al.} 2011), and the mean {\it Kp} magnitude given in the KIC.  Columns 4-5
contain our best estimate of the pulsation period, $P_{\rm puls}$, and a recent time of maximum light, $t_0$ (Barycentric Julian Date).  
Columns 6-8 contain the total amplitude, $A_{\rm tot}$, the Fourier $A_1$ amplitude coefficient,
and the Fourier $\phi_{\rm 31}^s$ phase parameter (defined as $\phi_{\rm 3}^s - 3 \phi_{\rm 1}^s$, where 
the superscript `s' indicates Fourier decomposition with a sine series), all derived from the $Kp$ photometry (see $\S3$).
Column~9 contains the photometric metallicity, [Fe/H]$_{\rm phot}$, derived in $\S5$ using
new bivariate linear regression analyses computed with the {\it Kepler} photometry ($\S3$) and 
the new spectroscopic metallicities ($\S4$).  

In Table~1 the number of non-Blazhko stars is two greater than the 19 stars studied by N11:
KIC\,9658012, KIC\,9717032 and V839~Cyg (KIC\,10136603) are new additions, while V838~Cyg
(KIC\,10789273) has been moved to the Blazhko list.   KIC\,9658012 and KIC\,9717032 were discovered
during the course of the BOKS survey (Feldmeier {\it et al.} 2011); both stars were observed with
{\it Kepler} for the first time in Q10, and neither star shows any evidence for amplitude
modulations.  A $g$-band light curve for KIC\,9717032 is shown in Fig.17 of the Feldmeier paper.
V839~Cyg is new and was discovered as a by-product of the ASAS-North survey (Pigulski {\it et al.}
2013, in preparation).      

The Blazhko variables listed in Table~1 include three stars in addition to those in the B10 list:
KIC\,7257008 and KIC\,9973633 (also discovered by Pigulski {\it et al.} 2013, in preparation) which
show amplitude modulations in the Q10 and Q11 {\it Kepler} data;  and V838~Cyg which was previously
thought to be a non-Blazhko variable but now appears to be the lowest-amplitude Blazhko star yet
discovered (see below).  Wils {\it et al.} (2006) listed NR~Lyr (KIC\,3733346) as a Blazhko star with
an uncertain modulation period of 27 days; however, our analysis of the Q1-Q13 long cadence data (51883 data points)
and the Q11.1 short cadence data (45512 data points) showed no evidence for amplitude modulations, thus
supporting the B10 and N11 conclusion that NR~Lyr is a non-Blazhko RRab variable.   
The faint star V349~Lyr (KIC\,7176080) is in the B10 Blazhko list, but was classified by N11 as a probable
non-Blazhko star; our analysis of the Q1-Q13 long cadence data (51877 data points) and the
Q9 short cadence data (51878 data points) again shows no evidence for amplitude modulations.      

To date only four {\it bona fide} c-type RR~Lyrae (RRc) stars have been identified in the {\it Kepler}
field.  Two were observed at CFHT and two at Keck: the two shorter period stars are metal rich, and
the two longer period stars are metal poor (see $\S4-5$ below).  All four stars have been found
to be multiperiodic and are being studied in detail by Moskalik {\it et al.} (2012; 2013, 
in preparation).  Owing to the challenges of distinguishing RRc stars from high-amplitude $\delta$~Sct (HADS) stars
and some close eclipsing binaries, all of which have similar light curves, the RRc section of
Table~1 probably is more incomplete than the RRab sections.  
KIC\,10063343 ({\it Kp}=13.16) has a pulsation period appropriate for an RRc star, $P_{\rm
puls}=0.332764$\,d, but the variable light curve and Fourier parameters raise serious doubts that it
is an RR~Lyrae star.  Other candidate RRc stars are being sought by Kinemuchi {\it et al.} (2013, in
preparation).  

The relatively bright star KIC\,3868420 ({\it Kp}=10.11) is of particular interest.  The early {\it Kepler}
photometry suggested that it might be an RRc star, and as such a single spectrum was taken at CFHT.
Analysis of this spectrum  revealed it to be a hot, high surface gravity star with 
spectroscopic [Fe/H]=$-0.32\pm0.13$\,dex;  however, with a dominant period of 
only 0.208~day,  a light curve that is more symmetric and more sharply peaked than the light curves 
of the RRc stars, and a relatively small $RV$, it more probably is a HADS star.  

Three additional RRab stars recently have been identified in the {\it Kepler} field,
in the directions of KIC\,3448777, KIC\,4917786 and KIC\,7295372. 
All have $P_{\rm puls}\sim0.5$\,d, low amplitudes resulting from `crowding' by brighter
nearby stars\footnote{The `crowding' here is determined by the 3.98 arcsec pixel$^{\rm -1}$ image
scale of the CCD chips on the {\it Kepler} telescope.}, and unknown metal abundances.  These newly
discovered RR~Lyr stars continue to be observed by {\it Kepler}. 




\section{{\it KEPLER} PHOTOMETRY}

\begin{figure}
\plotone{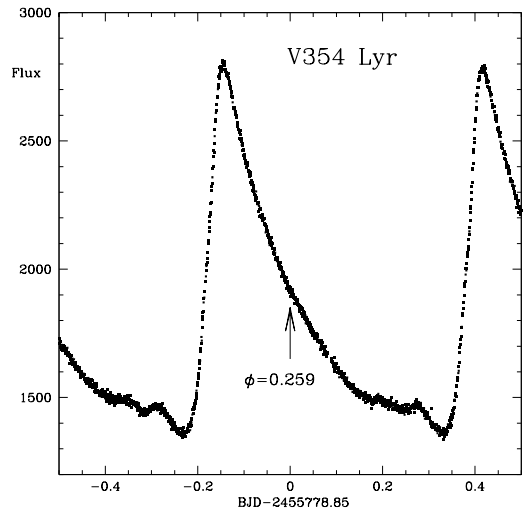}
\caption {A typical {\it Kepler} light curve, in this case for V354~Lyr (KIC\,6183128), a Blazhko
RRab star with $P_{\rm puls} = 0.561691\pm0.000001$\thinspace d and $P_{\rm BL}=723\pm12$ days.  The
data plotted are a one-day segment of short cadence flux measurements made simultaneously with three
Keck-I 10-m spectra.  The abscissa is time (Barycentric Julian Date), starting 0.5 day before and
ending 0.5 day after the mid time of the spectra;  and the ordinate is the {\it Kepler} raw flux
(counts per second).  The pulsation phase at the mid-time of the spectroscopic observations is
indicated by a labelled arrow.  } 
\label{fig:V354LyrBL_LC} \end{figure}

An unprecedented amount of high-precision {\it Kepler} photometry is currently available for
deriving the photometric characteristics (including photometric metal abundances) for the majority
of the stars in Table~1.  Most of the program RR~Lyr stars have been observed every quarter since the first {\it Kepler}  flux
measurements were made in May 2009.  Up to and including Q11 a typical RRab star with a pulsation
period of 0.5 day had over 3.5 years of quasi-continuous high-precision photometric data.  Long
cadence observations (LC, {\it i.e.}, 29.4 minute flux integrations) over this 3.5-year interval
amount to $\sim$350,000 data points spread over 2500 pulsation cycles, with $\sim$24 brightness
measurements per cycle.  For most of the stars the CFHT and Keck spectra had simultaneously-acquired
{\it Kepler} photometry.   The CFHT spectroscopic observations were made during the sixth and seventh quarters
(Q6,Q7) of the {\it Kepler} observations, and the Keck spectra were acquired during the first month
of the tenth quarter (Q10.1).   

Five of the RR~Lyr stars in Table~1 were not observed with {\it Kepler} prior to  the start of Q10:
KIC\,7257008, KIC\,9658012, KIC\,9717032, KIC\,9973633 and V839~Cyg.  Fortunately, the Keck observations
for the last four  stars were made in August 2011, which occurs in Q10.1 of the {\it Kepler}
observations.  KIC\,7257008 ({\it Kp}=16.54), KIC\,7021124 ({\it Kp}=13.55) and V349~Lyr ({\it
Kp}=17.43) have yet to be observed spectroscopically. 

In addition to the LC data most of the program stars also have been observed at short cadence (SC,
{\it i.e.}, a flux measurement every 1 minute) for at least one or two quarters.  The available SC
data for these stars usually amounts to almost half a million observation points per star.  RR~Lyrae
itself has been observed at SC every quarter since Q5, the total number of data points being just
under $9\times10^5$. 

A typical RRab star with SC data has $\sim$720 brightness measurements per pulsation cycle, which is
sufficient for defining extremely-well the shape of the  light curve (as illustrated in {\bf
Fig.1}), and for detecting and defining from precise Fourier parameters the smallest amplitude and
frequency modulations. 

The task of processing the {\it Kepler} raw flux data has been discussed by Koch {\it et al.}
(2010), Jenkins {\it et al.} (2010) and Gilliland {\it et al.} (2010).  Of particular concern was
the detrending of data within quarters, the `stitching together' of the data from the different
quarters, and the merging of long and short cadence observations.   The reduction methods
(transforming the raw flux counts to {\it Kp} magnitudes) that were used  here were similar to those
described in N11.   To achieve the highest levels of photometric precision multiple detrending
passes were found to be necessary to reduce the residuals about the non-Blazhko light curves to
$\sim$0.2 mmag.

\begin{deluxetable*}{llllll}
\tabletypesize{\scriptsize}
\tablewidth{0pt}
\tablecaption{Amplitude and period (phase) modulations of 16 Blazhko stars in the {\it Kepler} field }
\label{tab:Table2}
\tablehead{
\colhead{Star}&\colhead{ $\Delta A_{\rm tot}$ [mag]  } & \colhead{ $\Delta P_{\rm puls}$ [day] } &  
\colhead{ $\Delta A_1$ [mag] } &  \colhead{ $\Delta \phi_1$ [rad] }     & \colhead{$\Delta \phi_{\rm 31}^s$ [rad] }    \\
\colhead{(1)} & \colhead{(2)}        &\colhead{(3)}       &\colhead{(4)}        & \colhead{ (5) }  & \colhead{(6) }  }
\startdata
V445~Lyr     & 0.80\,(0.20-1.00) & 0.013\,(0.509-0.522)   & 0.380\,(0.018-0.398)   & 3.12\,(1.48-4.60)       & 6.28\,(0.00-6.28) \\
V2178~Cyg    & 0.84\,(0.30-1.14) & 0.0014\,(0.4859-73)    & 0.279\,(0.162-0.441)   & 1.09\,(3.91-5.00)       & 3.55\,(3.41-6.96) \\
V450~Lyr     & 0.56\,(0.57-1.13) & 0.0007\,(0.5042-49)    & 0.240\,(0.207-0.447)   & 0.44\,(4.74-5.18)       & 1.12\,(4.53-5.65) \\ 
KIC\,7257008  & 0.68\,(0.44-1.12) & 0.0030\,(0.5105-35)    & 0.211\,(0.179-0.390)   & 0.60\,(3.00-3.60)       & 1.05\,(4.77-5.82) \\  
V354~Lyr     & 0.60\,(0.49-1.09) & 0.0003\,(0.5616-19)    & $>$0.20\,($<$0.19-0.39)& 1.26\,(2.51-3.77)       & 0.70\,(4.85-5.55) \\ 
V360~Lyr     & 0.21\,(0.54-0.75) & 0.0010\,(0.5570-80)    & 0.162\,(0.176-0.338)   & 0.37\,(0.99-1.36)       & 0.82\,(4.78-5.60) \\
V808~Cyg     & 0.35\,(0.71-1.06) & 0.0020\,(0.5467-87)    & 0.130\,(0.237-0.367)   & 0.67\,(0.28-0.95)       & 0.64\,(4.96-5.60) \\
RR~Lyrae     & 0.23\,(0.59-0.82) & 0.0024\,(0.5657-81)    & 0.127\,(0.186-0.313)   & 0.65\,(4.77-5.42)       & 1.21\,(4.85-6.06) \\
V366~Lyr     & 0.18\,(0.77-0.95) & 0.0003\,(0.5269-72)    & 0.127\,(0.248-0.375)   & 0.14\,(1.21-1.35)       & 0.98\,(4.75-5.73) \\ 
KIC\,9973633  & 0.39\,(0.61-1.00) & 0.0009\,(0.5105-14)    & 0.119\,(0.225-0.344)   & 0.22\,(4.00-4.22)       & 0.78\,(4.80-5.58) \\ 
V355~Lyr     & 0.15\,(0.87-1.02) & 0.0008\,(0.4733-41)    & 0.110\,(0.313-0.423)   & 0.15\,(1.05-1.20)       & 0.62\,(4.67-5.29) \\ 
V353~Lyr     & 0.24\,(0.71-0.95) & 0.0004\,(0.5566-70)    & 0.082\,(0.250-0.332)   & 0.16\,(0.59-0.75)       & 0.24\,(5.01-5.25) \\ 
V1104~Cyg    & 0.17\,(1.03-1.20) & 0.0003\,(0.4363-66)    & 0.062\,(0.365-0.427)   & 0.070\,(0.165-0.235)    & 0.16\,(4.78-4.94) \\
V783~Cyg     & 0.08\,(0.76-0.84) & 0.0005\,(0.6204-09)    & 0.033\,(0.253-0.286)   & 0.08\,(1.82-1.90)       & 0.14\,(5.44-5.58) \\ 
KIC\,11125706 & 0.03\,(0.45-0.48) & 0.0004\,(0.6128-32)    & 0.008\,(0.176-0.184)   & 0.031\,(5.452-5.483)    & 0.112\,(5.772-5.884)  \\ 
V838~Cyg     & 0.02\,(1.09-1.11) & 0.0002\,(0.4801-03)    & 0.0016\,(0.3871-87)    & 0.0010\,(3.3577-87)     & 0.003\,(4.855-4.858)  
\enddata
\tablecomments{The columns contain:  
(1) star name; 
(2) range of total amplitude ({\it Kp}-system), with minimum and maximum values in parentheses;
(3) range of the pulsation period (min and max in parentheses);
(4) range of the Fourier $A_1$ amplitude coefficient (min and max in parentheses) -- note that $\Delta A_1$ here is the $A_1$ range, which is not the same 
as the $\Delta A_1$ defined by B10; 
(5) range of the first Fourier phase coefficient, $\phi_1$ (min and max in parentheses); 
(6) range of the Fourier phase-difference parameter $\phi_{\rm 31}^s$ (min and max in parentheses).
}
\end{deluxetable*}
\vskip0.5cm


\subsection{Pulsation Periods, Amplitudes and Phases }

Pulsation periods, $P_{\rm puls}$, times of maximum light, $t_0$, and total amplitudes on the {\it
Kepler} photometric system, $A_{\rm tot}$, are given for  the program RR~Lyrae stars in columns 4-6 
of Table~1.  All values are newly derived  based on analyses of the 
available {\it Kepler} Q0-Q11 long and short cadence data, which in most cases comprises almost 1000 days of
quasi-continuous high-precision photometry. 
  
The  $P_{\rm puls}$ are mean values of the dominant oscillation frequencies, ignoring the secondary
periods that are present in some non-Blazhko stars (V350 Lyr, KIC\,7021124) and in all the Blazhko stars,
ignoring changing periods, and averaging over many modulation cycles for the Blazhko stars.   The periods were computed with the
`Period04' package of Lenz \& Breger (2005) using the Monte Carlo mode for error estimation, 
Stellingwerf's (1978) `PDM2' program\footnote{PDM2 is
freely available at http://www.stellingwerf.com. }, updated to deal with rich data sets,
and the period-finding program of Kolaczkowski (see Moskalik \& Kolaczkowski 2009).

The accuracy of the pulsation periods reported in Table~1 is $\sim$few units in the last decimal place given,  
with the periods for the  Blazhko and RRc stars tending to be less accurate than those  for the
non-Blazhko stars.  The uncertainties depend on several factors: the particular data set (or subset)
that was analyzed, the search method, the constancy of the primary pulsation period, and whether or
not amplitude or frequency modulations are present.  In general the periods agree well  with those given by B10
for 15 non-Blazhko and 14 Blazhko stars, and with the more precise values derived by N11 for 19
non-Blazhko stars.   Since the new periods were computed with longer time baselines than these
earlier studies, and are based on both long and short cadence data, they are the best available
values yet reported for these stars.  In general, the accuracy of the  $P_{\rm puls}$ values is more
than sufficient for estimating the phases at which the spectroscopic observations were made.

The $A_{\rm tot}$ in Table~1 are the trough-to-peak {\it Kp} amplitudes.  For the non-Blazhko stars
they are time invariant values obtained by inspection of the complete light curve and by averaging
$A_{\rm tot}$ calculated from individual  light curves for successive  time segments (see Fig.2).
For the Blazhko variables and the four RRc stars, all of which are multiperiodic and show amplitude
variations, $A_{\rm tot}$ is the time-averaged value.  According to the transformation given in
equation 2 of N11 the $A_{\rm tot}$({\it Kp}) are $\sim0.14$ mag smaller than the $A_{\rm
tot}$($V$), where $V$ is for the Johnson $UBV$ system. 

The zero points $t_0$ were estimated from the mean light curves and correspond to times of
zero-slope at maximum light.  For the non-Blazhko stars they were usually chosen  to approximately
coincide with the time of the first {\it Kepler} observations  ({\it i.e.}, just after BJD 2454953),
and for the Blazhko variables the $t_0$  generally coincide with a time of maximum light occuring
during a period of maximum amplitude.  Thus, $t_0$  serves as the zero point for both the pulsation
phase,  $\phi_{\rm puls}$, defined to be the fractional part of $(t-t_0)/P_{\rm puls}$, and for the
Blazhko phase, $\phi_{\rm BL}$, defined to be the fractional part of $(t-t_0)/P_{\rm BL}$.

\subsection{Blazhko Period and Amplitude Modulations}

The most noticeable signature of the Blazhko effect is variation in the light curve shape, in particular amplitude modulations,
over time scales ranging from several days to several years.  The amplitude that is usually 
considered is the total amplitude ({\it i.e.}, from trough to peak) through the $V$ or $B$ filter, but may also be one of the Fourier amplitude coefficients.  
In the absence of $BV$ photometry for most of the {\it Kepler}-field RR~Lyr stars we focus in this paper on the $A_{\rm tot}$ and $A_1$
amplitudes derived from the {\it Kp} photometry.  The $A_1$ values are from Fourier decompositions of the {\it Kp} data  
performed using the following sine-series sum: 
\begin{equation}
m(t) = A_0 + \sum\limits^{N}_{k=1} A_k   \sin(k 2 \pi t / P + \phi_k),
\end{equation}
where m(t) is the apparent {\it Kp} magnitude as a function of time,
$A_0$ is the mean {\it Kp} magnitude (assumed to be the value given in the KIC catalog),
$N$ is the adopted number of terms in the Fourier series,  
the $A_k$ and $\phi_k$ are the Fourier amplitudes and
phases, and $P$ is the pulsation period.  

Blazhko stars are also known to exhibit period (frequency, or phase) modulations, although usually these are less noticeable than the 
amplitude modulations.  Frequency modulations were detected previously in the {\it Kepler}-field Blazhko stars 
studied by B10,  in RR~Lyrae by K10 and K11, and in V445~Lyr by G12.  The 16 {\it Kepler}-field Blazhko 
stars studied here all exhibit both amplitude and frequency modulations.

While the detection of amplitude and frequency modulations is usually straightforward, the
physical mechanism responsible remains unknown, and Blazhko stars are almost as
enigmatic now (see Kov\'acs 2009) as when they were discovered over 100 years ago (Blazhko 1907). 
The problem in recent years has taken on new significance with the discovery that 
almost half of all RRab stars are Blazhko variables (Jurcsik {\it et al.} 2009c), 
and that possibly all RRc stars are multiperiodic (Moskalik {\it et al.} 2012).  
These discoveries raise the question whether all RR~Lyrae stars might be amplitude and frequency modulators.  

The measured amplitude and frequency modulations for the Blazhko RRab stars in our sample are summarized in {\bf Table~2}, where
the stars have been ordered from largest to smallest amplitude modulations, $\Delta A_1$ ({\it i.e.}, from the most extreme Blazhko variability to the least variability).
All the quantities are newly-derived from both the long and short cadence Q0-Q11 photometry, and 
the uncertainties are at the level of the precisions given.
Columns 2-6 contain, respectively, the measured ranges (minimum to maximum values) for:  the total amplitude in the {\it Kp}-passband,  $\Delta A_{\rm tot}$;
the pulsation period,  $\Delta P_{\rm puls}$;   
the Fourier amplitude coefficient, $\Delta A_1$;
the first Fourier phase coefficient, $\Delta \phi_1$;
and the Fourier phase-difference parameter $\Delta \phi_{\rm 31}^s$,
where  $\phi_{\rm 31}^s = \phi_3 - 3\phi_1$ (adjusted to 
the value closest to the mean value of 5.8) and the superscript `s' indicates that a 
sine (rather than a cosine) series was used for the Fourier decomposition (see equation 1).    

The amplitude modulations, $\Delta A_{\rm tot}$, range from as large as $\sim$0.82 mag (V445~Lyr, V2178~Cyg)
to as small as $\sim$0.02 mag  (V838~Cyg, KIC\,11125706), corresponding to 2-80\% of $A_{\rm tot}$(max).   
The stars with the largest amplitude variations tend also to exhibit the largest period variations.   
The average period range $\Delta P_{\rm puls}$ is 0.0010\,d. 
Plots of $\Delta A_{\rm tot}$ versus period, and $\Delta\phi_{\rm 31}^s$ versus period,  
show an approximately normal distribution with the largest modulations occuring for those
stars with pulsation periods around 0.53\,d. In the period-$\Delta\phi_{\rm 31}^s$ diagram the extreme Blazhko stars V445~Lyr and V2178~Cyg (see Fig.\,3) are clearly
outliers with $\Delta\phi_{\rm 31}^s$ values considerably larger than for all the other Blazhko stars.
Estimates of $\Delta P_{\rm puls}$ and $\Delta A_1$ of RR~Lyrae by K11, and V445\,Lyr by G12, agree well with the values
reported in Table~2.

\subsubsection{V838~Cyg and KIC\,11125706}

\begin{figure*}
\epsscale{1.15}
\plottwo{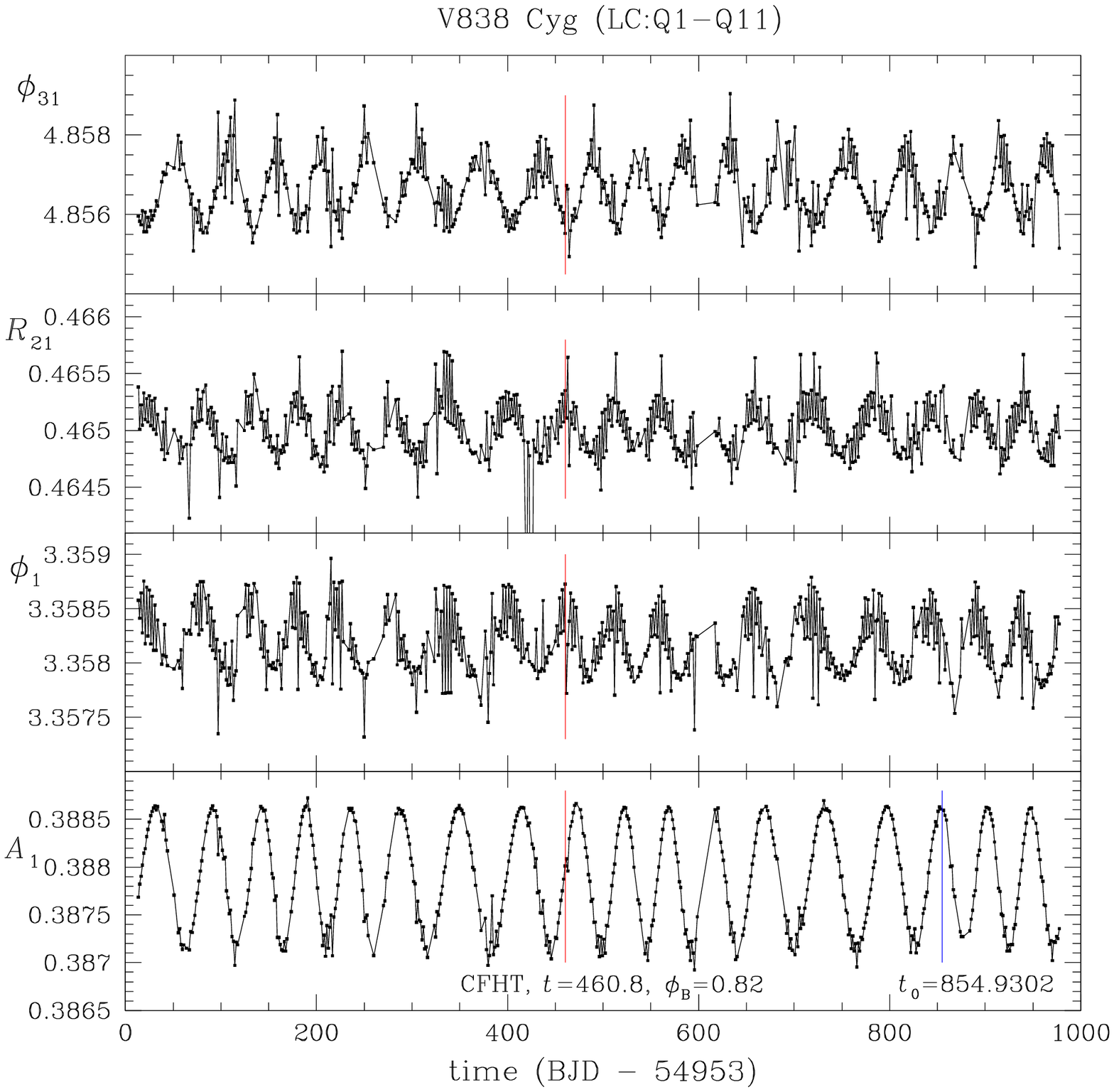}{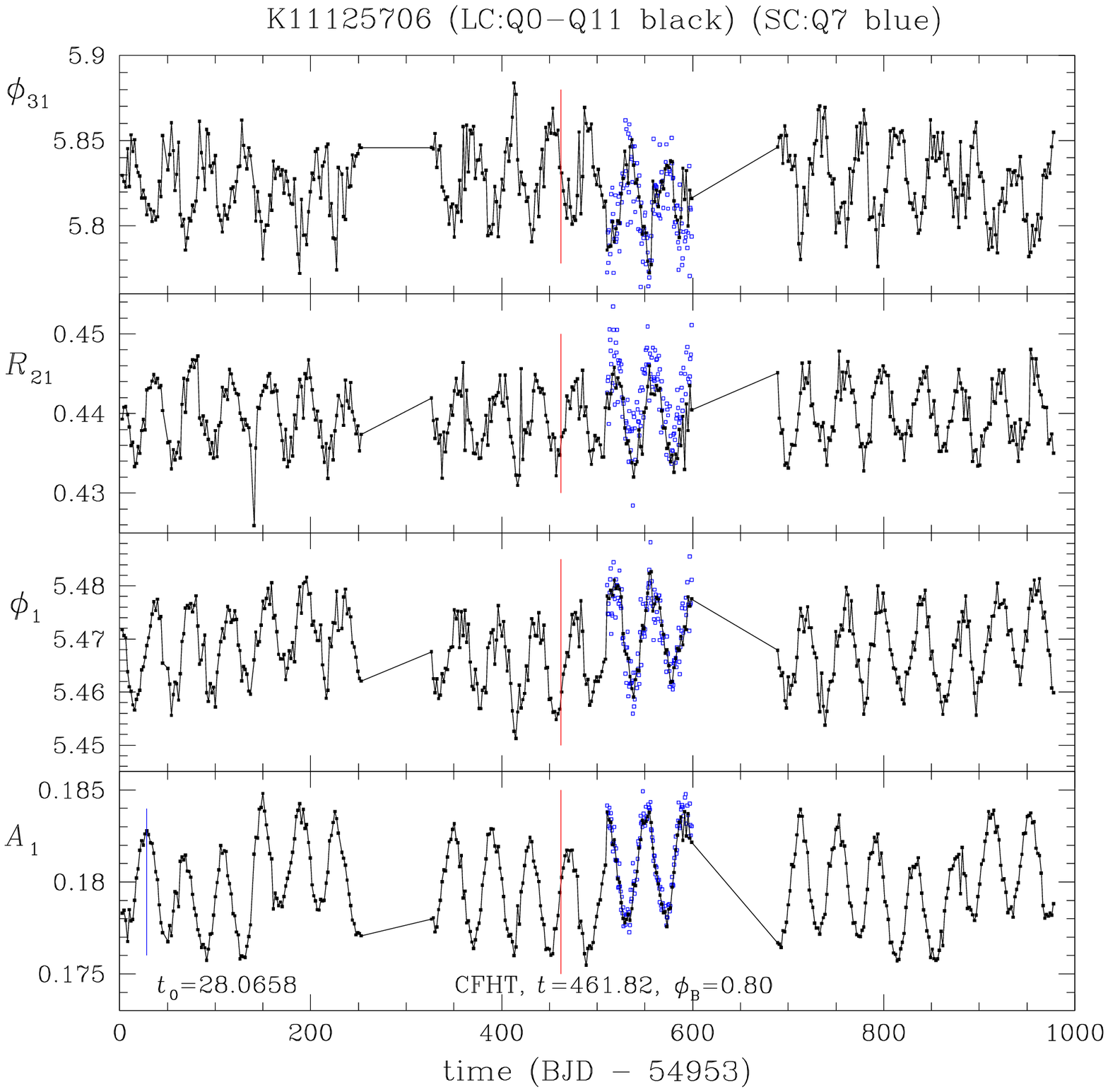}
\caption{Amplitude and frequency modulations for V838~Cyg (left) and KIC\,11125706 (right), as seen in time series graphs for four Fourier parameters.
The results from analyses of both the long cadence {\it Kepler} photometry (black connected points) and the short cadence {\it Kepler} photometry (blue open squares) are plotted.  
Vertical lines (labelled) are drawn at the mid-times of the CFHT spectroscopic observations (red) and at the adopted $t_0$ zero points (blue). 
 A color version of this figure is given in the on-line version of the paper.    }
\label{fig:V838Cyg_K11125706}
\end{figure*}

When the amplitude modulations of Blazhko stars are very small careful analysis is required to ensure that the fluctuations 
are not artificial, perhaps caused by the data reduction methods.  Such borderline Blazhko stars among the {\it Kepler}-field RR~Lyr 
stars include KIC\,11125706, V838~Cyg and V349~Lyr (discussed by N11).  
Previously, KIC\,11125706 was found by B10 to have ``the lowest amplitude modulation ever detected in an RR~Lyrae star'',
even smaller than the small amplitude modulations seen in RR~Gem (Jurcsik {\it et al.} 2005) and in SS~Cnc (Jurcsik {\it et al.} 2006);
and V838~Cyg was classified as non-Blazhko by both B10 and N11.
We confirm the low amplitude modulations of KIC\,11125706, but also find that V838~Cyg exhibits amplitude and 
frequency modulations and that these are even smaller than those of KIC\,11125706 ($\Delta A_1 = 0.0016$\,mag vs. 0.008 mag, and $\Delta P_{\rm puls}$=0.0002\,d vs. 0.0004\,d).  
The two stars are considerably different in that 
V838~Cyg has a much higher amplitude and shorter period than KIC\,11125706  ($A_{\rm tot} = 1.10$\,mag vs. 0.47\,mag; 
$P_{\rm puls} = 0.48$\,d vs. 0.61\,d;  see Fig.~4 below).

{\bf Fig.~2} shows for V838~Cyg (left) and KIC\,11125706 (right) four Fourier variables plotted as time series.  
The plotted variables are (from bottom to top): the amplitude $A_1$ and the phase $\phi_1$, both from the first term in the Fourier sine series;  
the ratio $R_{\rm 21}$ of the second and first Fourier amplitudes;  and the Fourier phase difference parameter, $\phi_{\rm 31}^s$.    
The four variables were obtained by
dividing the long cadence data from Q1 to Q11 into time segments, each segment corresponding to three pulsation periods ({\it cf.} Fig.5 of N11 where 
the {\it Kepler} stars NQ~Lyr and V783~Cyg are plotted).  Approximately 40000 data points were analyzed for each star, and 15 terms were used 
for the Fourier decompositions. 

For both stars the mid-times and Blazhko phases of the respective CFHT spectra are indicated (labelled red vertical lines), as are 
the adopted $t_0$ zero points (labelled blue vertical lines).   For KIC\,11125706 the large gaps at $t$=253-323 and 
at $t$=600-688 are because no observations were made in 
the second two months of Q4 and in Q8.  For the same star the 
blue open squares result from analysis of the 126955 short cadence photometric measurements made in Q7 -- agreement of the short and long cadence results is excellent\footnote{Such agreement is not always the
case owing to different normalizations used during the detrending procedure, but the differences appear always to be small (less than 1\% for each variable) and do not 
affect our main results.}.
  
The modulation periods are obvious and seen to be close to 55\,d for V838~Cyg, and 40\,d for KIC\,11125706  -- these
are the basic Blazhko periods.  In addition to the amplitude modulations seen in the $A_1$ and $R_{\rm 21}$ panels, the 
variation of  $\phi_1$  is indicative of frequency modulation.    
That the time-series plots for $\phi_1$ and $\phi_{\rm 31}^s$ (which measures the pulse shape) are mirror images of each other is not
uncommon (but not always the case, as is seen for V2178~Cyg in Fig.3 below).   

On relatively short time scales, such
as the $\sim$90 day baseline of the available SC:Q10 data for V838~Cyg (not plotted), the modulations look sinusoidal;  
however, closer inspection shows that they are not.  
That the $A_1$ waves are compressed around time $t=200$ and 550 but are 
wider around $t=400$ and 750 proves that the modulation period for V838~Cyg is varying.    
Thus the amplitude and frequency modulations are complex.  
A ``Period04'' frequency analysis of the long cadence data identified multiple Blazhko periods 
of 54, 64 and 47 days, the first being the most significant.
Finally, the cycle-to-cycle amplitude variations seen in the bottom right panel suggest that additional amplitude modulations 
may be present for KIC\,11125705.

Many other {\it Kepler}-field Blazhko stars in addition to V838~Cyg and KIC\,11125706 have been found to exhibit interesting and complex modulation behaviours.  
In the next subsection two quite different Blazhko variables (V1104~Cyg and V2178~Cyg) are used to illustrate the time- and frequency-domain 
methods that were used for deriving the Blazhko periods and Blazhko types.

\subsubsection{Blazhko Periods and Types from Power Spectra} 

Blazhko periods can also be inferred from power spectra.  In fact, the Blazhko classification scheme 
of Alcock {\it et al.} (2000, 2003) is based solely on characteristic power-spectra patterns.  
When the Fourier transform of prewhitened photometry shows a single major feature at $f_1$,  offset from the fundamental pulsation 
frequency $f_0$, the Blazhko type is said to be `RR0-BL1'.  In this case the Blazhko frequency is  $f_{\rm BL}$ = $\mid f_1-f_0 \mid$, 
and $P_{\rm BL} = 1/f_{\rm BL}$.
When the Fourier transform of the prewhitened data shows two major features, 
symmetric about the fundamental frequency, the type is said to be `RR0-BL2'.    
For such stars $P_{\rm BL}$ is equal to the reciprocal of the mean frequency shift (assuming that
the two features are equidistant from $f_0$). 
For other types found in the MACHO surveys see Table~1 of Alcock {\it et al.} (2003).

\begin{figure*}
\plottwo{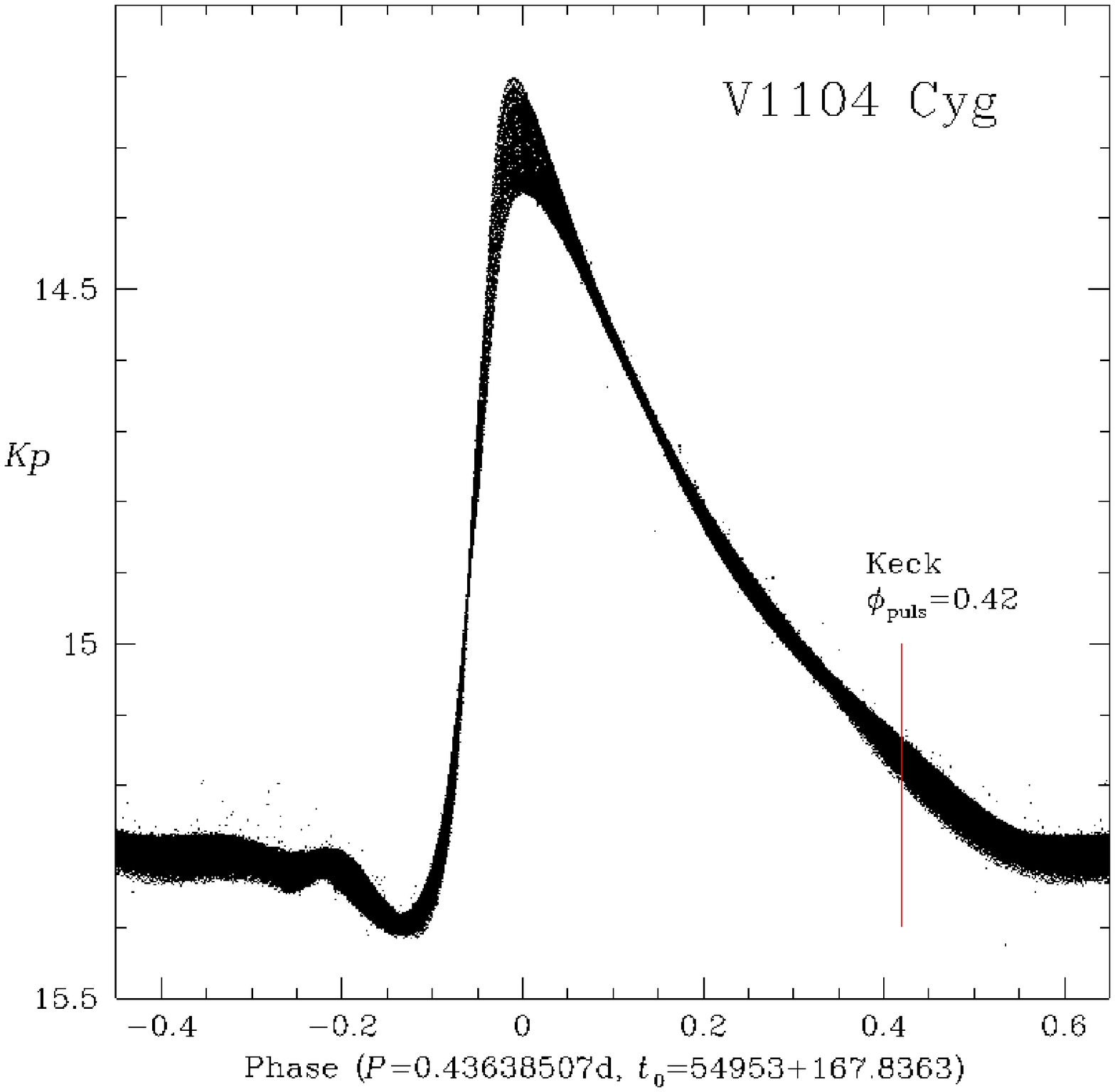}{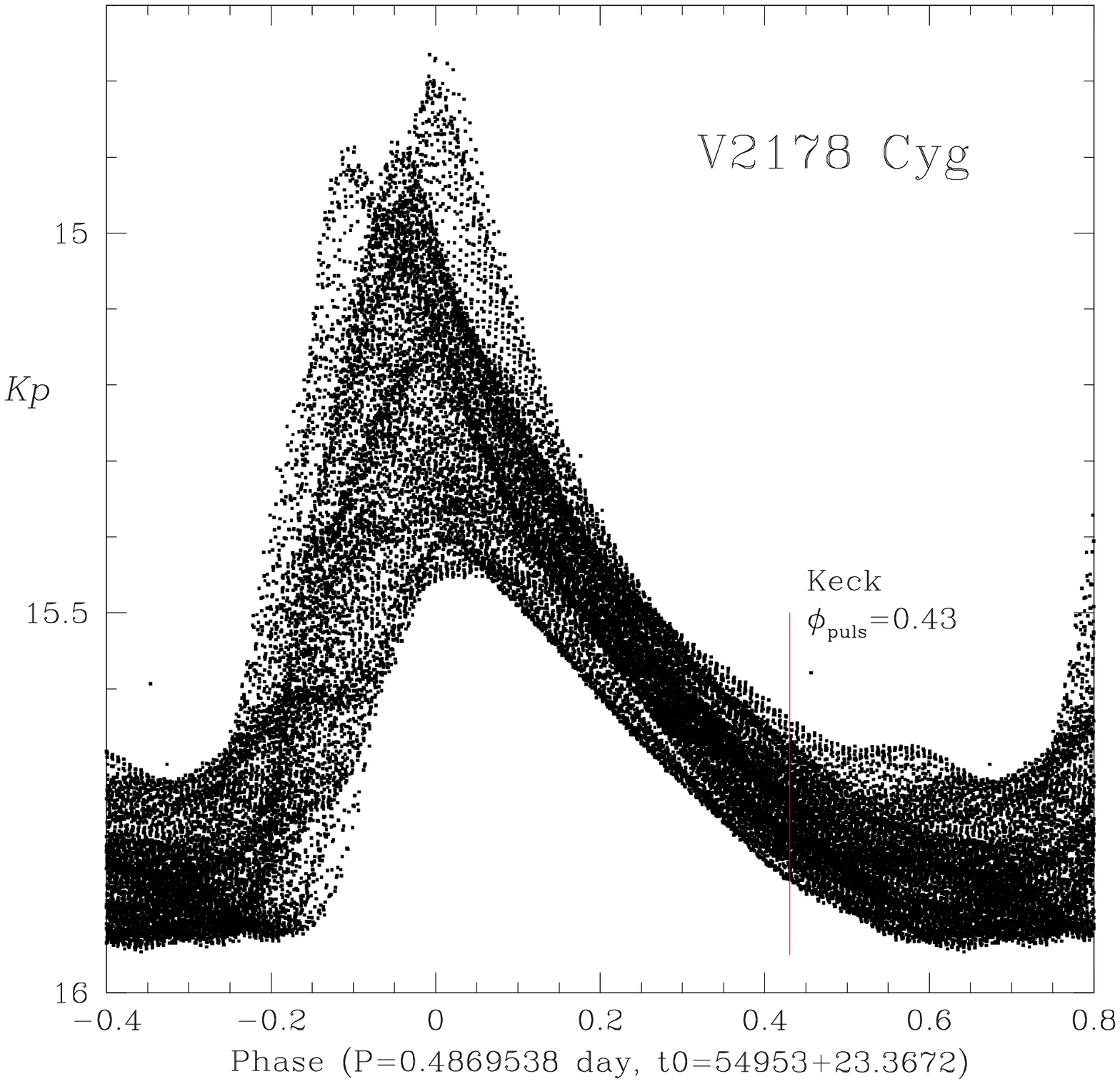}
\plottwo{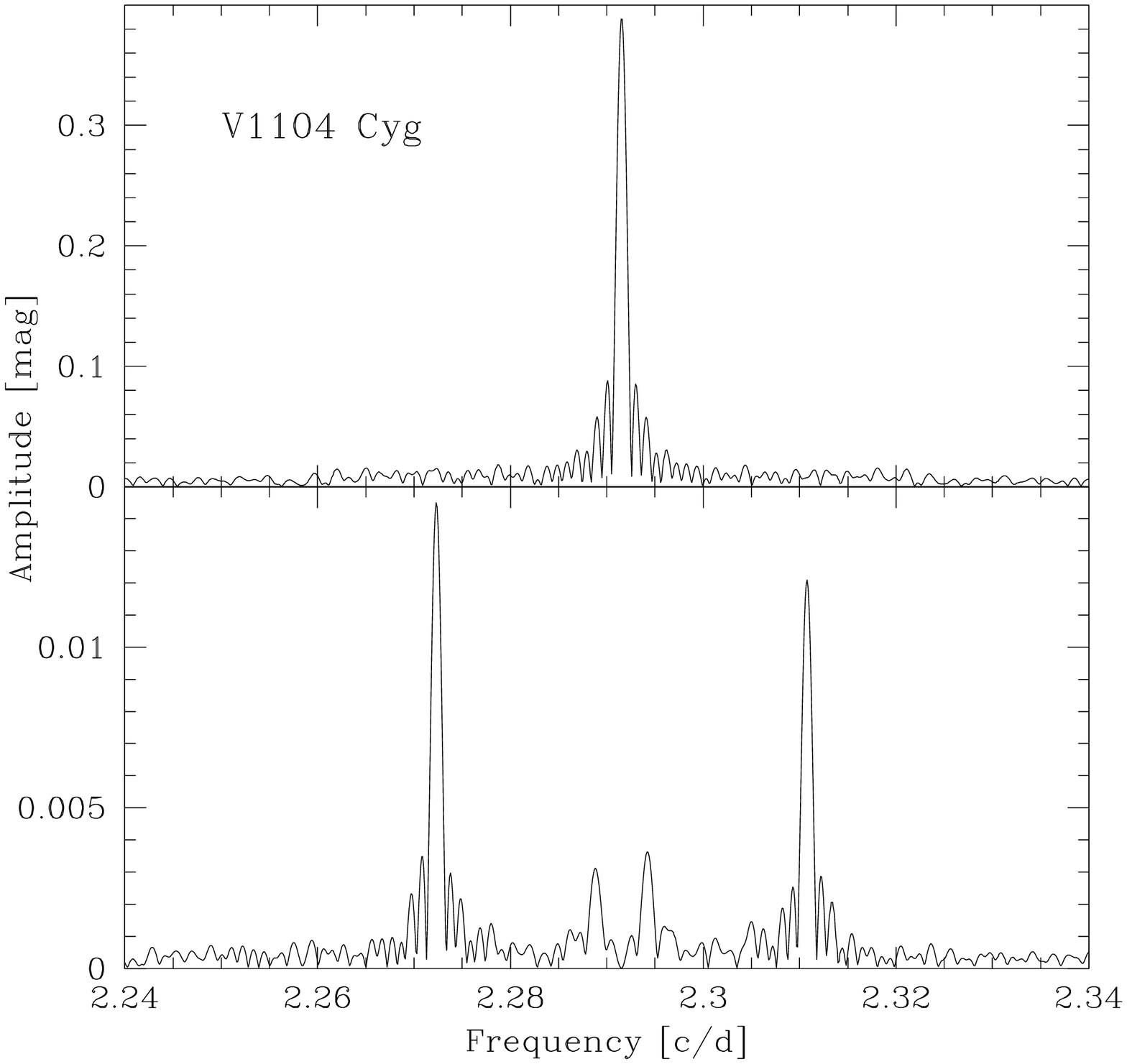}{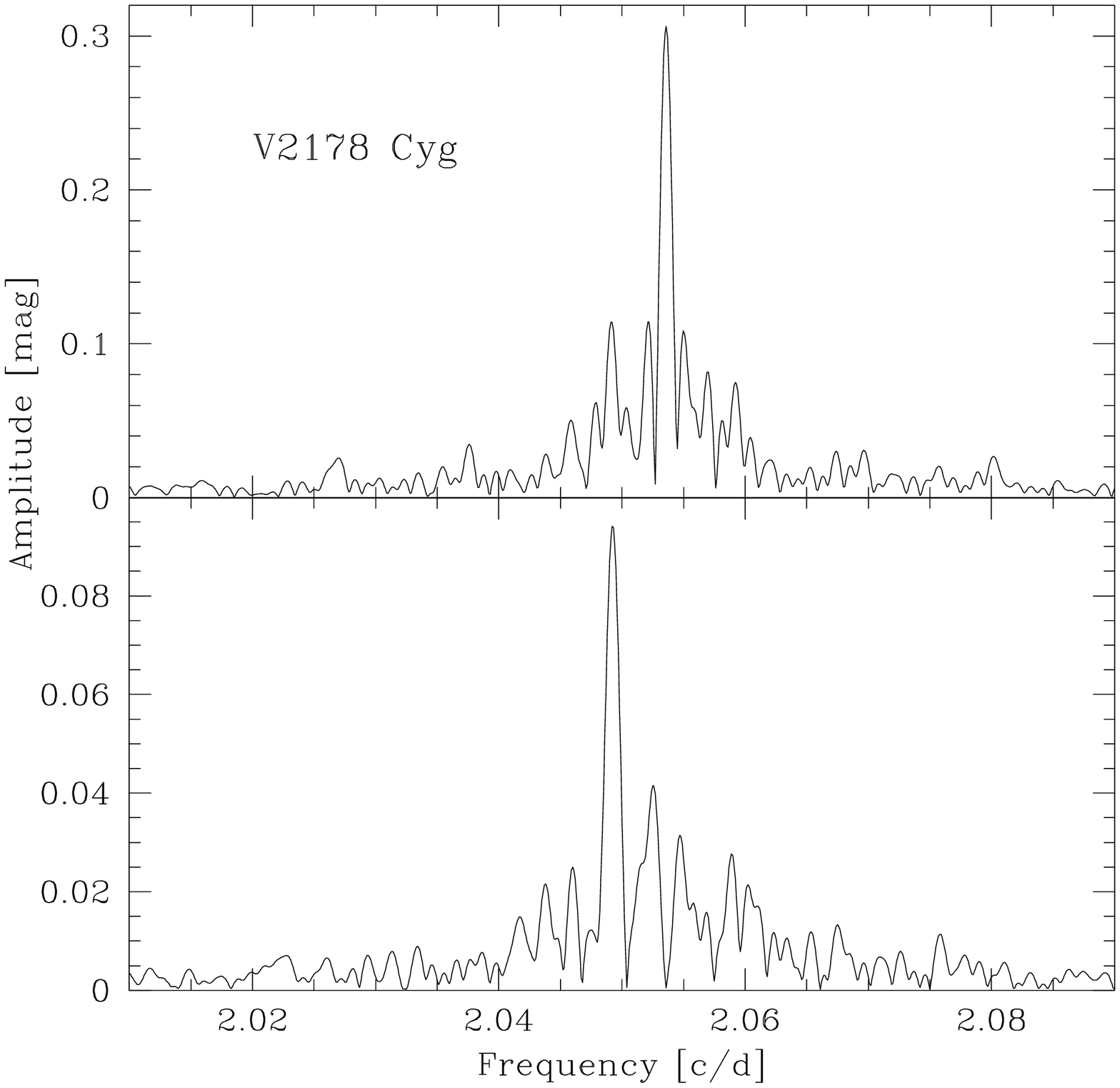} 
\plottwo{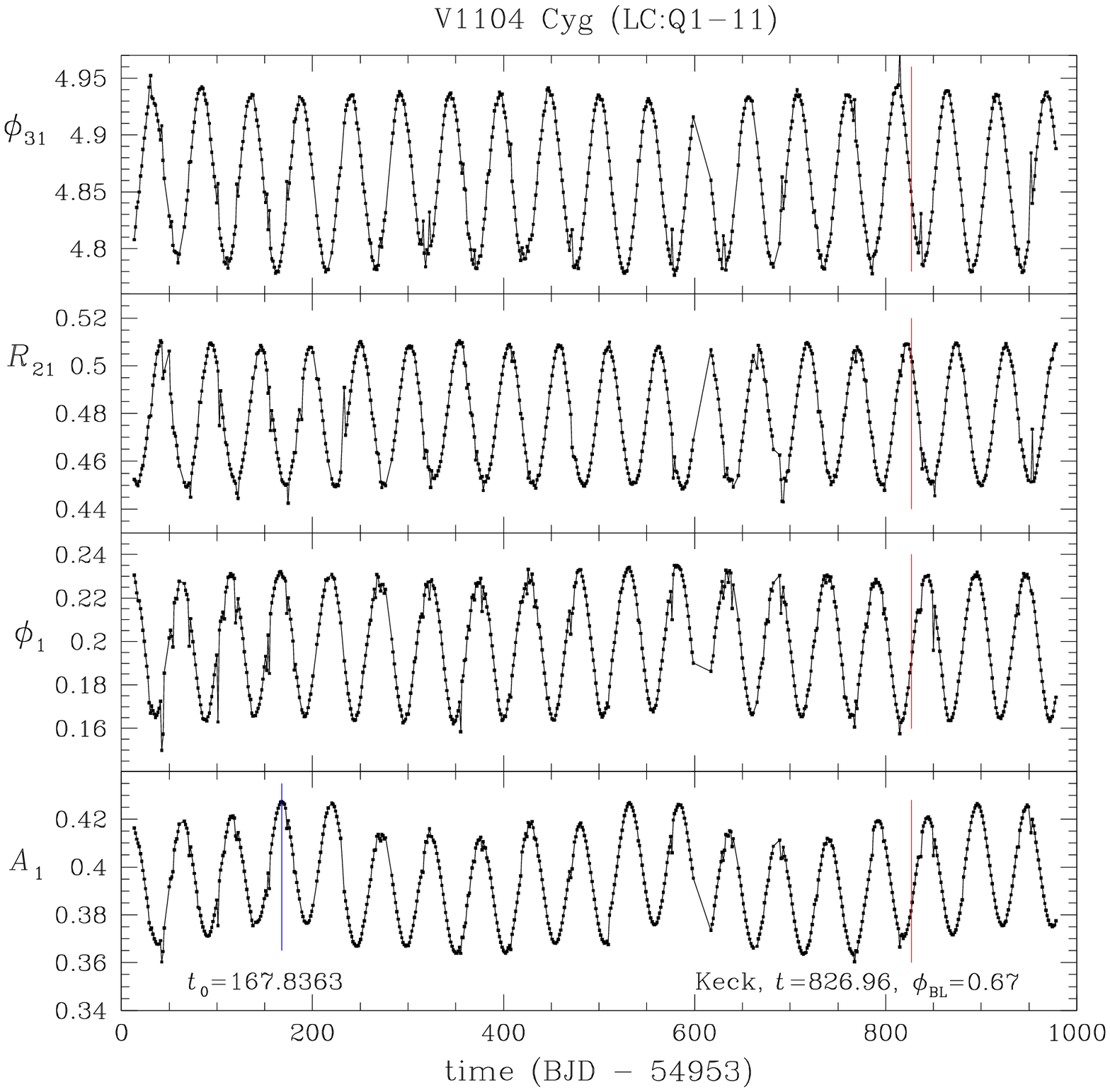}{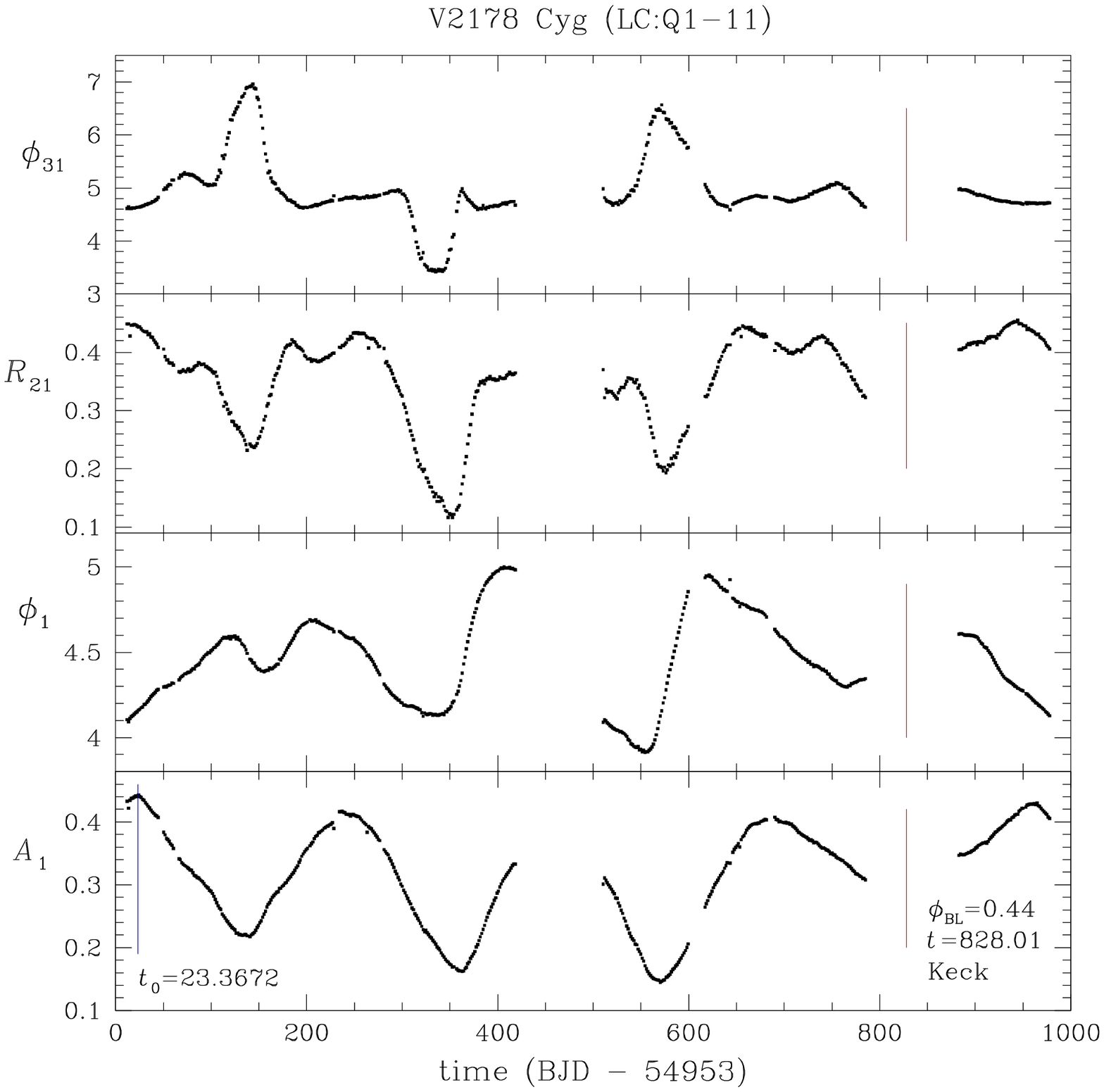}
\caption{({\bf Top})  Phased light curves for V1104~Cyg (KIC\,12155928, left) and V2178~Cyg (KIC\,3864443, right) 
- the mean pulsation phases of the
Keck spectroscopic observations are indicated by a vertical line and labelled;   
({\bf Middle}) Power spectra,  where the upper panel shows the Fourier transform in the vicinity 
of the primary frequency, and the lower panel shows the Fourier transform after prewhitening with the primary frequency and 
its harmonics.   V1104~Cyg is characterized by a symmetric triplet (RR0-BL2) and V2178~Cyg by an possible asymmetric doublet (RR0-BL1?). 
({\bf Bottom})  Time-series graphs for four Fourier parameters - the vertical lines indicate the assumed times of maximum amplitude
and light, $t_0$, and the mean times, $t$, and Blazhko phases, $\phi_{\rm BL}$, at which the Keck observations were made. 
 A color version of this figure is given in the on-line version of the paper.
 }
\label{fig:V1104Cyg_V2178Cyg}
\end{figure*}

{\bf Fig.\,3}  illustrates the time- and frequency-domain methods used here. 
V1104~Cyg (left panels) is a typical Blazhko star having relatively-low 
amplitude modulation (although much larger than that of KIC\,11125706 and V838~Cyg) and a $P_{\rm BL}\sim53$\,d (18 cycles in $\sim$950\,d);
and V2178~Cyg (right panels) exhibits extreme amplitude modulation (similar to but not as extreme as V445~Lyr) with 
$P_{\rm BL}\sim235$\,d (four cycles in $\sim$940\,d).
The top panel for each cluster shows the folded light curves phased with the mean pulsation period (Table~1), where
amplitude variations and phase shifting (indicative of the frequency modulation) are clearly seen.  
The middle panels show portions of the power spectra for frequencies near 
the primary pulsational frequency before (upper graph) and after prewhitening with the primary frequency and its harmonics (lower graph).
And the bottom multipanels are time series plots similar to those shown in Fig.~2.

The V1104~Cyg light curve (top left panel of Fig.3) is derived from the Q9 short cadence photometry (138540 data points), and the power spectra
and time series plots are based on  the Q1-Q11 long cadence photometry (43341 points).
A nonlinear least squares fit to the long cadence data gave an estimated primary frequency  
of $f_0 = 2.29155411$($\pm5$)\,c/d (radial fundamental mode), corresponding to pulsation period $P = 0.43638507$($\pm5$)\,d. 
The secondary frequencies are symmetric about $f_0$ and occur at
$f_0 - f_{\rm BL} = 2.2723$ c/d and $f_0 + f_{\rm BL} = 2.3108$ c/d, from which we derive the Blazhko frequency, $f_{\rm BL}=0.01925$ c/d 
and $P_{\rm BL} = 51.995 \pm 0.005$\,d, and classify the star as type RR0-BL2. 
In the time series plots the variations of  all four Fourier parameters are seen to be nearly sinusoidal, with an
estimated mean Blazhko period of 52.003\,d. 
The agreement with the $P_{\rm BL}$ derived from the Fourier transform is excellent, and we adopt the mean of the two values, $51.999 \pm 0.005$ days.
The flatness of the $\phi_1$ graph lends supports to the derived $P_{\rm puls}$.  An upwards sloping line
would have resulted if the assumed period had been longer; in this sense the $\phi_1$ panels are analogous to O-C diagrams.    
Finally, the $A_1$ panel shows amplitude maxima around $t=200$, 550 and\phn900 days, corresponding to an additional period around 1~year; it is quite
possible that this is an artifact of the pre-processing procedure.

The V2178\,Cyg light curve (top right panel of Fig.\,3) is derived from the long cadence photometry from Q1-Q11 (34994 data points).
The primary frequency, $f_0= 2.053586 \pm0.000004$\,c/d, corresponds to $P_{\rm puls} = 0.486954 \pm0.000004$\,d.
After prewhitening with this frequency and its harmonics, a single strong sidepeak is seen (mag$\sim$0.09), accompanied by several
weaker sidepeaks (mag$\sim$0.02-0.04).  Thus we have what
appears to be an asymmetric doublet pattern, and a classification of RR0-BL1.
However, this star, and to a lesser extent V354~Lyr which has an uncertain Blazhko type owing to its long $P_{\rm BL}$,
are the only such possible RR0-BL1 types in our sample and it is possible that both are BL2 stars with very asymmetric sidelobes.  
Such borderline classifications are quite common:  Alcock {\it et al.} (2003) found that ``160 out of the 400 BL1 stars [in the
MACHO LMC sample] could be classified as BL2''. 
In any case, the Blazhko period that follows from the frequency difference, $f_0 - f_{\rm BL} = 0.00433$ c/d, is $P_{\rm BL} = 234 \pm 10 $ days.  
The time-series plots are unusual and in general the variations are seen to be considerably more complex than for V1104 Cyg 
(the gaps in the time-series plots near $t\sim$450 and 800 are due to no observations having been made in Q6 and in Q10).
The alternating up-down bumps seen in the $\phi_{\rm 31}^s$ panel is, to our knowledge, unique among Blazhko stars.     

Diagrams similar to those plotted in Figs.\,2 and 3 were constructed for all the program stars.
While each of the stars seems to be of considerable interest in its own right, many of the findings are 
only indirectly pertinent to the subject of the metal abundances.  
Where the diagrams are most relevant is in the adopted mean $\phi_{\rm 31}^s$ values that go into the 
photometric [Fe/H] analysis (see $\S5$), in the pulsation phases that are related to the effective temperatures (and are used to calculate $\gamma$-velocities), 
and in the Blazhko phases that affect luminosities and effective temperatures needed for deriving 
spectroscopic [Fe/H] values.  Detailed discussions of our findings concerning modulation characteristics 
of the individual Blazhko stars will be presented elsewhere.

{\bf Table~3} gives the estimated Blazhko periods and types of the 16 {\it Kepler}-field Blazhko stars.   
The $P_{\rm BL}$ estimates were derived by fitting sine functions to time series plots of the amplitudes (see bottom panels of Figs.\,2 and 3), 
and from power spectra of the raw and prewhitened photometry, with both methods producing nearly identical estimates.   
The final $P_{\rm BL}$ values range from 27.667$\pm$0.001 days (V783~Cyg) to 723$\pm$12 days (V354~Lyr), 
with an average $\langle P_{\rm BL} \rangle = 111$ days.  
It is somewhat surprising that none of the stars have Blazhko periods shorter than 27 days, the median $P_{\rm BL}$ for 
the 14 Blazhko stars discovered in the Konkoly Blazhko Survey (see Table 2 of Jurcsik {\it et al.} 2009c).  
None of the stars has $P_{\rm BL}$ as long as the  
four OGLE-III Galactic Bulge stars which have  $P_{\rm BL}$ up to $\sim$3000 days (see Fig.5 of Soszynski {\it et al.} 2011 for light curves).    
V353~Lyr seems to exhibit doubly-periodic modulation with periods of $71.64\pm0.06$\,d and $132.6\pm0.7$\,d. 

For those stars with  $P_{\rm BL}$$<$100\,d and in common with B10 our estimates of $P_{\rm BL}$ agree well with the values given
in their Table~2.  The estimated uncertainties from our analyses are considerably smaller (by factors $\sim$10-100 times) owing to the much longer time baseline and the inclusion of the
short cadence photometry.  The longer baseline also allowed us to derive Blazhko periods for those stars with amplitude modulations
that are occuring on time scales longer than 100 days.  For example, we derived the following estimates of $P_{\rm BL}$ 
for V2178~Cyg, V808~Cyg, V354~Lyr, V450~Lyr:  234$\pm$10, 92.14$\pm$0.01, 723$\pm$12 and 123.7$\pm$1.6 days, respectively.   Table~3 also includes  
the first estimates of  $P_{\rm BL}$ for  three stars not in the B10 list: KIC\,7257008, KIC\,9973633 and V838~Cyg.

\begin{deluxetable}{llrcl}
\tabletypesize{\scriptsize}
\tablewidth{0pt}
\tablecaption{Blazhko Periods, Phases and Types }
\label{tab:Table2}
\tablehead{
\colhead{Star} & \colhead{ $P_{\rm BL}$ [days] }  & \colhead{$t_0$(BL)} & \colhead{ $\phi_{\rm BL}$ }  & \colhead{Type}   \\
\colhead{(1)}  & \colhead{(2)}                   & \colhead{(3)}       & \colhead{(4)}                & \colhead{ (5) }   }  
\startdata
V2178~Cyg   & 234$\pm$10        &  23.3672\phn & 0.44 & BL1?  \\
V808~Cyg    &  92.14$\pm$0.01   &  17.2834\phn & 0.82 & BL2   \\
V783~Cyg    &  27.667$\pm$0.001 &  22.5439\phn & 0.93 & BL2   \\ 
V354~Lyr    & 723$\pm$12        & 292.1590\phn & 0.74 & unc.  \\  
V445~Lyr    &  54$\pm$1         & 207.5957\phn & 0.32 & BL2   \\
RR~Lyrae    &  39.20$\pm$0.10   & 325.2263\phn & 0.77 & BL2   \\
KIC\,7257008 &  39.56$\pm$0.08   & 805.5859\phn & \phn-& BL2   \\        
V355~Lyr    &  31.05$\pm$0.04   & 171.7072\phn & 0.64 & BL2   \\ 
V450~Lyr    & 123.7$\pm$1.6     &  43.3226\phn & 0.34 & BL2   \\ 
V353~Lyr    &  [71.6, 132.6]    & 129.6820\phn & 0.74 & BL2x2 \\ 
V366~Lyr    &  62.84$\pm$0.03   & 373.1915\phn & 0.24 & BL2   \\ 
V360~Lyr    &  52.07$\pm$0.02   &  35.9332\phn & 0.17 & BL2   \\ 
KIC\,9973633 &  73.0$\pm$0.6     & 827.3655\phn & 0.99 & BL2   \\ 
V838~Cyg    & [54,64,47]        & 854.9302\phn & 0.82 & BL2x3 \\ 
KIC\,11125706&  40.23$\pm$0.03   &  28.0658\phn & 0.76 & BL2   \\ 
V1104~Cyg   & 51.999$\pm$0.005  & 167.8363\phn & 0.67 & BL2   
\enddata
\tablecomments{The columns contain:  
(1) star name; 
(2) Blazhko period (=period of amplitude modulation);
(3) time of maximum amplitude (=BJD$-$2454953); 
(4) Blazhko phase at the mid-time of our spectroscopic observations;           
(5) RR0 Blazhko type, as defined by Alcock {\it et al.} (2000, 2003); the type for V354~Lyr is uncertain owing to its
long $P_{\rm BL}$.  }
\end{deluxetable}
\begin{figure*}
\epsscale{1.09}
\plottwo{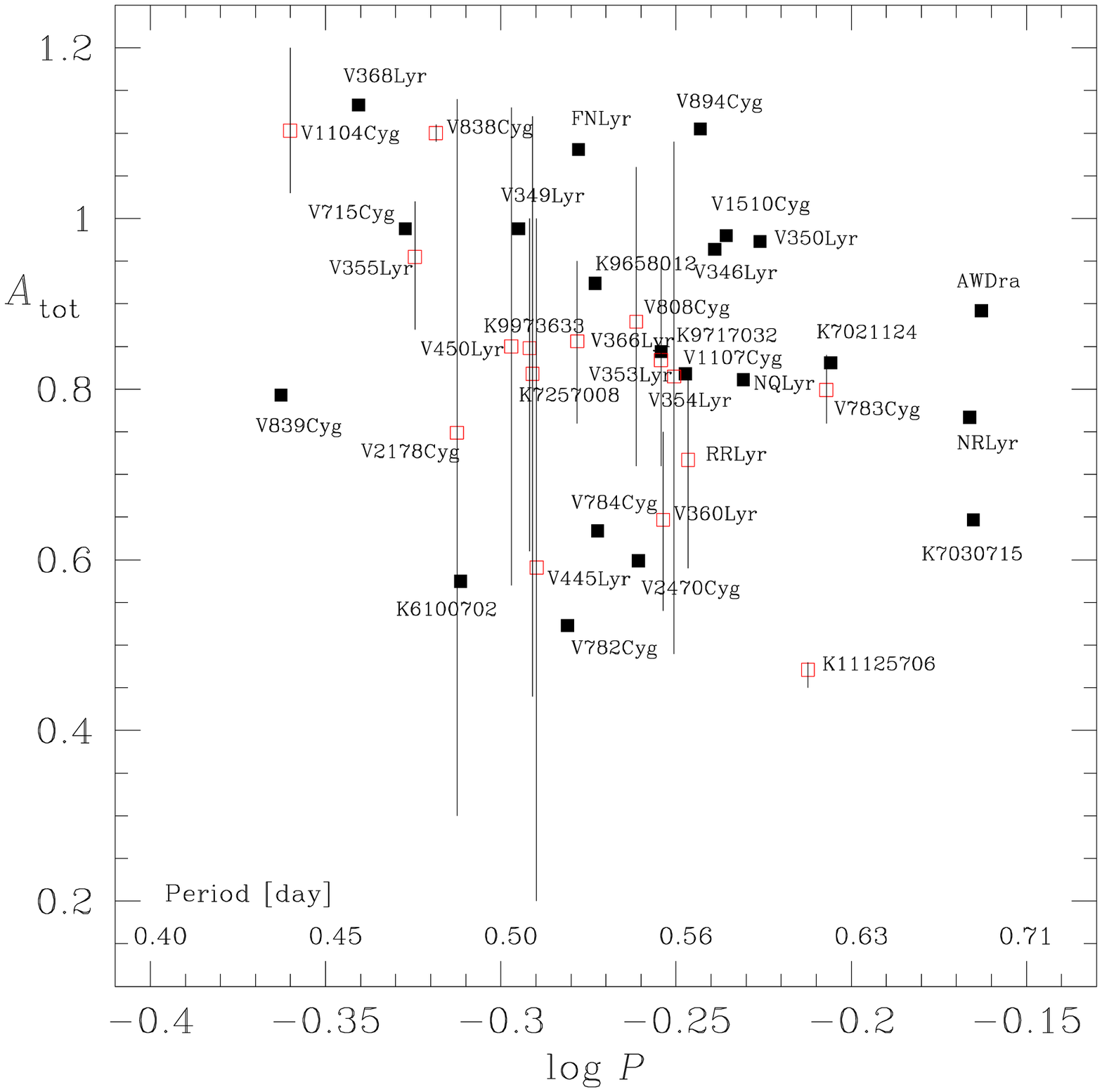}{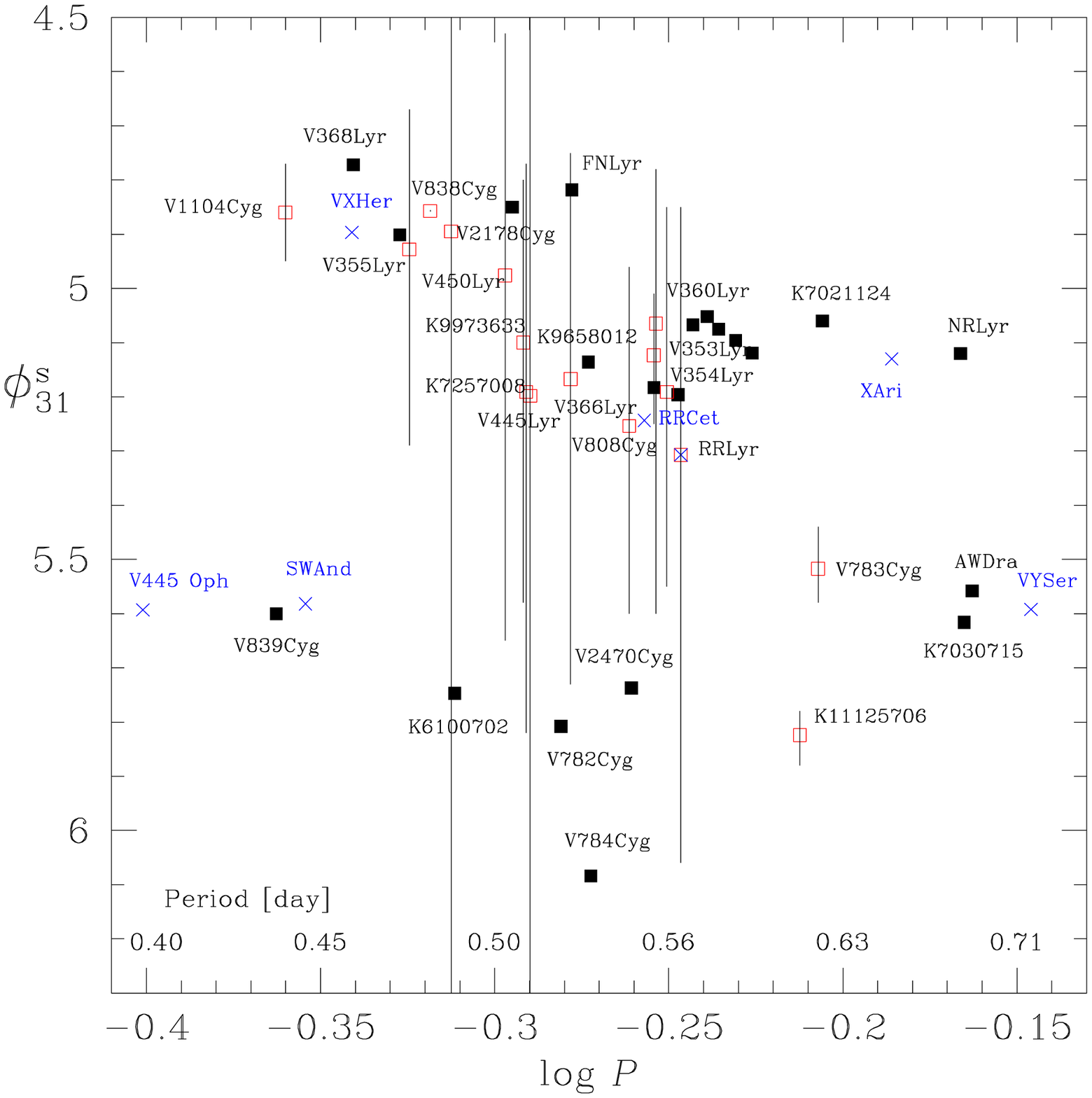}
\caption {Period-amplitude diagram (left) and period-$\phi_{\rm 31}^s$ diagram (right) for the non-Blazhko (black squares) and Blazhko (red open squares) RRab stars in the {\it Kepler} field.
Also represented in the right panel are several Keck spectroscopic standard stars (blue crosses).
The periods are pulsation periods, the $A_{\rm tot}$ are on the {\it Kp}-system, and the Fourier phase parameters $\phi_{\rm 31}^s$ are from sine-series Fourier decomposition of the {\it Kepler} photometry. 
The $\phi_{\rm 31}^s$ axis has been reversed to emphasize the similarity of the two diagrams.
For the Blazhko stars the vertical lines indicate the measured ranges in their $A_{\rm tot}$ and $\phi_{\rm 31}^s$ values (see Table~2),
and the (red open) boxes are plotted at the mean $\phi_{\rm 31}^s$ values.  A color version of this figure is given in the on-line version of the paper.
  }
\label{fig:PA}
\end{figure*}

\subsubsection{Blazhko Phases}

The Blazhko phase at which the spectra were taken, $\phi_{\rm BL}$,
and the adopted time of zero phase for the Blazhko cycles, $t_0$(BL), assumed to be when the pulsation amplitude was largest,
are also given in Table~3.   
The $t_0$(BL) are identical to the values given in Table~1, but are more conveniently expressed to correspond with the 
time-series graphs (see Figs.2-3). For a given star the Blazhko phases were necessarily random.   

If the $P_{\rm BL}$ and $A_{\rm tot}$ were time invariant then any time of maximum amplitude could have been chosen as the zero-point for the Blazhko variations.
However, this is not the case (see Table~2) since many of the Blazhko stars also have multiple Blazhko frequencies and complex power spectra.
For such stars the $\phi_{\rm BL}$ depend on many factors, and in these cases local maxima before and after the times of the observed
spectra were identified and these were used to compute the Blazhko phase. 

The time series plots shown in Figs.\,2-3 serve to illustrate the $\phi_{\rm BL}$ determinations. 
Four spectra of V838~Cyg were taken at CFHT, between BJD-54953=460.79 and 460.82, with a mid-time of 460.80.
The times of maximum amplitude immediately preceeding and following 461 are at 415 and 472, respectively.  Thus $\phi_{\rm BL} = 0.82$ for V838~Cyg.
V1104~Cyg and V2178~Cyg were both observed at Keck, with mid-times of the observations at 826.96 and 828.01, respectively.
Since the times of maximum amplitude immediately before and after 826.96~d occur at 791.1 and 844.8 the $\phi_{\rm BL}$ for V1104~Cyg is 0.67.
A similar calculation made for V2178~Cyg gave $\phi_{\rm BL}= 0.44$.

\subsection{Period-Amplitude and Period-$\phi_{\rm 31}$ Diagrams}

The early studies by Oosterhoff (1939, 1944), Arp (1955) and Preston (1959) established that in the period-amplitude diagram for RR~Lyrae stars there is a 
separation by metal abundance, with metal-poor RRab stars tending to have longer pulsation periods at a given amplitude than metal-rich RRab stars.
This effect has been used to better understand the Oosterhoff dichotomy and to derive metallicities for RR Lyrae stars found in different environments, 
such as in globular clusters (Sandage 1981, 1990, 2004; Cacciari {\it et al.} 2005), in different fields of our Galaxy, and in other galaxies.   
The effect is now known to be due to the lower metal abundance stars having greater luminosities and therefore longer periods 
(see Sandage 2010; Bono {\it et al.} 2007), a result which is consistent with the recent Warsaw convective pulsation hydro-models (see Fig.14 of N11).  

In the 1980's Simon and his collaborators introduced Fourier decomposition techniques for describing the shapes of 
RR~Lyr light curves, and established useful correlations between various Fourier parameters and [Fe/H], mass, luminosity, and other  physical parameters 
(Simon \& Lee 1981; Simon \& Teays 1982; Simon 1985, 1988; Simon \& Clement 1993).
Since the mid-1990's Kov\'acs and Jurcsik and their collaborators (Kov\'acs \& Zsoldos 1995; Jurcsik \& Kov\'acs 1996, hereafter JK96; 
Kov\'acs \& Jurcsik 1996;  Kov\'acs \& Walker 2001; Kov\'acs 2005) have expanded upon these Fourier ideas and have produced useful empirical 
equations that describe the relationships between RRab light curve parameters and physical characteristics.  In particular,
JK96 derived a  $P$-$\phi_{\rm 31}^s$-[Fe/H] relation (their equation 3) which is often used to estimate the metallicities of RRab stars.  More recently, 
Morgan {\it et al.} (2007, hereafter M07) derived analogous equations for RRc stars.

In {\bf Fig.~4} the  $\log P$-$A_{\rm tot}$  and  $\log P$-$\phi_{\rm 31}^s$  diagrams are plotted for the {\it Kepler}-field RRab stars.  
The pulsation periods,  $A_{\rm tot}$({\it Kp}) and $\phi_{\rm 31}^s$({\it Kp}) are the new values given in Table~1,  and
the diagrams include both non-Blazhko (black squares) and Blazhko stars (red open squares).
By reversing the direction of the $\phi_{\rm 31}^s$ axis one observes directly the 
similarity of the two diagrams (see Sandage 2004) and that those stars with the largest amplitudes tend to have the most 
asymmetric light curves, {\it i.e.}, the smallest $\phi_{31}^s$ values.
In both diagrams all but a few of the most crowded points have been labelled with the star names. 
For the two most extreme Blazhko stars, V445~Lyr and V2178~Cyg,  the $\phi_{\rm 31}^s$ ranges exceed the range of the plotted y-axis.
The RRc stars in the {\it Kepler} field are off-scale to the left of both graphs, with small amplitudes, large $\phi_{\rm 31}^s$ values, and 
periods shorter than 0.4\,d. 

Several trends are apparent in Fig.\,4, the most obvious being the similarity of the two diagrams. 
There is also an apparent separation of the stars into two groups,  discriminated  better by  $P_{\rm puls}$ and $\phi_{\rm 31}^s$ than by  $P_{\rm puls}$ and $A_{\rm tot}$. 
Most of the stars are observed to have $A_{\rm tot}$$>$0.8\,mag and $\phi_{31}^s$$<$5.4\,radians.  According to the metallicity correlations mentioned above these are 
metal-poor stars with [Fe/H]$<$$-1.0$\,dex. 
The other group, those shorter-period stars with lower amplitudes and higher $\phi_{\rm 31}^s$ values, are expected to be more metal-rich, with [Fe/H]$>$$-1.0$\,dex.
For a fixed metallicity as the pulsation period increases the stars tend to have smaller amplitudes and more sinusoidal light curves ({\it i.e.}, smaller $\phi_{31}^s$ values).

When all the stars in a given sample are suspected of having the same or similar metallicities (e.g., in globular clusters with narrow red giant branches),
or when there exist independent spectroscopic metal abundances for the sample stars, then diagonal lines can be fit to the observational 
data (e.g., Sandage 2004, Cacciari {\it et al.} 2005, N11).  Such diagonal trends are evident in the Fig.\,4 $\log P$-$\phi_{\rm 31}^s$ diagram,
especially when the spectroscopic standard stars (blue crosses) are included in the diagram. This diagram 
also suggests that the new non-Blazhko star V839~Cyg, and the low-amplitude-modulation star KIC\,11125706 (Fig.\,2) are metal rich.  
Indeed, we show below that V839~Cyg is metal-rich, with  [Fe/H]=$-0.05\pm0.14$ dex;  however, the
CFHT spectra of KIC\,11125706, with  [Fe/H]=$-1.09\pm0.08$ dex (see Table~7), indicate that it probably belongs to the metal-poor group.

The Blazhko stars require special consideration.  They are represented in Fig.\,4 by their mean pulsation periods and their mean $A_{\rm tot}$ or $\phi_{\rm 31}^s$ values (red open squares), 
with vertical lines spanning the measured ranges of $A_{\rm tot}$ and $\phi_{\rm 31}^s$.   
If metal abundances are derived by substituting the mean $\phi_{\rm 31}^s$ values given in Table~1 into the JK96 $P$-$\phi_{\rm 31}^s$-[Fe/H] relation, then the resulting
photometric [Fe/H] values suggest that almost all of the Blazhko variables are metal-poor; indeed, this is supported by the spectroscopic metallicities (next section), and 
runs contrary to the suggestion by Moskalik \& Poretti (2003) that the incidence of Blazhko variables increases with [Fe/H]. 
Also,  since the Blazhko stars appear to have, at a given period, lower mean $A_{\rm tot}$ values 
and higher mean $\phi_{31}^s$ values  than the non-Blazhko stars,  it appears that the {\it Kepler} Blazhko stars must be more metal-rich,
on average, than the non-Blazhko stars. 
The validity of these conclusions depends on the applicability of the JK96 formula, and on the appropriateness of the assumed $\phi_{\rm 31}^s$ values,
the latter being particularly true for the most extreme Blazhko stars (V445~Lyr, V2178~Cyg) where the $\phi_{\rm 31}^s$ show large modulations.  
These topics will be discussed further in $\S5$.

\section{CFHT AND KECK SPECTROSCOPY}

High-resolution spectroscopic observations were made of the  {\it Kepler}-field RR~Lyrae stars using the 
Canada-France-Hawaii 3.6-m telescope (CFHT) and the Keck-I 10-m telescope (W.M. Keck Observatory), both 
located on Mauna Kea, Hawai'i.  A total of 123 spectra were taken of 42 stars and in most cases the derived 
spectroscopic metal abundances are the first available estimates.  

Since a primary goal of this study  was  derivation of [Fe/H]$_{\rm spec}$ values,  spectra 
were taken at pulsation phases between $\sim$0.2 and $\sim$0.5, at the changeover from outflow to infall.   
Away from these phases the spectra are 
increasingly affected by velocity-gradients in the atmosphere and shock waves that broaden the spectral lines 
(see Preston 2009, 2011; K11;  For {\it et al.} 2011).
The pulsation phases  for the mid-times of the individual spectra are given in Tables 4-5 below and 
were derived   using the updated pulsation periods and $t_0$ values that were calculated using 
Q0-Q11 photometry.  
Nearly $80\%$ of the spectra were acquired at phases between 0.25 and 0.55 (see upper panel of {\bf Fig.\,5}).  
The overall median and mean $\phi_{\rm puls}$ are 0.35 and 0.38, respectively, with $\sigma=0.13$;  the 
mean $\phi_{\rm puls}$ for the 54 Keck spectra is larger than that for the 68 CFHT spectra (0.48 vs. 0.30).

\begin{figure}
\figurenum{5}
\epsscale{1.0}
\plotone{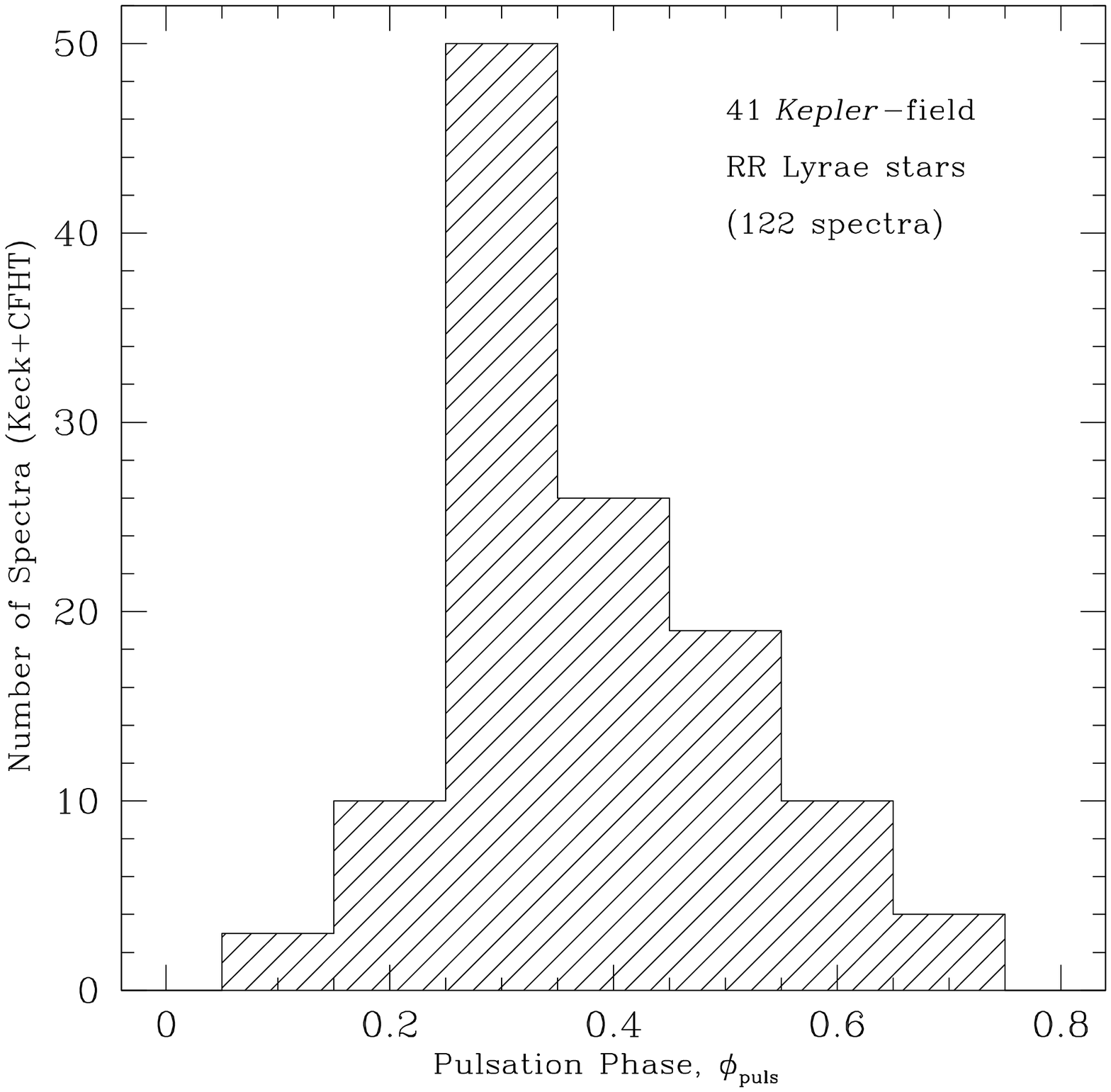}
\plotone{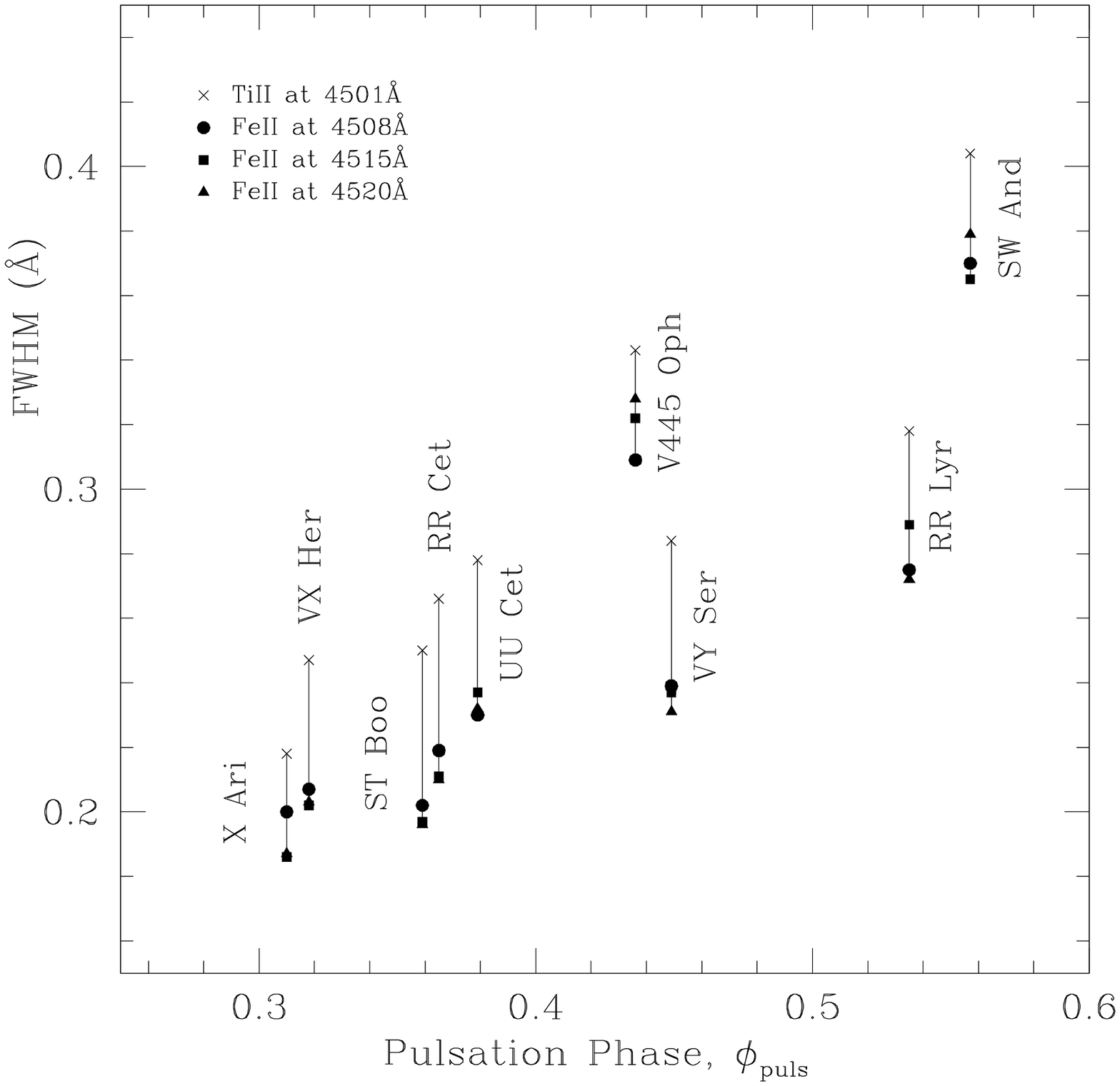}
\caption{(Top) Histogram of the observed pulsation phases for the 41 {\it Kepler}-field RR~Lyrae stars, 
based on the 122 spectra taken with the CFHT 3.6-m and Keck-I 10-m telescopes.
(Bottom) Full-width half minimum (FWHM) values for four spectral lines near 450~nm, as a function of the pulsation phase.  
The points plotted are for the bright RR~Lyrae standard stars observed with Keck, and the FWHM values were measured
using the IRAF `splot' routine.  The narrowest lines
are seen to occur at phases between 0.3 and 0.4, and 
for a given phase the three Fe~II lines (at 450.8, 451.5 and 452.0 nm) are narrower than the Ti~II line (at 450.1 nm). 
 }
\label{fig:PulsationPhaseHistogram}
\end{figure}

\subsection{CFHT 3.6-m ESPaDOnS Spectra}

\begin{deluxetable*}{lcccclccrr}
\tabletypesize{\scriptsize}
\tablewidth{0pt}
\tablecaption{CFHT ESPaDOnS Spectra of {\it Kepler} RR Lyrae Stars 
\label{tab:Table4}}
\tablehead{
\colhead{Star}  & 
\colhead{Spectrum} & 
\colhead{$t_{\rm exp}$} & 
\colhead{Observation} & 
\colhead{S/N} &
\colhead{FWHM} &
\colhead{AM} &
\colhead{HJD\,(mid)}   &
\colhead{$\phi_{puls}$ } & 
\colhead{$RV$ (fxcor)} \\
\colhead{    } &  \colhead{No.}  &  \colhead{ [\,s\,] }  &  \colhead{date (UT)}  &
\colhead{  } &  \colhead{ [\,m\AA\,] } &  \colhead{   }  & \colhead{2400000+} & \colhead{ }   &  \colhead{ [km/s] }   \\
\colhead{(1)} & \colhead{(2)} & \colhead{(3)} & 
\colhead{(4)} & \colhead{(5)} & \colhead{(6)} & \colhead{(7)} & \colhead{(8)} &
\colhead{(9)} & \colhead{(10)} }   
\startdata 
KIC\,6100702  & 1218711 &\phn900   & 2010-07-30 &23& 308\,(10) &1.13& 55407.9264 & 0.227     & $ -54.16\,(20)$  \\  
              & 1218712 &\phn900   & 2010-07-30 &22& 269\,(08) &1.17& 55407.9406 & 0.256     & $ -52.38\,(17)$ \\
              & 1218713 &\phn900   & 2010-07-30 &21& 283\,(06) &1.20& 55407.9515 & 0.279     & $ -50.54\,(17)$  \\
              & 1218714 &\phn900   & 2010-07-30 &20& 249\,(20) &1.24& 55407.9625 & 0.301     & $ -47.85\,(17)$  \\
AW\,Dra       & 1218715 &\phn900   & 2010-07-30 &32& 226\,(23) &1.34& 55407.9754 & 0.287     & $-197.85\,(18)$  \\  
              & 1218716 &\phn900   & 2010-07-30 &31& 210\,(16) &1.34& 55407.9863 & 0.302     & $-196.25\,(21)$  \\
              & 1218717 &\phn900   & 2010-07-30 &32& 200\,(03) &1.44& 55407.9973 & 0.318     & $-194.65\,(21)$  \\
              & 1218718 &\phn900   & 2010-07-30 &31& 198\,(08) &1.51& 55408.0082 & 0.334     & $-193.70\,(22)$  \\
FN\,Lyr       & 1218719 &\phn900   & 2010-07-30 &32& 226\,(21) &1.49& 55408.0204 & 0.265     & $-261.20\,(23)$   \\  
              & 1218720 &\phn900   & 2010-07-30 &33& 203\,(03) &1.57& 55408.0313 & 0.286     & $-258.57\,(21)$  \\
              & 1218721 &\phn900   & 2010-07-30 &32& 236\,(12) &1.68& 55408.0423 & 0.307     & $-256.85\,(23)$  \\
              & 1218722 &\phn900   & 2010-07-30 &29& 217\,(18) &1.81& 55408.0533 & 0.328     & $-254.67\,(16)$ \\
V894\,Cyg     & 1219066 &\phn900   & 2010-08-01 &27& 178\,(18) &1.63& 55409.7372 & 0.354     & $-234.14\,(21)$ \\  
              & 1219067 &\phn900   & 2010-08-01 &29& 214\,(28) &1.54& 55409.7482 & 0.383     & $-232.47\,(18)$ \\
              & 1219068 &\phn900   & 2010-08-01 &29& 189\,(18) &1.47& 55409.7591 & 0.402     & $-230.38\,(18)$ \\
              & 1219069 &\phn900   & 2010-08-01 &29& 196\,(11) &1.40& 55409.7702 & 0.422     & $-228.90\,(21)$\\  
NQ\,Lyr       & 1219070 &\phn900   & 2010-08-01 &27& 198\,(09) &1.23& 55409.7836 & 0.301     & $ -72.45\,(18)$  \\  
              & 1219071 &\phn900   & 2010-08-01 &27& 183\,(34) &1.19& 55409.7945 & 0.320     & $ -70.47\,(15)$ \\  
              & 1219072 &\phn900   & 2010-08-01 &27& 185\,(12) &1.16& 55409.8054 & 0.339     & $ -68.43\,(22)$ \\
              & 1219073 &\phn900   & 2010-08-01 &26& 152\,(12) &1.14& 55409.8164 & 0.357     & $ -66.97\,(13)$ \\
NR\,Lyr       & 1219477 &\phn900   & 2010-08-02 &35& 222\,(28) &1.24& 55410.9703 & 0.274     & $-124.30\,(26)$  \\  
              & 1219478 &\phn900   & 2010-08-02 &35& 147\,(24) &1.29& 55410.9813 & 0.290     & $-123.20\,(19)$  \\
              & 1219479 &\phn900   & 2010-08-02 &36& 129\,(05) &1.34& 55410.9922 & 0.306     & $-121.82\,(15)$  \\
              & 1219480 &\phn900   & 2010-08-02 &35& 094\,(04) &1.41& 55411.0032 & 0.322     & $-119.42\,(28)$  \\
V355\,Lyr     & 1219876 &   1200   & 2010-08-05 &17& 309\,(05) &1.37& 55413.7353 & 0.144     & $-240.73\,(16)$  \\  
              & 1219877 &   1200   & 2010-08-05 &16& 161\,(34) &1.29& 55413.7497 & 0.175     & $-238.03\,(34)$  \\
              & 1219878 &   1200   & 2010-08-05 &15& 195\,(50) &1.24& 55413.7641 & 0.205     & $-233.78\,(17)$  \\
              & 1219879 &   1200   & 2010-08-05 &16& 219\,(13) &1.19& 55413.7785 & 0.236     & $-231.82\,(11)$  \\
V838\,Cyg     & 1219880 &\phn900   & 2010-08-05 &14& 173\,(34) &1.23& 55413.7924 & 0.328     & $-209.05\,(21)$  \\  
              & 1219881 &\phn900   & 2010-08-05 &14& 231\,(47) &1.21& 55413.8033 & 0.351     & $-206.91\,(25)$  \\
              & 1219882 &\phn900   & 2010-08-05 &13& 140\,(60) &1.18& 55413.8143 & 0.374     & $-205.73\,(26)$  \\
              & 1219883 &\phn900   & 2010-08-05 &13& 166\,(06) &1.17& 55413.8252 & 0.396     & $-202.29\,(19)$  \\
V2470\,Cyg    & 1220136 &\phn900   & 2010-08-06 &23& 250\,(13) &1.37& 55414.7545 & 0.289     & $ -57.06\,(18)$  \\  
              & 1220137 &\phn900   & 2010-08-06 &23& 225\,(17) &1.32& 55414.7654 & 0.309     & $ -55.16\,(22)$  \\
              & 1220138 &\phn900   & 2010-08-06 &23& 255\,(07) &1.28& 55414.7764 & 0.329     & $ -53.38\,(22)$  \\
              & 1220139 &\phn900   & 2010-08-06 &23& 257\,(16) &1.24& 55414.7874 & 0.349     & $ -51.59\,(25)$  \\
KIC\,11125706 & 1220140 &\phn900   & 2010-08-06 &57& 228\,(04) &1.19& 55414.8033 & 0.308     & $ -69.42\,(20)$  \\  
              & 1220141 &\phn900   & 2010-08-06 &59& 229\,(07) &1.17& 55414.8142 & 0.326     & $ -68.08\,(22)$  \\
              & 1220142 &\phn900   & 2010-08-06 &59& 223\,(05) &1.16& 55414.8251 & 0.343     & $ -66.58\,(22)$  \\
              & 1220143 &\phn900   & 2010-08-06 &61& 236\,(03) &1.15& 55414.8361 & 0.361     & $ -65.19\,(22)$  \\
KIC\,3868420\tablenotemark{\dag} &1220147&\phn600 &2010-08-06&96&359\,(09)&1.06& 55414.9016 & 0.173 & $ -27.63\,(16)$  \\
KIC\,9453114  & 1220148 &\phn900   & 2010-08-06 &29& 283\,(13) &1.16& 55414.9120 & 0.241     & $-152.09\,(20)$ \\  
(RRc)         & 1220149 &\phn900   & 2010-08-06 &30& 320\,(60) &1.19& 55414.9230 & 0.271     & $-153.53\,(22)$ \\
              & 1220150 &\phn900   & 2010-08-06 &30& 375\,(42) &1.22& 55414.9339 & 0.301     & $-152.31\,(19)$ \\
              & 1220151 &\phn900   & 2010-08-06 &29& 267\,(42) &1.25& 55414.9449 & 0.331     & $-151.70\,(15)$ \\
V1104\,Cyg    & 1259187 &   1200   & 2010-11-16 &19& 316\,(57) &1.33& 55516.6860 & 0.117     & $-325.49\,(63)$ \\  
              & 1259188 &   1200   & 2010-11-16 &20& 294\,(13) &1.39& 55516.7004 & 0.150     & $-320.87\,(69)$ \\
              & 1259189 &   1200   & 2010-11-16 &19& 287\,(49) &1.46& 55516.7148 & 0.184     & $-318.40\,(24)$ \\
              & 1259190 &   1200   & 2010-11-16 &15& 380\,(10) &1.55& 55516.7293 & 0.217     & $-313.73\,(38)$ \\
V1510\,Cyg    & 1259908 &   1200   & 2010-11-18 &20& 250\,(50) &1.19& 55518.6891 & 0.316     & $-350.35\,(40)$ \\  
              & 1259909 &   1200   & 2010-11-18 &20& 220\,(40) &1.24& 55518.7035 & 0.341     & $-348.22\,(18)$ \\ 
              & 1259910 &   1200   & 2010-11-18 &19& 250\,(50) &1.29& 55518.7179 & 0.366     & $-345.90\,(53)$ \\
              & 1259911 &   1200   & 2010-11-18 &19& \hfil - \hfil &1.36& 55518.7323 & 0.391     & $-343.55\,(20)$ \\
V783\,Cyg     & 1261460 &   1200   & 2010-11-25 &17& 200\,(60) &1.24& 55525.6867 & 0.324     & $-197.26\,(27)$ \\ 
              & 1261461 &   1200   & 2010-11-25 &16& 250\,(100)&1.30& 55525.7011 & 0.347     & $-195.38\,(22)$ \\
              & 1261462 &   1200   & 2010-11-25 &16& 250\,(70) &1.38& 55525.7156 & 0.370     & $-193.16\,(22)$ \\
              & 1261463 &   1200   & 2010-11-25 &17& 250\,(100)&1.48& 55525.7300 & 0.394     & $-190.49\,(36)$ \\
KIC\,8832417  & 1262172 &\phn900   & 2010-11-27 &30& 243\,(74) &1.59& 55527.7291 & 0.223     & $-28.52\,(15)$ \\  
(RRc)         & 1262173 &\phn900   & 2010-11-27 &29& 293\,(09) &1.69& 55527.7400 & 0.267     & $-27.07\,(16)$ \\
              & 1262174 &\phn900   & 2010-11-27 &28& 324\,(33) &1.82& 55527.7509 & 0.311     & $-25.86\,(17)$ \\
              & 1262175 &\phn900   & 2010-11-27 &27& 351\,(28) &1.96& 55527.7619 & 0.355     & $-25.41\,(51)$ \\
V808\,Cyg     & 1265907 &   1200   & 2010-12-15 &\phn9& \hfil - \hfil &1.71& 55545.6950 & 0.256     & {\hfil - \hfil} \\  
              & 1265908 &   1200   & 2010-12-15 &\phn8& \hfil - \hfil &1.89& 55545.7094 & 0.283     & {\hfil - \hfil} \\
              & 1265909 &   1200   & 2010-12-15 &\phn8& \hfil - \hfil &2.14& 55545.7238 & 0.309     & {\hfil - \hfil} \\
              & 1265910 &   1200   & 2010-12-15 &\phn7& \hfil - \hfil &2.47& 55545.7382 & 0.335     & {\hfil - \hfil} \\
KIC\,7030715  & 1266124 &\phn900   & 2010-12-16 &23& 330\,(140)&1.89& 55546.6890 & 0.238     & $-373\,(8)$\phn\phn   \\   
              & 1266125 &\phn900   & 2010-12-16 &26& 222\,(34) &2.06& 55546.6999 & 0.254     & $-365.41\,(17)$ \\
              & 1266126 &\phn900   & 2010-12-16 &23& 208\,(02) &2.27& 55546.7109 & 0.270     & $-364.41\,(48)$ \\
              & 1266127 &\phn900   & 2010-12-16 &14& \hfil - \hfil &2.53& 55546.7218 & 0.286     & $-361.98\,(62)$ 
\enddata
\tablecomments{The columns contain:  (1)  star name (upper) and metal abundance (lower, from Table 7);  
(2) ESPaDOnS spectrum number;  
(3) exposure time [s];  
(4) Observation date (UT);
(5) Signal-to-noise ratio per CCD pixel, at $\lambda\sim570$\,nm and computed by {\it Upena} at CFHT;  
(6) Average FWHM (full-width at half-minimum) value for the three Fe~II lines at 4508, 4515 and 4520 \AA, where the uncertainty (given in parentheses) is approximated by 
the standard deviation of the mean ({\it e.g.}, {10\,m\AA} for spectrum number 1218711); 
(7) airmass at start of observation; 
(8) heliocentric Julian date at mid-exposure;   
(9) pulsation phase at mid-exposure time;  
(10) heliocentric radial velocity, calculated using the IRAF `fxcor' routine. 
\vskip0.2cm
\dag KIC\,3868420 is either a rare short-period multiperiodic RRc (RRe?) star, or a possible high-amplitude $\delta$\,Scuti (HADS) star; in either 
case our VWA and SME analyses indicate that it is metal rich.}
\end{deluxetable*}



\begin{figure*} \figurenum{6} \epsscale{1.2} \plotone{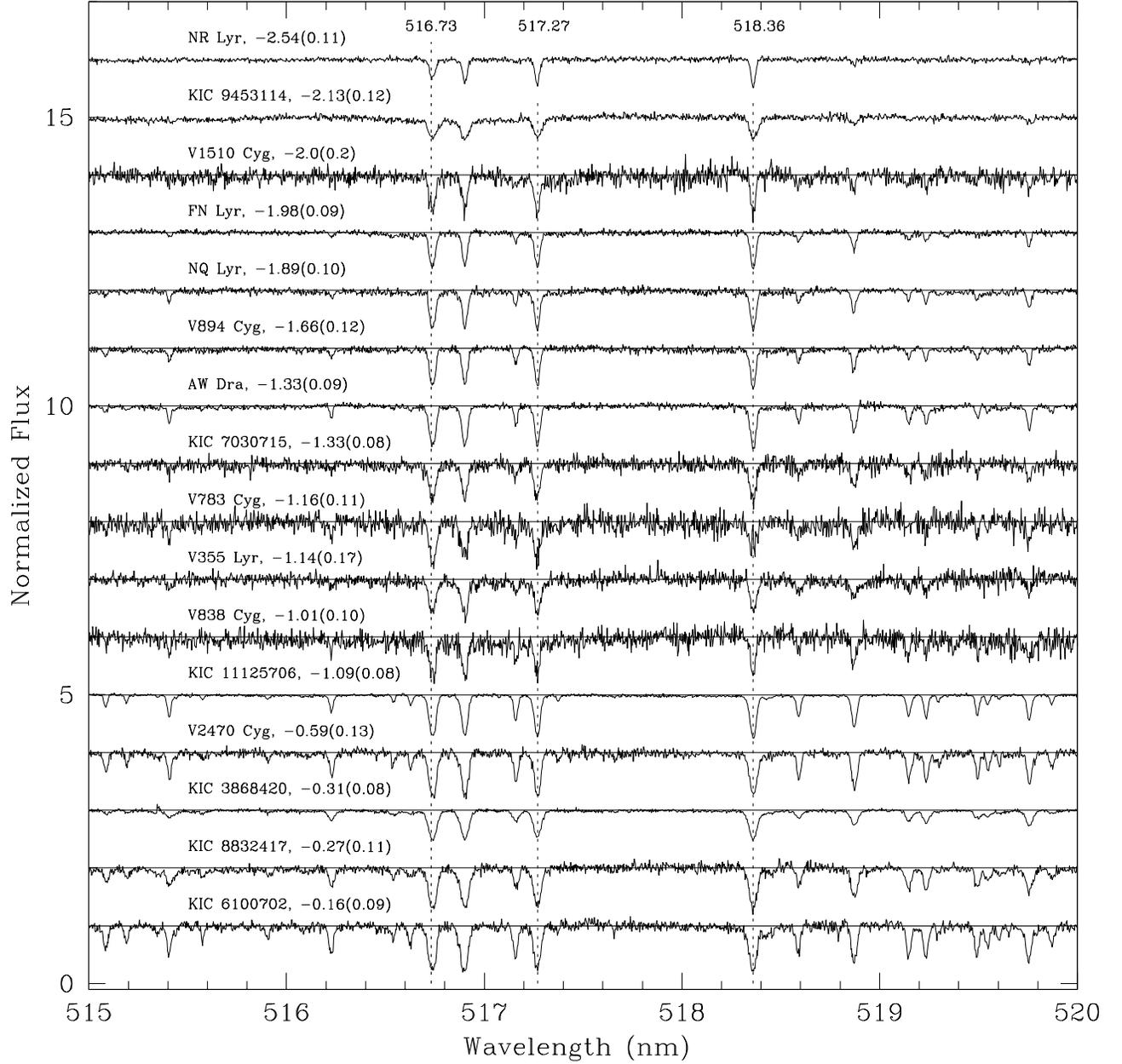} 
\caption{Mosaic showing the 517~nm Mg~I triplet region of the CFHT spectra for the 16 stars for
which metallicities could be derived.  The rest wavelengths of the Mg~I triplet lines are
given at the top of the graph.  The enhanced line strengths of the most metal-rich stars (bottom)
clearly distinguish them from the more metal-poor stars (top).  The probable HADS star, KIC\,3868420,
is third from the bottom.  } 
\label{fig:517nm_MgI_mosaic} \end{figure*}

The CFHT spectra were acquired  
from July-December 2010  using the ESPaDOnS prism cross-dispersed \'echelle spectrograph 
in the `star+sky' non-polarimetric mode.
Two  optical fibers fed the star and sky images from the Cassegrain focus to the Coud\'e focus, then onto a Bowen-Walraven image slicer at the entrance of the 
spectrograph.   
The  resolving power was  $R$\,$=\lambda/\Delta\lambda \sim65000$, which corresponds to a 
spectral resolution of $\Delta \lambda \sim 0.008$ nm at $\lambda=517$\, nm (Mg triplet).
The spectra are spread over 40 grating orders (on an EEV1 CCD chip), with wavelength coverage from 370 to 1048 nm, an 
average reciprocal dispersion  of 0.00318 nm/pixel, and $\sim$3 pixels 
per resolution element.

Eighteen of the brightest RR~Lyrae stars in the {\it Kepler} field were observed at CFHT (Service Observing). 
The total observing time amounted to 20 hours. 
The observations were made  under photometric skies for all but three of the stars (V355~Lyr, V838~Cyg, KIC\,7030715), and 
at various hour angles and airmasses.
The seeing was rarely better than 1 arcsec.
Multiple spectra were taken  for most stars, usually with exposure times of $4\times900$~s; for the 
faintest observed stars  (14.5$<${\it Kp}$<$15.0) the exposure times were increased to 4$\times$1200~s.

The `Upena' software, which uses `Libre-ESpRIT'\footnote{The original `ESpRIT' program is described by Donati {\it et al.} (1997); `Libre-ESpRIT' is the current release 
used at CFHT and documented at http://www.cfht.hawaii.edu/Instruments/Spectro-
scopy/Espadons/Espadons\_esprit.html.
}, was used to pre-process the CFHT spectra -- {\it i.e.}, perform bias subtraction and flat fielding, 
subtract the sky and scattered light, identify the echelle orders and perform optimal image extraction, 
make the wavelength calibrations (using spectra of a Th-Ar lamp), calculate the signal-to-noise ratios (S/N per pixel), and make the necessary heliocentric corrections for the
Earth's motion. 
The individual spectra were normalized and the overlapping \'echelle orders merged using the  VWA `rainbow' widget (Bruntt {\it et al.} 2010a,b). 
For those stars with multiple observations the normalized spectra were co-added to increase the signal-to-noise ratios. 
Details of the individual spectra are given in {\bf Table~4}.  
The S/N per pixel values (column 5) were 
measured at 570\,nm\footnote{The S/N ratio is largest in the interval 550-580 nm, but does not vary by more than
$\sim$20\%  over the interval 500-900 nm.  Since one resolution element is approximately 1.4 pixels, or 2.6 km/s, the S/N  per resolution element are $\sim$1.2 times larger.}
 and range from 7-96, with typical values $\sim$20-30.  The full-width at half-minimum (FWHM) values (column 6)  were measured using the 
Fe\,II lines at 450.8, 451.5 and 452.0 nm, and range from $\sim$100-400\,m\AA\footnote{The 
measured FWHM values are related to the true (or instrument-corrected) values according to 
${\rm FWHM}_{\rm true} = ({\rm FWHM}_{\rm meas}^2 - {\rm FWHM}_{\rm instr}^2)^{1/2} $, where 
for the ESPaDOnS spectra the FWHM$_{\rm instr}$ was measured to be $\sim$2.0\,pixels, or 64\,m\AA.}. 

Co-added spectra for the 16 CFHT stars for which metallicities could be derived are plotted in {\bf Fig.\,6}, where the wavelength range 
515$<$$\lambda$$<$520\,nm contains the green Mg\,I triplet lines.
The spectra have been corrected for Doppler shifts, and, to illustrate the effect of increasing metallicity, are ordered by [Fe/H].
The spectra of V1104~Cyg ({\it Kp}=15.03) and V808~Cyg ({\it Kp}=15.36) were of insufficient quality to permit the derivation of 
metal abundances (but were sufficient for deriving {\it RVs} for V1104~Cyg) and are not shown;  both stars were
re-observed with the Keck telescope.

\begin{deluxetable*}{lcccccccr}
\tabletypesize{\scriptsize}
\tablewidth{0pt}
\tablecaption{Keck-I 10-m HIRES Spectra of Program and Standard Stars 
\label{tab:Table5}}
\tablehead{
\colhead{Star}  & 
\colhead{Spec.} & 
\colhead{$t_{\rm exp}$} & 
\colhead{S/N} &
\colhead{FWHM} &
\colhead{AM} &
\colhead{HJD\,(mid)} & 
\colhead{$\phi_{\rm puls}$ }   & 
\colhead{$RV$\,(fxcor)}  \\
\colhead{  } &
\colhead{No.} &
\colhead{[\,s\,]} &
\colhead{       } &
\colhead{[\,m\AA\,] } &
\colhead{  } &
\colhead{2400000+} &
\colhead{      } &
\colhead{[\,km/s\,] }  \\
\colhead{(1)} & \colhead{(2)} & \colhead{(3)} & 
\colhead{(4)} & \colhead{(5)} & \colhead{(6)} &
\colhead{(7)} & \colhead{(8)} & \colhead{(9)}  } 
\startdata
\multicolumn{9}{c}{(a) 22 {\it Kepler}-field RR\,Lyrae program stars (excluding RR~Lyrae) } \\[0.1cm]
V839~Cyg    & 5951  & 1200    &  51 & 379\,(01) & 1.40 & 55778.7493 & 0.100  &  $ -97.48\,(14)$  \\ 
            & 5952  & 1200    &  50 & 368\,(02) & 1.30 & 55778.7639 & 0.133  &  $ -93.78\,(14)$  \\ 
V360~Lyr    & 5953  & 1200    &  17 & 348\,(11) & 1.24 & 55778.7786 & 0.495  &  $-170.80\,(26)$  \\
            & 5954  & 1200    &  18 & 353\,(04) & 1.20 & 55778.7931 & 0.521  &  $-168.85\,(14)$  \\
            & 5955  & 1200    &  19 & 343\,(15) & 1.16 & 55778.8076 & 0.547  &  $-167.95\,(17)$  \\
V354~Lyr    & 5957  & 1200    &  20 & 272\,(05) & 1.08 & 55778.8347 & 0.354  &  $-207.67\,(14)$  \\
            & 5958  & 1200    &  19 & 327\,(11) & 1.08 & 55778.8492 & 0.380  &  $-205.35\,(14)$  \\
            & 5959  & 1200    &  18 & 282\,(05) & 1.08 & 55778.8637 & 0.405  &  $-202.95\,(17)$  \\
V368~Lyr    & 5960  & 1200    &  15 & 161\,(08) & 1.09 & 55778.8786 & 0.401  &  $-267.41\,(14)$  \\
            & 5961  & 1200    &  11 & 195\,(11) & 1.11 & 55778.8931 & 0.432  &  $-264.59\,(22)$  \\
            & 5962  & 1200    &  12 & 201\,(13) & 1.18 & 55778.9076 & 0.464  &  $-261.43\,(99)$  \\
V353~Lyr    & 5963  & 1200    &\phn9& 199\,(18) & 1.18 & 55778.9220 & 0.415  &  $-199.42\,(53)$  \\
            & 5964  & 1200    &\phn8& 145\,(17) & 1.22 & 55778.9366 & 0.443  &  $-196.91\,(21)$  \\
            & 5965  & 1200    &\phn7& 204\,(51) & 1.27 & 55778.9510 & 0.467  &  $-194.38\,(29)$  \\
            & 5966  & 1200    &\phn7&  -        & 1.33 & 55778.9656 & 0.493  &  $-192.30\,(17)$  \\
            & 6000  & 1200    &  13 & 192\,(07) & 1.39 & 55779.9752 & 0.307  &  $-210.26\,(11)$  \\
            & 6001  & 1200    &  12 & 219\,(35) & 1.48 & 55779.9897 & 0.332  &  $-207.77\,(11)$  \\
V782~Cyg    & 5967  & 1200    &  22 & 277\,(19) & 1.19 & 55778.9814 & 0.418  &  $ -60.97\,(15)$  \\
            & 5968  & 1200    &  22 & 287\,(06) & 1.24 & 55778.9962 & 0.446  &  $ -58.88\,(14)$  \\
V784~Cyg    & 5969  & 1200    &  19 & 306\,(13) & 1.31 & 55779.0128 & 0.462  &  $ -10.31\,(17)$  \\
            & 5970  & 1200    &  19 & 304\,(09) & 1.39 & 55779.0273 & 0.489  &  $  -8.81\,(16)$  \\
V1107~Cyg   & 5983  & 1200    &  15 & 215\,(20) & 1.46 & 55779.7425 & 0.475  &  $-117.07\,(12)$  \\
            & 5984  & \phn800 &  14 & 221\,(27) & 1.38 & 55779.7549 & 0.497  &  $-115.08\,(23)$  \\
KIC\,9973633 & 5985  & 1200    &\phn8&  -        & 1.47 & 55779.7679 & 0.576  &  $-206.84\,(48)$  \\
            & 5986  & 1200    &\phn8&  -        & 1.38 & 55779.7837 & 0.606  &  $-203.70\,(23)$  \\
            & 5987  & 1200    &\phn8&  -        & 1.31 & 55779.7982 & 0.635  &  $-202.93\,(35)$  \\
            & 5988  & 1200    &\phn9&  -        & 1.26 & 55779.8127 & 0.664  &  $-205.54\,(32)$  \\
KIC\,4064484 & 5990  & \phn800 &  37 & 288\,(11) & 1.10 & 55779.8324 & 0.586  &  $-290.79\,(12)$  \\
KIC\,9658012 & 5991  & 1200    &  21 & 369\,(31) & 1.15 & 55779.8457 & 0.499  &  $-312.52\,(10)$  \\
            & 5992  & 1200    &  20 & 403\,(29) & 1.13 & 55779.8602 & 0.527  &  $-310.41\,(09)$  \\
            & 5993  & 1200    &  21 & 423\,(09) & 1.12 & 55779.8748 & 0.554  &  $-309.32\,(11)$  \\
V445~Lyr    & 5996  & 1200    &  14 & 319\,(24) & 1.13 & 55779.9167 & 0.285  &  $-392.79\,(24)$  \\
            & 5997  & 1200    &  14 & 237\,(25) & 1.17 & 55779.9312 & 0.314  &  $-390.40\,(17)$  \\
            & 5998  & 1200    &  14 & 245\,(21) & 1.21 & 55779.9457 & 0.342  &  $-388.18\,(11)$  \\
V1104~Cyg   & 5999  & 1000    &  37 & 174\,(03) & 1.31 & 55779.9599 & 0.420  &  $-293.96\,(10)$  \\
KIC\,5520878 & 6002  & 1000    &  45 & 319\,(02) & 1.47 & 55780.0041 & 0.459  &  $  -0.70\,(29)$  \\
KIC\,9717032 & 6003  & 1200    &  13 & 303\,(19) & 1.46 & 55780.0179 & 0.321  &  $-457.11\,(21)$  \\
            & 6004  & 1200    &  14 & 281\,(31) & 1.57 & 55780.0324 & 0.347  &  $-454.63\,(11)$  \\
V450~Lyr    & 6017  & \phn900 &  11 & 273\,(42) & 1.37 & 55780.7406 & 0.546  &  $-269.48\,(19)$  \\
            & 6018  & 1200    &  13 & 240\,(07) & 1.32 & 55780.7535 & 0.571  &  $-267.65\,(10)$  \\
            & 6019  & 1200    &  16 & 254\,(10) & 1.25 & 55780.7703 & 0.606  &  $-265.84\,(10)$  \\
V366~Lyr    & 6022  & 1200    &  15 & 301\,(01) & 1.13 & 55780.8367 & 0.658  &  $ -68.51\,(16)$  \\
            & 6023  & 1200    &  16 & 357\,(05) & 1.12 & 55780.8512 & 0.685  &  $ -68.47\,(18)$  \\
            & 6024  & 1200    &  16 & 350\,(28) & 1.12 & 55780.8657 & 0.712  &  $ -67.61\,(18)$  \\
V346~Lyr    & 6027  & 1200    &  10 & 293\,(08) & 1.18 & 55780.9138 & 0.619  &  $-272.88\,(14)$  \\  
V808~Cyg    & 6028  & 1200    &  19 & 473\,(16) & 1.10 & 55780.9338 & 0.630  &  $ +32.40\,(15)$  \\
            & 6029  & \phn600 &\phn4&  -        & 1.12 & 55780.9449 & 0.649  &  $ +31.83\,(33)$  \\
V350~Lyr    & 6032  & 1200    &  22 & 283\,(25) & 1.44 & 55780.9775 & 0.523  &  $ -91.41\,(14)$  \\
            & 6033  & 1200    &  23 & 324\,(17) & 1.54 & 55780.9920 & 0.547  &  $ -89.56\,(15)$  \\
V2178~Cyg   & 6034  & 1200    &  26 & 212\,(13) & 1.36 & 55781.0085 & 0.241  &  $-116.89\,(20)$  \\
            & 6035  & 1200    &  26 & 213\,(09) & 1.46 & 55781.0230 & 0.270  &  $-115.04\,(16)$  \\
V715~Cyg    & 6036  & 1200    &  15 & 238\,(17) & 1.57 & 55781.0387 & 0.494  &  $ -78.36\,(18)$  \\
            & 6037  & 1200    &  15 & 229\,(09) & 1.73 & 55781.0532 & 0.525  &  $ -75.87\,(20)$  \\[0.1cm] 
\multicolumn{9}{c}{(b) Nine bright RR~Lyrae stars (including RR~Lyrae) } \\[0.1cm]
ST Boo      & 5950  & \phn600 & 160 & 198\,(02) & 1.04 & 55778.7289 & 0.359  &  $   1.23\,(10)$  \\ 
V445 Oph    & 5956  & \phn600 & 125 & 315\,(05) & 1.24 & 55778.8211 & 0.436  &  $ -11.96\,(13)$  \\ 
RR~Lyr      & 5971  & \phn200 & 324 & 276\,(06) & 1.68 & 55779.0361 & 0.535  &  $ -56.48\,(06)$  \\ 
SW And      & 5972  & \phn300 & 184 & 367\,(06) & 1.04 & 55779.0416 & 0.557  &  $  -4.32\,(07)$  \\ 
X Ari       & 5973  & \phn300 & 168 & 188\,(05) & 1.63 & 55779.0441 & 0.310  &  $ -42.01\,(13)$  \\ 
VX Her      & 5982  & \phn200 & 109 & 202\,(02) & 1.01 & 55779.7287 & 0.318  &  $-381.72\,(20)$  \\ 
VY Ser      & 5989  & \phn300 & 135 & 232\,(03) & 1.40 & 55779.8218 & 0.449  &  $-137.41\,(10)$  \\ 
UU Cet      & 6006  & \phn300 &\phn74&226\,(03) & 1.25 & 55780.0649 & 0.379  &  $-112.28\,(12)$  \\ 
RR Cet      & 6042  & \phn300 & 226 & 212\,(03) & 1.06 & 55781.1104 & 0.365  &  $ -77.06\,(09)$  \\[0.1cm] 
\multicolumn{9}{c}{(c) Three red giants in the globular cluster M92} \\[0.1cm]
M92-XII-34  & 5994  & 1200    &  77 & 178\,(02) & 1.28 & 55779.8894 &  -     &  $-113.86\,(26)$  \\   
            & 5995  & \phn800 &  59 & 173\,(03) & 1.35 & 55779.9017 &  -     &  $-114.23\,(26)$  \\
M92-IV-79   & 6025  & 1200    &  70 & 158\,(04) & 1.27 & 55780.8828 &  -     &  $-120.74\,(13)$  \\
            & 6026  & 1200    &  59 & 156\,(05) & 1.33 & 55780.8975 &  -     &  $-120.92\,(13)$  \\
M92-IV-10   & 6030  & 1200    &  41 & 171\,(06) & 1.78 & 55780.9545 &  -     &  $-120.31\,(19)$  \\
            & 6031  & \phn500 &  39 & 182\,(12) & 1.98 & 55780.9650 &  -     &  $-120.04\,(19)$           
\enddata
\tablecomments{The columns contain:  (1) star name;  (2) spectrum number;  (3) exposure time;  (4) SNR per CCD pixel, at $\lambda\sim517$\,nm (CCD1, 23rd order)
and computed using MAKEE -- for SNR per resolution element multiply the given values by 2.5;   
(5) Measured mean FWHM values for the three Fe~II lines at 4508, 4515 and 4520 \AA, where the uncertainty is approximated by 
the standard deviation of the mean ({\it e.g.}, {1m\AA} for spectrum 5951);  
(6) airmass (1.4 corresponds to a zenith angle of 45$^{\circ}$); 
(7)  heliocentric JD at mid-exposure;   
(8) puls.phase at mid-exposure;  (9) heliocentric RV, derived using `fxcor' - the uncertainties 
are in units of 0.01 km/s.}
\end{deluxetable*}
 
\subsection{Keck-I 10-m HIRES Spectra}

\begin{figure*}
\figurenum{7a}
\epsscale{1.2}
\plotone{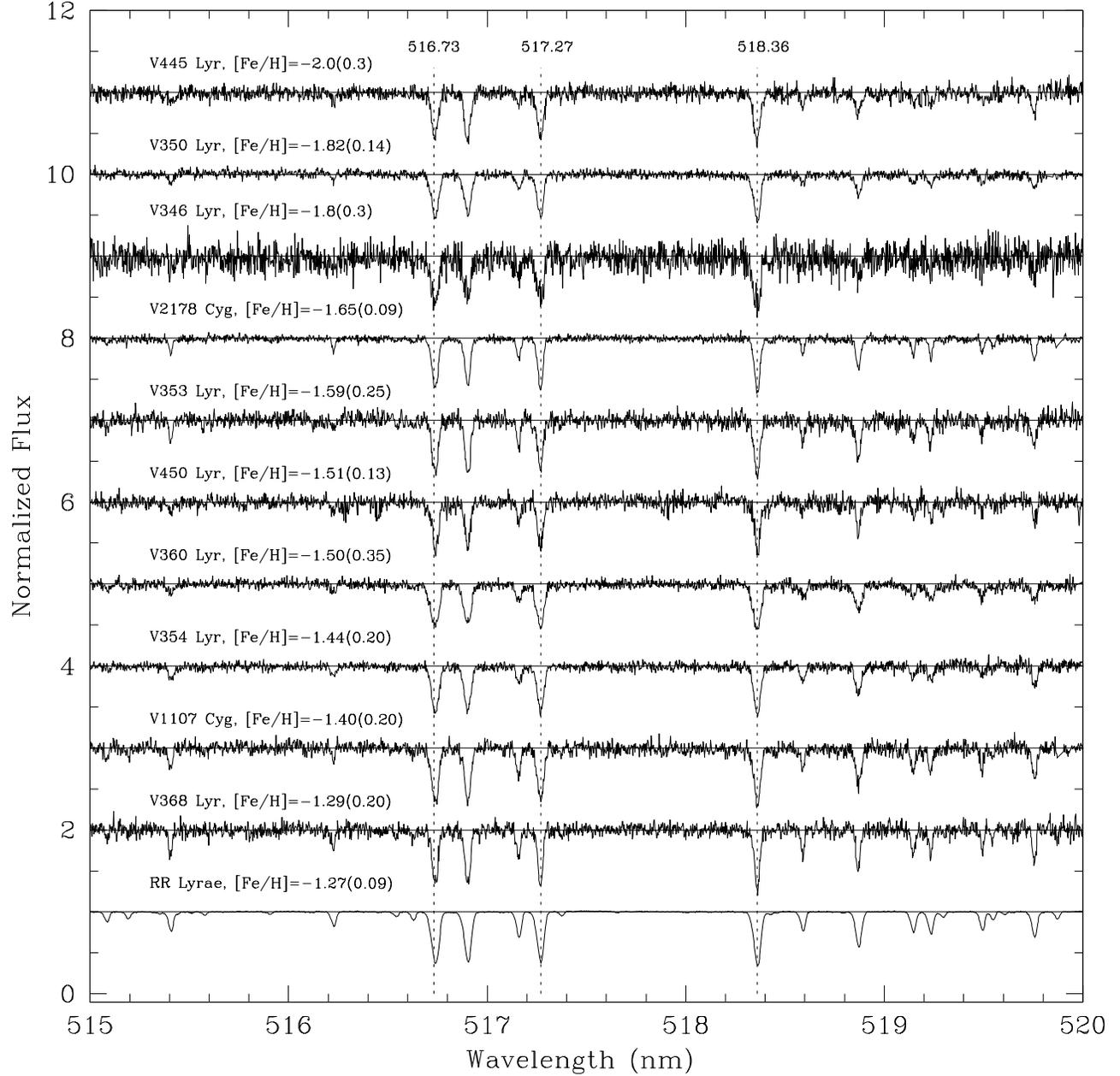}
\caption{Mosaic showing the 517nm Mg\,I triplet region of the Keck spectra for ten {\it
Kepler}-field RRab stars more metal poor than RR~Lyrae itself.  (Note: this region is a small
fraction of the available $\lambda$=389-838 nm range of the HIRES spectra).    The rest wavelengths
of the three Mg~I lines are given at the top of the graph.   As in Fig.6 the stars have been ordered
according to increasing metal abundance.   The Keck spectrum of RR~Lyrae (S/N$\sim$250) is plotted
at the bottom of this mosaic.   } \label{fig:515-520nm_MgI_mosaic1} \end{figure*}

\begin{figure*}
\figurenum{7b}
\epsscale{1.2}
\plotone{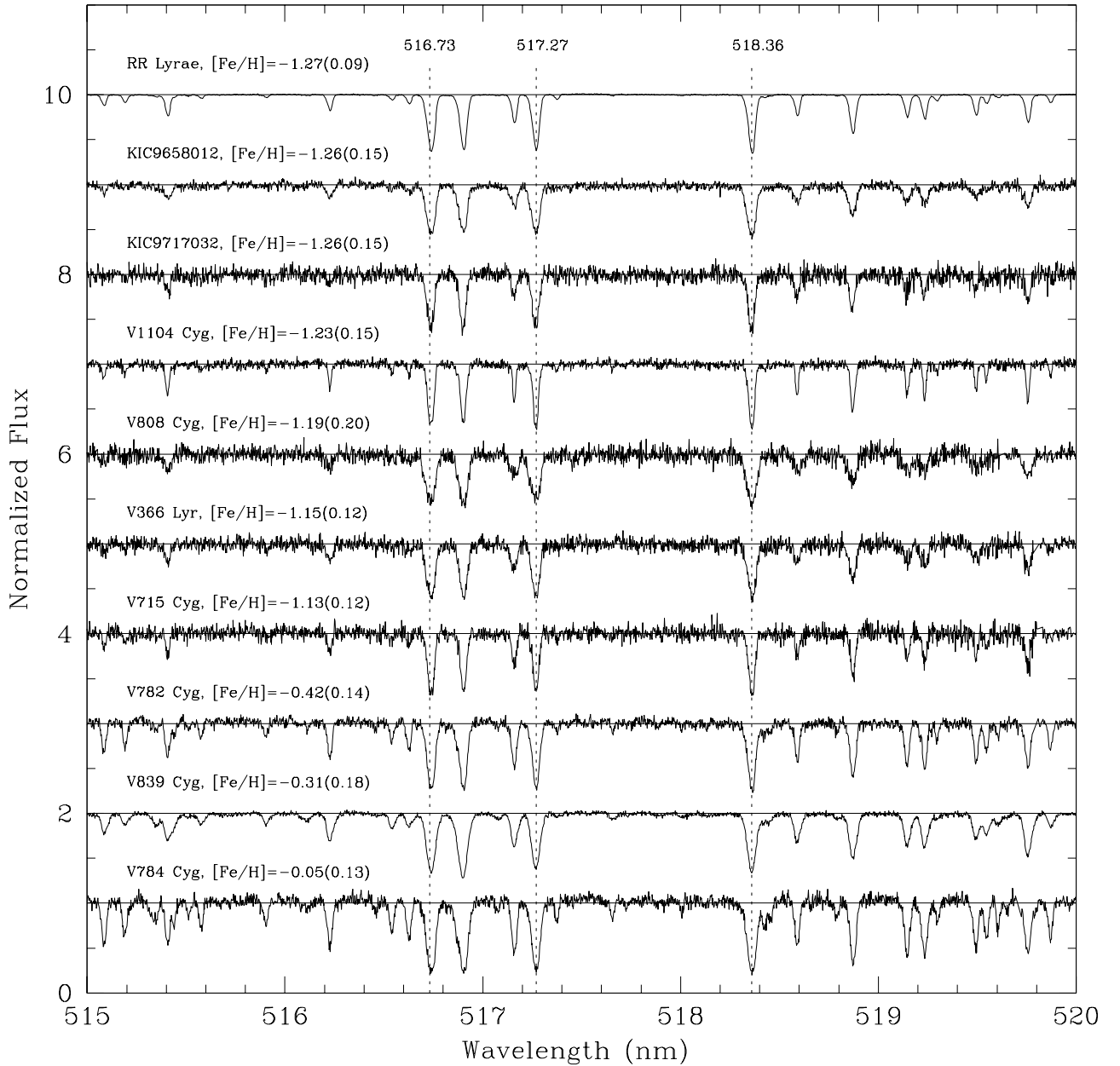}
\caption{Continuation of mosaic, in this case for  nine {\it Kepler}-field RR~Lyr stars more metal rich than RR~Lyrae itself (shown at
top).  The three metal-rich RR~Lyrae stars (bottom) are clearly distinguishable from the more
metal-poor stars. } \label{fig:515-520nm_MgI_mosaic2} \end{figure*}

\begin{figure*}
\figurenum{7c}
\epsscale{1.2}
\plotone{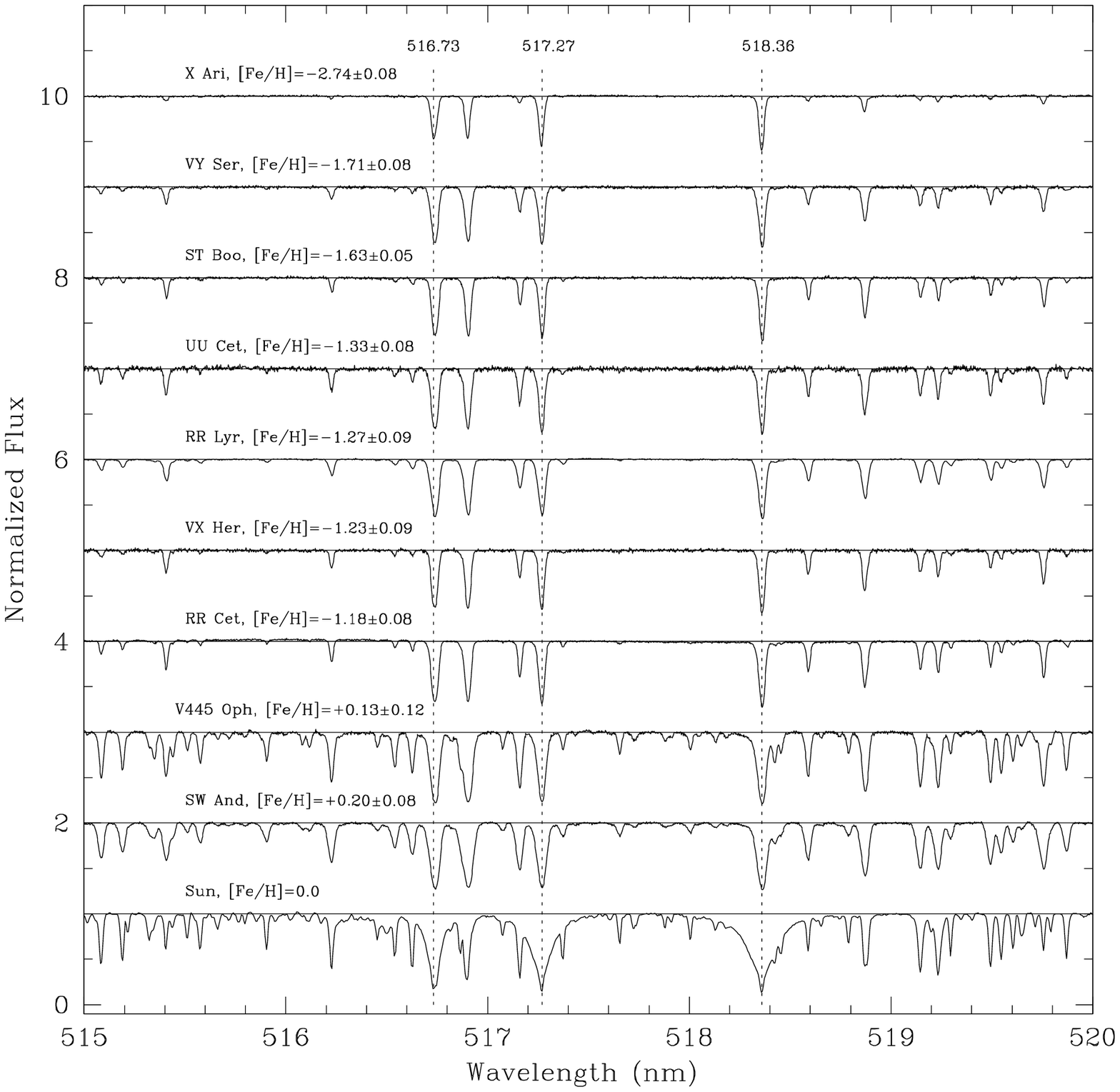}
\caption{Continuation of mosaic, in this case for the nine bright RR~Lyr standard stars
(including RR~Lyrae), sorted in order of increasing metal abundance, and for the Sun.  The  broad Mg~I lines
in the solar spectrum readily show the greater pressure broadening in the Sun than in the RR~Lyr stars. } 
\label{fig:KeckStds} \end{figure*}

The faintest {\it Kepler}-field RR~Lyrae stars were observed on the nights of 2011 August 4/5, 5/6
and 6/7 with the HIRES echelle spectrograph (Vogt {\it et al.} 1994) mounted on the Keck-I 10-m
telescope (see Cohen {\it et al.} 2005a,b; 2009) -- see  {\bf Table~5} for details.
The spectrograph was used in the HIRES-r configuration and thus optimized for long wavelength
observations (QE$>$60\% between 400 and 900 nm).  A slit of width 1.15 arcsec and length 7 arcsec
(C5 decker, no filters) was used, resulting in resolving power $R\sim36,000$ 
($\Delta\lambda\sim0.014$\,nm at 517\,nm) over the range 389-836 nm.
The average reciprocal dispersion is
0.00230 nm/pixel (140\,nm spread over 60882 pixels) and the number of pixels per resolution
element is $\sim$6.  The spectra were spread over a mosaic of three
MIT/Lincoln\,Labs 2048$\times$4096 CCD chips\footnote{The blue chip (CCD1) covers the range 389\,nm
to 529\,nm (24 spectral orders) and includes H$\beta$, H$\gamma$, H$\delta$ and the H and K lines.
The green chip (CCD2) covers the range 538-689 nm (15 orders) and includes H$\alpha$ -- for the last
few orders there are small gaps in the spectrum.  And the red chip (CCD3) covers the range 698-836
nm (9 orders), with between-order gaps ranging from 2-5 nm, the largest gaps occuring at the longest
wavelengths (e.g., at 734-736\,nm, 749-752\,nm, 799-803\,nm, and 817-822 nm).   The largest gaps in
the spectral coverage occur between the chips (9.5\,nm between CCD1 and CCD2, and 9\,nm between CCD2
and CCD3). }.  The $\lambda$-scale was calibrated using Th-Ar lamp exposures
taken at the beginning and end of each night; measurements of the widths of typical emission lines
suggest an instrumental FWHM of $5.6\pm0.3$\,pixels, corresponding to $129\pm7$\,m\AA.  
To reduce phase-smearing and potential cosmic ray problems exposure times for the program stars 
were never longer than 1200~sec, and usually multiple exposures were taken and
co-added.  The Mauna Kea Echelle Extraction (MAKEE\footnote{MAKEE was written by T.A.~Barlow
specifically for the reduction of HIRES spectra, and is publicly available at
http://spider.ipac.caltech.edu/staff/tab/makee/.}) reduction package was used to pre-process the data.  

In addition to the program stars we observed nine bright  RR~Lyr standard stars 
studied by Layden (1994, hereafter L94) and Clementini {\it et al.} (1995, hereafter C95), and
three red giant stars in the very metal-poor globular cluster M92 (mean [Fe/H]$=-2.33$ dex, with little
within-cluster variation - Cohen 2011).   All of the RR\,Lyr standard stars are RRab pulsators and include RR~Lyrae itself ($Kp=7.862$), 
which happens to be located in the {\it Kepler} field and has been the subject of several recent detailed investigations
(K10, K11,  Benedict {\it et al.} 2011).  
The M92 stars served primarily as $RV$ standards and the bright RR~Lyr stars as metallicity standards.
The most metal poor and metal rich standard stars are X\,Ari and SW\,And, respectively, for  which  
[Fe/H]=$-2.74\pm0.09$\,dex and $+0.20\pm0.08$\, dex  (see Table~7), and the  corresponding
{\it Kepler}-field RR~Lyr stars are NR~Lyr and V784~Cyg, for which [Fe/H]=$-2.54 \pm 0.11$ dex and $-0.05\pm0.10$ dex.   
Although  X~Ari and NR~Lyr are very metal
poor, and have metallicities similar to those of the most metal-poor RR~Lyr stars known in the Large
and Small Magellanic Clouds (see Haschke {\it et al.} 2012), they are not the most metal poor stars
known in our Galaxy (see Sch\"ork  {\it et al.} 2009).

Sample Keck spectra are plotted in {\bf Fig.\,7}, for the same  wavelength range (515-520\,nm) as  Fig.~6.  
The three panels, each of which includes the high S/N spectrum of RR~Lyrae, show  program stars more metal-poor than RR~Lyrae (Fig.\,7a),
program stars more metal-rich than RR~Lyrae (Fig.\,7b), and Keck RR~Lyr standard stars (Fig.\,7c).  
The high-resolution, S/N$\sim$1000, solar spectrum  supplied with VWA (Fig.\,7c) was acquired with the 
KPNO Fourier Transform Spectrometer (Kurucz {\it et al.} 1984; Hinkle {\it et al.} 2000). 

Typical FWHM values were measured for  individual and co-added Keck spectra using the same
three Fe\,II spectral lines that were used for the CFHT spectra.   
The measured FWHM values for the individual spectra (see Table\,5) and for the co-added spectra (see Table 7) range 
from 150 to 470\,m\AA, which is similar to the range  for the CFHT spectra.
As expected the measured FWHM values tend to be largest for  phases away from maximum radius.  
This effect is illustrated in the lower panel of Fig.5, where the  FWHM for
the three Fe\,II lines near 450\,nm and for the Ti\,II line at 450.1 nm are plotted as a function 
of the pulsation phase for the nine bright RR~Lyr standard stars.  
The narrowest lines are seen to occur at the earliest phases ($\sim$0.3-0.4),
and at all phases the Fe\,II lines (solid symbols) are significantly narrower than the Ti\,II line (crosses).  
The same pattern for these four lines is also seen for the {\it Kepler} stars 
(where phase range and scatter are larger),  for RR~Lyrae (Fig.\,4 of K10),   
and for XZ~Apr (Fig.\,18 of For {\it et al.} 2011). 



\subsection{Radial Velocities}

It is well known  that the $RVs$ derived from spectra of RR~Lyr stars
are phase dependent and are not necessarily the same as  the velocities that the stars
would have if they were not pulsating (Oke 1966, Liu \& Janes 1989, 1990a,b). 
Furthermore, the derived {\it RVs} depend on the spectral lines used for their measurement: {\it RV} curves calculated with metal lines tend to have
smaller amplitudes than those calculated using Balmer-series lines (H$\alpha$, H$\beta$, H$\gamma$) -- a  
difference that has recently been quantified by Sesar (2012).
A nice illustration of both effects is given in Fig.2 of Preston (2011), which shows multiple {\it RV} curves for the southern RRab stars Z~Mic and RV~Oct. 

In general, the {\it RVs} of RR~Lyr stars are most negative near maximum light ({\it i.e.}, at $\phi_{\rm puls}\sim0$) and
most positive near minimum light (which, depending on the risetime, occurs for RRab stars around $\phi_{\rm puls}\sim0.80-0.95$ and at earlier phases for RRc stars). 
Fundamental-mode pulsators also tend to show larger {\it RV} variations than first-overtone pulsators (as is also the case for the light variations), 
with double-mode (RRd) and other multi-mode pulsators showing the most complex variations.  
If a star exhibits the Blazhko effect then the {\it RV} curve will reflect the amplitude and phase
variations of such stars, and if  pulsational velocities are required (e.g., Baade-Wesselink studies) then the observed {\it RV} must be corrected for projection effects (see Nardetto {\it et al.} 2004, 2009).

For our spectra the heliocentric {\it RVs} given in Tables 4-5 were derived using metal lines and the IRAF
`fxcor' routine. The heliocentric corrections were made during standard pipeline reductions, by  {\it Upena} for the CFHT data 
and by MAKEE for the Keck data.  
For the ESPaDOnS spectra only the limited wavelength range 511-518\,nm was used for the cross-correlations and 
a typical {\it RV} uncertainty is $\sim \pm0.2$ km/s.  
The HIRES  flux spectra were normalized using MAKEE  and very accurate pixel shifts were 
calculated by cross-correlating all pairs of spectra that were taken over the three night run.
The results were used  to derive very precise and accurate relative velocities that were then shifted to the
mean of the {\it RVs} computed relative to the known wavelengths. 

The derived  heliocentric {\it RVs}  for the three M92 red giant standard stars (see Table~5) 
are all close to the cluster mean of $-120$~km/s, suggesting that all are cluster members (see Roederer \& Sneden 2011, Cohen 2011).
Comparison with the {\it RVs} derived by Drukier {\it et al.} (2007) gives  a median difference of $\sim$0.5 km/s and an
RMS difference  of 0.2 km/s.

\begin{deluxetable}{lcrrr}
\tabletypesize{\scriptsize}
\tablewidth{0pt}
\tablecaption{$\gamma$-Velocities for the RR~Lyr Standard Stars}
\label{tab:Table6}
\tablehead{
\colhead{Star} & \colhead{$\phi_{\rm puls}$ } & \multicolumn{3}{c}{ $\gamma$-velocity [km/s] }    \\
\cline{3-5}
\colhead{ }    & \colhead{                  } & \colhead{ Layden}  & \colhead{ This paper }  & \colhead{ diff. }  \\
\colhead{(1)}  & \colhead{(2)}   & \colhead{(3)} & \colhead{(4)} & \colhead{5}  }  
\startdata
ST Boo    & 0.359 & $+13\pm4\phn$   & $  +1.7\pm3$& $11\pm5\phn$   \\ 
V445 Oph  & 0.436 & $-22\pm5\phn$   & $ -22.6\pm4$& $ 1\pm6\phn$   \\ 
RR~Lyr    & 0.535 & $-63\pm8\phn$   & $ -72.0\pm2$& $ 9\pm8\phn$  \\
SW~And    & 0.557 & $-21\pm2\phn$   & $ -14.1\pm3$& $ 7\pm4\phn$  \\ 
X Ari     & 0.310 & $-35\pm3\phn$   & $ -41.6\pm3$& $-7\pm4\phn$  \\  
VX Her    & 0.318 & $-375\pm40$     & $-361.5\pm5$& $-14\pm40$  \\ 
VY Ser    & 0.449 & $-145\pm1\phn$  & $-142.8\pm3$& $-2\pm3\phn$  \\ 
UU Cet    & 0.379 & $-116\pm20$     & $-110.3\pm5$& $-6\pm21$   \\ 
RR Cet    & 0.365 & $-75\pm1\phn$   & $ -74.4\pm3$& $-1\pm3\phn$
\enddata
\tablecomments{The columns contain: (1) star name; (2) pulsation phase at the mid-time of the spectrum; 
(3) the $\gamma$-velocity from L94; (4) the $\gamma$-velocity from the present paper; and
(5) the $\gamma$-velocity difference, in the sense Layden minus this paper.  Note also that
RR~Lyr, SW~And and ST~Boo are Blazhko variables with modulation 
periods of  $39.6$\,d (see K11),  $36.8$\,d (Szeidl 1988; see also Le Borgne {\it et al.} 2012),
and  $284$\,d  (see Le Borgne {\it et al.} 2012), respectively.
}

\end{deluxetable}


The  $\gamma$-velocities for the nine bright RR~Lyrae standard stars  were computed using Liu's {\it RV} template and are given in {\bf Table\,6}.
The last column contains the differences between our estimates and those of Layden (1994), where the mean difference is 1.3 km/s.

\begin{deluxetable*}{llcrcllll}
\tabletypesize{\scriptsize}
\tablewidth{0pt}
\tablecaption{Spectroscopic Iron-to-Hydrogen Abundances (VWA) for the {\it Kepler}-field RR Lyrae Stars} 
\label{tab:Table9}
\tablehead{
\colhead{Star}                  & \colhead{Obs. }               & \colhead{$\phi_{\rm puls}$}   &  \colhead{S/N }   & \colhead{FWHM}  &
\colhead{$T_{\rm eff}$\,/\,log$g$\,/\,$\xi_t$ } & \multicolumn{3}{c}{ [Fe/H]$_{\rm spec}$ }   \\
\cline{7-9}
\colhead{ } & \colhead{ } & \colhead{ } & \colhead{ } & \colhead{ } & \colhead{ } &  \colhead{ Fe\thinspace I lines } & \colhead{ Fe\thinspace II lines } & \colhead{adopted} \\
\colhead{(1)}                   & \colhead{(2)}                  & \colhead{(3)} & 
\colhead{(4)}                   & \colhead{(5)}                  & \colhead{(6)}  & \colhead{(7)   }  & \colhead{(8)  } & \colhead{(9)} } 
\startdata
\multicolumn{9}{c}{(a) 17 Non-Blazhko RRab-type stars} \\[0.1cm]
NR\,Lyr       & CFHT         &  0.29 &  37 & 0.163\,(02)  & 6500/ 2.6/ 2.5    & $-2.54\pm0.13$\,(6)  & $-2.53\pm0.17$\,(5)  & $-2.54\pm0.11$ \\ 
FN\,Lyr       & CFHT         &  0.28 &  32 & 0.209\,(19)  & 6300/ 2.44/ 2.5   & $-2.00\pm0.13$\,(10) & $-1.97\pm0.12$\,(12) & $-1.98\pm0.09$ \\ 
NQ\,Lyr       & CFHT         &  0.32 &  30 & 0.197\,(15)  & 6000/ 1.80/ 2.5   & $-1.88\pm0.19$\,(11) & $-1.90\pm0.10$\,(3)  & $-1.89\pm0.10$ \\ 
V350~Lyr      & Keck         &  0.53 &  34 & 0.293\,(11)  & 6180/ 2.9/ 2.6    & $-1.84\pm0.15$\,(12) & $-1.81\pm0.14$\,(10) & $-1.83\pm0.12$ \\ 
V894~Cyg      & CFHT         &  0.39 &  28 & 0.193\,(08)  & 6200/ 2.46/ 2.0   & $-1.65\pm0.18$\,(11) & $-1.66\pm0.16$\,(9)  & $-1.66\pm0.12$ \\ 
AW\,Dra       & CFHT         &  0.30 &  38 & 0.202\,(05)  & 6540/ 2.40/ 2.5   & $-1.33\pm0.13$\,(9)  & $-1.33\pm0.12$\,(6)  & $-1.33\pm0.09$ \\ 
KIC\,7030715  & CFHT         &  0.26 &  19 &0.26\phn\,(04)& 6500/ 2.64/ 2.5   & $-1.32\pm0.07$\,(6)  & $-1.34\pm0.15$\,(3)  & $-1.33\pm0.08$ \\ 
V1107~Cyg     & Keck         &  0.48 &  21 & 0.220\,(17)  & 6300/ 2.8/ 3.0    & $-1.23\pm0.42$\,(33) & $-1.31\pm0.18$\,(11) & $-1.29\pm0.23$ \\ 
V368 Lyr      & Keck         &  0.43 &  20 & 0.183\,(02)  & 6300/ 2.8/ 3.0    & $-1.33\pm0.31$\,(22) & $-1.25\pm0.24$\,(10) & $-1.28\pm0.20$ \\ 
KIC\,9658012  & Keck         &  0.84 &  37 & 0.413\,(16)  & 6500/ 2.8/ 3.0    & $-1.22\pm0.26$\,(26) & $-1.31\pm0.12$\,(11) & $-1.28\pm0.14$ \\ 
KIC\,9717032  & Keck         &  0.23 &  19 & 0.262\,(06)  & 6620/ 2.8/ 3.5    & $-1.27\pm0.17$\,(14) & $-1.26\pm0.17$\,(13) & $-1.27\pm0.15$ \\ 
V715~Cyg      & Keck         &  0.51 &  21 & 0.241\,(13)  & 6400/ 3.0/ 3.0    & $-1.13\pm0.12$\,(25) & $-1.14\pm0.14$\,(12) & $-1.13\pm0.09$ \\ 
V2470 Cyg     & CFHT         &  0.32 &  26 & 0.262\,(11)  & 6400/ 2.44/ 2.6   & $-0.59\pm0.14$\,(45) & $-0.59\pm0.21$\,(14) & $-0.59\pm0.13$ \\ 
V782\,Cyg     & Keck         &  0.42 &  34 & 0.290\,(12)  & 6525/ 3.2/ 2.9    & $-0.40\pm0.15$\,(102)& $-0.43\pm0.14$\,(19) & $-0.42\pm0.10$ \\ 
KIC\,6100702  & CFHT         &  0.26 &  23 & 0.281\,(10)  & 6500/ 2.50/ 2.5   & $-0.19\pm0.14$\,(36) & $-0.13\pm0.12$\,(6)  & $-0.16\pm0.09$ \\ 
V839\,Cyg     & Keck         &  0.12 &  70 & 0.375\,(03)  & 7200/ 3.1/ 4.0    & $+0.07\pm0.18$\,(137)& $-0.09\pm0.14$\,(16) & $-0.05\pm0.14$ \\ 
V784\,Cyg     & Keck         &  0.47 &  27 & 0.323\,(12)  & 6400/ 3.3/ 4.3    & $-0.05\pm0.15$\,(107)& $-0.06\pm0.13$\,(10) & $-0.05\pm0.10$ \\ [0.1cm] 
\multicolumn{9}{c}{(b) 12 Blazhko RRab-type stars (excluding RR~Lyrae)} \\[0.1cm]
V2178~Cyg     & Keck         &  0.43 &  41 & 0.211\,(09)  & 6450/ 2.8/ 3.0    & $-1.63\pm0.24$\,(17) & $-1.67\pm0.09$\,(11) & $-1.66\pm0.13$ \\ 
V450~Lyr      & Keck         &  0.57 &  21 & 0.236\,(17)  & 6450/ 3.0/ 3.0    & $-1.52\pm0.13$\,(13) & $-1.50\pm0.21$\,(9)  & $-1.51\pm0.12$ \\ 
V353 Lyr      & Keck         &  0.32 &  20 & 0.205\,(17)  & 6300/ 2.8/ 3.0    & $-1.48\pm0.20$\,(15) & $-1.50\pm0.35$\,(4)  & $-1.50\pm0.20$ \\ 
V360 Lyr      & Keck         &  0.52 &  29 & 0.378\,(20)  & 6400/ 2.8/ 10:   & $-1.46\pm0.37$\,(72) & $-1.54\pm0.44$\,(26)  & $-1.50\pm0.29$ \\ 
V354 Lyr      & Keck         &  0.26 &  30 & 0.298\,(14)  & 6400/ 2.8/ 3.0    & $-1.44\pm0.24$\,(20) & $-1.44\pm0.20$\,(13) & $-1.44\pm0.16$ \\ 
%
V1104~Cyg     & Keck         &  0.42 &  33 & 0.174\,(03)  & 6300/ 2.8/ 3.0    & $-1.25\pm0.25$\,(51) & $-1.21\pm0.15$\,(17) & $-1.23\pm0.15$ \\ 
V808~Cyg      & Keck         &  0.70 &  19 & 0.473\,(16)  & 6300/ 2.8/ 3.0    & $-1.20\pm0.27$\,(15) & $-1.18\pm0.25$\,(7)  & $-1.19\pm0.18$ \\ 
V783\,Cyg     & CFHT         &  0.58 &  16 & 0.23\phn\,(01)&6400/ 2.1/ 2.6    & $-1.19\pm0.18$\,(13) & $-1.13\pm0.14$\,(2)  & $-1.16\pm0.11$ \\ 
V366~Lyr      & Keck         &  0.69 &  21 & 0.332\,(24)  & 6430/ 3.0/ 3.0    & $-1.13\pm0.13$\,(16) & $-1.19\pm0.10$\,(10) & $-1.16\pm0.09$ \\ 
V355\,Lyr     & CFHT         &  0.19 &  17 & 0.247\,(16)  & 6900/ 3.28/ 2.5   & $-1.13\pm0.14$\,(10) & $-1.17\pm0.30$\,(1)  & $-1.14\pm0.17$ \\ 
KIC\,11125706 & CFHT         &  0.33 &  69 & 0.232\,(04)  & 6200/ 2.35/ 2.25  & $-1.09\pm0.13$\,(57) & $-1.09\pm0.10$\,(17) & $-1.09\pm0.08$ \\  
V838~Cyg      & CFHT         &  0.36 &  13 & 0.236\,(28)  & 6850/ 2.12/ 2.4   & $-1.02\pm0.19$\,(4)  & $-1.00\pm0.10$\,(3)  & $-1.01\pm0.10$ \\ [0.1cm] 
\multicolumn{9}{c}{(c) Five RRc-type stars  (including the RRc/HADS? star KIC\,3868420)} \\[0.1cm]
KIC\,9453114   & CFHT         & 0.28 &  29 & 0.278\,(27)  & 6500/ 2.5/ 2.5    & {\hfil \nodata \hfil} & $-2.13\pm0.12$\,(4)  & $-2.13\pm0.12$ \\ 
KIC\,4064484   & Keck         & 0.59 &  33 & 0.288\,(11)  & 6500/ 2.8/ 3.0    & $-1.65\pm0.22$\,(9)  & $-1.54\pm0.12$\,(6)  & $-1.58\pm0.13$ \\ 
KIC\,3868420   & CFHT         & 0.47 &  96 & 0.359\,(09) & 7480/ 3.9/ 2.0     & $-0.35\pm0.14$\,(13) & $-0.29\pm0.08$\,(6)  & $-0.31\pm0.08$ \\  
KIC\,8832417   & CFHT         & 0.29 &  30 & 0.349\,(16)  & 7000/ 3.30/ 2.5   & $-0.22\pm0.19$\,(27) & $-0.29\pm0.09$\,(7)  & $-0.27\pm0.11$ \\  
KIC\,5520878   & Keck         & 0.39 &  40 & 0.319\,(02)  & 7250/ 3.9/ 2.4   & $-0.18\pm0.18$\,(76) & $-0.19\pm0.12$\,(18) & $-0.18\pm0.10$ \\ [0.2cm] 
\multicolumn{9}{c}{(d) Nine bright RR~Lyrae metal abundance standards (including RR~Lyrae) }  \\[0.1cm]
X~Ari          & Keck         & 0.31 & 223 & 0.188\,(05) & 6075/ 2.4/ 2.0   & $-2.75\pm0.08$\,(9)  & $-2.74\pm0.17$\,(11) & $-2.74\pm0.09$ \\ 
VY~Ser         & Keck         & 0.45 & 120 & 0.232\,(03) & 6170/ 2.8/ 2.5   & $-1.73\pm0.12$\,(29) & $-1.69\pm0.08$\,(16) & $-1.71\pm0.07$ \\ 
ST~Boo         & Keck         & 0.36 & 135 & 0.198\,(02) & 6313/ 2.6/ 1.8   & $-1.66\pm0.12$\,(39) & $-1.60\pm0.04$\,(15) & $-1.62\pm0.06$ \\  
UU~Cet         & Keck         & 0.38 &  66 & 0.226\,(03) & 6150/ 2.6/ 1.6   & $-1.34\pm0.14$\,(68) & $-1.32\pm0.07$\,(19) & $-1.33\pm0.08$ \\  
RR~Lyr         & Keck         & 0.54 & 252 & 0.276\,(06) & 6400/ 2.8/ 3.0  & $-1.27\pm0.22$\,(95) & $-1.27\pm0.09$\,(15) & $-1.27\pm0.12$ \\ 
VX~Her         & Keck         & 0.32 &  97 & 0.202\,(02) & 6690/ 2.9/ 2.2   & $-1.22\pm0.17$\,(51) & $-1.23\pm0.17$\,(17) & $-1.23\pm0.12$ \\   
RR~Cet         & Keck         & 0.37 & 184 & 0.212\,(03) & 6420/ 2.9/ 1.9   & $-1.17\pm0.15$\,(65) & $-1.19\pm0.08$\,(17) & $-1.18\pm0.09$ \\  
V445~Oph       & Keck         & 0.44 & 106 & 0.315\,(05) & 6200/ 2.8/ 1.8   & $+0.15\pm0.14$\,(218)& $+0.11\pm0.13$\,(16) & $+0.13\pm0.10$ \\  
SW~And         & Keck         & 0.56 & 151 & 0.367\,(06) & 6660/ 3.6/ 2.2   & $+0.19\pm0.14$\,(225)& $+0.21\pm0.09$\,(19) & $+0.20\pm0.08$ \\ [0.1cm]  
\enddata
\tablecomments{The columns contain: (1) Star name;  
(2) Source of the spectra, either `Keck\,I 10m + HIRES' or `CFHT 3.6-m + ESPaDOnS';  
(3) Pulsation phase at mid-exposure;  
(4) Signal-to-noise ratio per pixel for the co-added spectra, measured at $\sim$570\,nm and calculated using the VWA `rainbow' widget;
(5) Average full-width at half-minimum value [\AA] for the three Fe\,II lines at 4508, 4515 and 4520 \AA, derived 
from the spectrum (either single or co-added exposures) analyzed by VWA -- the uncertainty (given in parentheses) 
is approximated by the standard deviation of the mean FWHM for the three lines; 
(6) The assumed values for the model effective temperature $T_{\rm eff}$ [K],  surface gravity log$g$ [cm s$^{-2}$]   
and microturbulent velocity $\xi_t$ [km/s] -- in general the $T_{\rm eff}$ have uncertainties $\sim\pm150$\,K, the
log$g$ are $\pm$0.10 to 0.20, and the $\xi_t$ are $\pm$0.3 to 0.5 km/s; 
(7) [Fe/H] abundance based on the Fe\,I lines (number of lines given in parentheses);   
(8) [Fe/H] abundance based on the Fe\,II lines (number of lines given in parentheses); 
(9) Adopted VWA spectroscopic metal abundance, equal to the weighted (in inverse proportion to uncertainty) average of the abundances in columns (7) and (8).  
}
\end{deluxetable*}

\subsection{\rm  [Fe/H] and Atmospheric Parameters}

Estimation of metal abundances and other atmospheric parameters from 
high dispersion spectra depends on  numerous assumptions regarding  the
stellar atmosphere where the spectral lines were formed.  In our
analyses, which were performed using the VWA, MOOG and SME packages, the lines
were assumed to form in a one-dimensional plane-parallel atmosphere under local
thermodynamic equilibrium (LTE) conditions.  For RR~Lyr stars (absolute
magnitudes $\sim$0.5) the atmospheric extensions are sufficiently small (less than 1\% according to
Fig.~1 of Heiter \& Eriksson 2006) that the plane-parallel assumption was judged  to be adequate 
for the derivation of iron-to-hydrogen ratios.

The observed {\it Kepler}-field  RR~Lyr stars are a mixture of halo
and old-disk stars, with distances ranging from 265$\pm$11\,pc for RR~Lyrae (Benedict {\it et al.}
2011) to  $\sim$25 kpc for the faintest stars (N11);   therefore they are expected to be old and to have a wide range
of chemical and kinematic characteristics.  In the H-R diagram they are horizontal branch stars
at various stages of evolution away from the zero-age horizontal branch (ZAHB), and over a pulsation cycle
$L$ and $T_{\rm eff}$ trace out a  loop in the instability strip.  The masses are expected to be in the range 0.5-0.8 $\cal M_{\sun}$,
and most of the stars are expected to be near the
ZAHB-level appropriate for the mass and metal abundance of the star (see Fig.~13 of N11). 

The Blazhko and non-Blazhko RR~Lyr stars in the {\it Kepler} field occur in approximately equal proportions.  
The pulsations
of the non-Blazhko stars are mainly  fundamental or first-overtone radial modes.  To date 
no `classical' double-mode RR~Lyrae stars ({\it i.e.}, RRd stars) pulsating simultaneously in both modes have 
been identified;  however, Molnar {\it et al.} (2012) recently detected  a very weak first-overtone
pulsation in RR~Lyrae.   When averaged over a
pulsation cycle the mean values of $T_{\rm eff}$ are expected to range from 6000-7500\,K, with the
ab-type stars having cooler mean surface temperatures than the c-type stars.  The ranges of the mean values of log\,$g$
and $L$ are expected to be 2-3 and 40-60 $L_{\sun}$.  The instantaneous
values for these quantities, which are phase dependent, span much wider ranges -- see, for example,
Jurcsik {\it et al.} (2009a,b,c), who suggested that  the $T_{\rm eff}$ range might be as large as 3500\,K
(5500-9000\,K), the range of log\,$g$$\sim$1.5 to 4, and the range of $L$  $\sim$30-90\,$L_{\sun}$.
For all three physical characteristics the maxima occur slightly before $\phi_{\rm puls}$=1.
Because the CFHT and Keck spectra were taken at pulsation phases away from maximum light the RRab
stars were expected to  have `instantaneous' $T_{\rm eff}$ values in the considerably smaller range 6000-6500~K (see
figs.7-10, 36-37 of For {\it et al.} 2011), with  log\,$g$ between 2 and 3.  An exception
is V839~Cyg, which was observed just after maximum light, at $\phi_{\rm puls}=0.12$, and therefore expected
to have a rather higher temperature.  Unlike solar-like stars which typically have a microturbulent parameter 
$\xi_t$ $\sim$1\,km/s, the $\xi_t$ for RR~Lyr stars
generally range from 2.5 to 4.0 km/s, with instantaneous values at maximum radius tending to occur
at the lower end of this range  ({\it i.e.}, $\xi_t\sim$2.5-3.0 km/s for phases between 0.2 and 0.6 -- see
Figs.~13-14 and 17 of For {\it et al.} 2011).  Other information is available for 
some of the brighter stars.  For example, for RR~Lyrae at
$\phi_{\rm puls}=0.54$ the log\,$g$ and $T_{\rm eff}$ are expected to be $\sim$2.75 and
6250$\pm$100\,K (see Table 5.2 of Kolenberg 2002), both of which vary from one Blazhko cycle to the
next.   

For the solar photosphere the abundance of iron relative to the number of hydrogen atoms per unit
volume was assumed to be  $\log \epsilon({\rm Fe}) = 7.50$, where  $\log \epsilon(Fe) \equiv \log
(N_{\rm Fe} /N_{\rm H} ) + 12.0$ and $\log N_{\rm H}=12.0$.  This value is consistent
with the iron abundance  $7.45\pm0.08$ recommended by Holweger (2001),  the value 7.52 adopted by MOOG
(Sneden 2002), and with the most recent value $7.50\pm0.04$ given in  the review by Asplund {\it et al.}
(2009).  For iron abundances relative to hydrogen {\it and} relative to the Sun we use  [Fe/H]$_\ast$ = $\log (N_{\rm
Fe} / N_{\rm H})_\ast - \log (N_{\rm Fe} / N_{\rm H})_{\rm solar}$.  When abundances are expressed
relative to the {\it total} number of atoms per unit volume (e.g., the values   output from VWA
and Kurucz's ATLAS9 program),  the photospheric `abundance' of Fe is denoted $A_{\rm Fe} \equiv \log
(N_{\rm Fe} / N_{\rm total})$, where  $N_{\rm Fe}$ is the number of Fe atoms per unit volume and
$N_{\rm Fe}$/$N_{\rm total}$ is the corresponding fraction of Fe relative to the total number
density of atoms.   If the number densities for  hydrogen and helium are assumed to be  $\log N_{\rm H}=12.0$ and
$\log N_{\rm He}=10.99$ then  these two elements are the only significant contributors to $N_{\rm
total}$ and we have $\log N_{\rm total}=12.04$.  Thus, the solar number fractions of H, He and Fe
are, respectively, $N_{\rm H}$/$N_{\rm total} = 91.1\%$,  $N_{\rm He}$/$N_{\rm total} = 8.9\%$,  and
$N_{\rm Fe}$/$N_{\rm total} = 0.0029\%$  ({\it i.e.},    the log of the Fe abundance relative to the
{\it total} number density is $\log (N_{\rm Fe} / N_{\rm tot}) = -4.54$, which follows from  $A_{\rm
Fe} = \log \epsilon(Fe) - \log N_{\rm total} = 7.50 - 12.04$).

\begin{deluxetable*}{llllll}
\tabletypesize{\scriptsize}
\tablewidth{0pt}
\tablecaption{Spectroscopic Iron-to-Hydrogen Abundances (MOOG) } 
\label{tab:Table}
\tablehead{
\colhead{Star}   &   
\colhead{$T_{\rm eff}$\,/\,log$g$\,/\,$\xi_t$\,/\,[Fe/H] } & 
\colhead{ log\,$\epsilon$(FeI) } & 
\colhead{ log\,$\epsilon$(FeII) }  & 
\colhead{ log\,$\epsilon$(Fe) }  &
\colhead{[Fe/H]$_{\rm spec}$ } \\
\colhead{(1)}                  & \colhead{(2)}                  & \colhead{(3)} & 
\colhead{(4)}                  & \colhead{(5)}                  & \colhead{(6)}     } 
\startdata
X~Ari    &  6000/ 3.0/ 2.0/ $-2.5$  &  $4.73\pm0.12$\,(105) & $5.19\pm0.18$\,(24) & $4.91\pm0.07$ & $-2.59\pm0.12$  \\ 
VY Ser   &  6250/ 3.0/ 2.0/ $-1.5$  &  $5.83\pm0.12$\,(147) & $6.01\pm0.12$\,(26) & $5.92\pm0.06$ & $-1.58\pm0.11$  \\  
ST~Boo   &  6250/ 3.0/ 2.0/ $-1.5$  &  $5.78\pm0.12$\,(177) & $6.09\pm0.12$\,(29) & $5.94\pm0.06$ & $-1.57\pm0.11$  \\ 
RR~Lyr   &  6000/ 3.0/ 2.0/ $-1.5$  &  $5.96\pm0.14$\,(145) & $6.25\pm0.09$\,(19) & $6.14\pm0.05$ & $-1.36\pm0.10$  \\ %
VX~Her   &  6750/ 3.0/ 2.0/ $-1.5$  &  $6.24\pm0.12$\,(169) & $6.35\pm0.13$\,(24) & $6.29\pm0.06$ & $-1.21\pm0.11$  \\  
SW~And   &  6250/ 3.0/ 2.0/ $+0.0$  &  $7.35\pm0.15$\,(177) & $7.42\pm0.15$\,(11) & $7.39\pm0.08$ & $-0.12\pm0.13$  \\   
V782~Cyg &  6250/ 3.0/ 2.0/ $+0.0$  &  $7.41\pm0.30$\,(162) & $7.45\pm0.28$\,(11) & $7.43\pm0.14$ & $-0.07\pm0.19$  \\
V445~Oph &  6250/ 3.0/ 2.0/ $+0.0$  &  $7.45\pm0.11$\,(188) & $7.61\pm0.13$\,(11) & $7.52\pm0.06$ & $+0.02\pm0.11$  \\   
V784~Cyg &  6250/ 3.0/ 2.0/ $+0.0$  &  $7.86\pm0.35$\,(142) & $7.92\pm0.34$\,(7)  & $7.89\pm0.17$ & $+0.39\pm0.22$    
\enddata
\tablecomments{The columns contain: (1) the star name;  
(2) the assumed values for the (WebMARCS) model effective temperature, $T_{\rm eff}$ [K],  surface gravity, log$g$ [cm s$^{-2}$],   
microturbulent velocity, $\xi_t$ [km/s], and metal abundance, [Fe/H];
(3) the [Fe\,I/H] abundance (based on the number of Fe\,I lines given in parentheses);   
(4) the [Fe\,II/H] abundance (based on the number of Fe\,II lines given in parentheses); 
(5) the weighted average of the  [Fe\,I/H] and [Fe\,II/H] abundances; 
(6) the adopted MOOG spectroscopic iron abundance.  }
\end{deluxetable*}

\subsubsection{Parameter Estimation with VWA and MOOG}

The bulk of the spectroscopic data analyses of both the CFHT and Keck data was carried out with
the spectral synthesis program VWA\footnote{The VWA software was written by Hans Bruntt
(University of Aarhus, Denmark), in IDL, and is publicly available at
https://sites.google.com/site/vikingpowersoftware/.}  (Bruntt {\it et al.} 2002, 2008, 2010a,b).   This
all-purpose program has the capability of normalizing and co-adding spectra, computing synthetic
spectra,  selecting suitable lines for abundance determination, measuring EWs, calculating expected
EWs for the assumed model parameters, and calculating abundances for a wide range of elements/ions.
The abundances follow from iteratively varying the model parameters, calculating synthetic spectra,
and then comparing the EWs calculated from the spectra with the measured EWs. 

For each RR~Lyrae star the following photospheric parameters were derived: the effective temperature,
$T_{\rm eff}$\,[K];  the surface gravity, $\log\,g$ [cm/s$^2$];
the microturbulent velocity, $\xi_t$\,[km/s]; the macroturbulent velocity, $v_{\rm
mac}$; the projected rotational velocity, $v$\,sin$i$; and the spectroscopic iron-to-hydrogen ratio,
[Fe/H]$_{\rm spec}$.  Because the RR~Lyr stars are undergoing (mainly) radial pulsations, these parameters are
`instantaneous' quantities, specific for the times (phases) at which the spectroscopic observations were
made.   The metal abundances, while depending on the assumed atmospheric conditions, should be
independent of the observed pulsation phases (see  Figs.13-14 of
For {\it et al.} 2011, which show the variation with pulsation phase of their derived $T_{\rm
eff}$, log$g$, [Fe/H] and $\xi_t$ values).  

The synthetic spectra were calculated within VWA using the SYNTH code 
(Piskunov 1992; Valenti \& Piskunov 1996), with atomic parameters (eg., log\,$g$$f$ values) and line-broadening coefficients
retrieved from the Vienna atomic line database, VALD (Kupka {\it et al.} 1999)\footnote{See
http://www.astro.uu.se/vald.}.  Two sets of model stellar atmospheres were used.  For
all the stars the model atmosphere grid of ATLAS9 models provided by Heiter {\it et al.} (2002) was
used.  For this grid the step-increments in effective temperature, surface gravity and metal
abundance are  $\Delta T_{\rm eff} = 250$\,K,  $\Delta$log$g$=0.5, and $\Delta$[Fe/H]=0.25 dex. 
For the bright RR~Lyrae stars both ATLAS9 and MARCS models were used.    

The derivation of [Fe/H]$_{\rm spec}$ values followed from analysis of the normalized flux spectra.
For each star  an initial set of values for the atmospheric parameters was assumed, and a synthetic spectrum was calculated.   
The usual starting values were: $T_{\rm eff}=6500$~K, log$g$=2.50, $\xi_t = 2.50$~km/s and $v_{\rm mac} = 2$~km/s.   
After adjusting the projected rotational velocity, $v$sin$i$  and
the resolving power ($R=65000$ for the CFHT spectra and 36000 for the Keck spectra) for each analyzed spectrum, the 
spectrum was Doppler shifted according to the $RV$ (see Tables 4-5) and aligned with the synthetic template spectrum.


The final set of fitted lines was inspected and poor fits rejected.  
The microturbulent velocity, $\xi_t$ was determined by requiring that the correlation between
the derived Fe abundances and the reduced EWs (i.e., EW/$\lambda$) was near zero;     
$T_{\rm eff}$ was estimated by requiring that the correlation of Fe abundance and the lower excitation potential (EP) was near zero; 
and the surface gravities were calculated by requiring good agreement between the iron abundances determined from lines of the 
Fe~I and Fe~II ionization stages (which depended on the assumed $T_{\rm eff}$ and log $g$ values).
Although LTE conditions are assumed by VWA, the effect of departures from LTE on the  [Fe/H] values
derived using the Fe~I lines were considered, and where appropriate the 
corrections calculated by Rentzsch-Holm (1996) were applied;  these were usually smaller than $\sim$0.10 dex.
The Fe\,II lines were assumed to be unaffected by deviations from LTE.    

{\bf Table~7}  summarizes the basic spectroscopic information derived from the VWA analyses of the CFHT and Keck spectra, where the stars 
are sorted by type (non-Blazhko, Blazhko, RRc, etc.), and within a given type  by spectroscopic metal abundance.
The pulsation phases (column\,3) are mid-exposure values (i.e., average values of the mid-times when multiple spectra were taken),
and the  S/N per pixel (column 4) are for the co-added normalized spectra\footnote{A  
comparison of the MAKEE and `rainbow' S/N values for the seven stars with only single exposures shows the MAKEE values to be
systematically larger (by $\sim$16\%) than the `rainbow' values.}.  
The FWHM values (column\,5) are for the co-added spectra and the  $T_{\rm eff}$, log$g$ and $\xi_t$ values (column 6) 
are either those assumed for the adopted VWA runs,
or are interpolated values for two similar atmosphere models.  The value $v_{\rm mac}$ was arbitrarily set to either 2 or 3 for all the stars, 
while the projected rotational velocity, $v$\,sin$i$, was set to 2 km/s for the Keck spectra and 7-10 km/s for the CFHT spectra; 
the impact on these choices on the derived abundances seems  to have been minimal.
The iron-to-hydrogen abundances (columns 7-8) are for the neutral and singly-ionized ions, the weighted average of which, [Fe/H], is given in column 9.
The derived metallicities range from a low of $-2.54\pm0.11$\,dex (for NR~Lyr) to a high
of $-0.05\pm0.10$\,dex (for V784~Cyg). The same two stars were the extremes identified  by 
N11 in their analysis of the Q1-Q5 {\it Kepler} photometry
of non-Blazhko variables, and the  four non-Blazhko stars that were suspected (from their locations in Fig.9 of N11) of being metal rich 
(V782~Cyg, V784~Cyg, KIC\,6100702 and V2470~Cyg) are confirmed as such.  
Not included in Table~7 are four stars for which the VWA analysis of the co-added spectrum proved inconclusive or impossible: 
(1) V1510~Cyg, a non-Blazhko star  ({\it Kp}=14.5 mag); 
(2) the faint non-Blazhko star V346~Lyr ({\it Kp}=16.4 mag);  
(3) the faint Blazhko star KIC\,9973633 ({\it Kp}=17.0 mag), for which the S/N of the co-added Keck spectrum (not plotted in Fig.7) was insufficient for estimating [Fe/H]; and
(4) the faint ({\it Kp}=17.4 mag) extreme-Blazhko star V445~Lyr (see Guggenberger {\it et al.} 2012).
In all cases except  KIC\,9973633 we were able to derive [M/H] values using SME (see Table 9). 

The EWs derived using VWA are in good agreement  with the values measured by hand using the
IRAF `splot' routine, with values calculated with ARES, and with the values calculated from the synthetic spectra.  
The EWs of the  standard stars  also agree (to within the uncertainties arising from the observations being made at different phases) 
with the values listed in Table 3(b) of C95\footnote{For RR~Lyrae itself the Clementini {\it et al.} EWs are larger 
than the ARES EWs for lines with EWs$\geq$100; however, such lines are not used here.}. 
To minimize the effects of line saturation only Fe~I and Fe~II lines with EW $<90$\,m\AA \, were used in our  analysis.
A table of measured and calculated EWs (m\AA) for the Fe~I and Fe~II lines (program and standard stars), including 
the assumed lower excitation potentials and log\,$g$$f$ values (from the VALD database), 
is available from the first author. 

As a check on the VWA results [Fe/H]  abundances for several of the stars  were also derived using the 
LTE Stellar Line Analysis program MOOG (Sneden 1973, 2002; Sneden {\it et al.} 2011)\footnote{MOOG is publicly available at
www.as.utexas.edu/$\sim$chris/
moog.html (2010 version).}.  
In this case EWs were calculated using the highly-automated program ARES (Sousa {\it et al.} 2007, 2008).  
A comparison of the EWs derived by ARES, by VWA, and by the IRAF `splot' routine, showed  excellent agreement for EWs between 5 and 100 m\AA.
The WEBMARCS grid of model atmospheres (Gustafsson {\it et al.} 1975, 2008)\footnote{The WEBMARCS models, also known 
as `MARCS 2008', or `MARCS 35', or `new MARCS' models, are available on-line at http://marcs.astro.uu.se}
was used for the MOOG calculations. These models used the Grevesse {\it et al.} (2007) compositions for the solar photosphere, and the available 
grid had $T_{\rm eff}$ values up to 7000~K with a step size of 250~K.  
Unfortunately the smallest surface gravities available were for log\,$g$=3, which is larger than the value appropriate for most RR~Lyrae stars.
  
The MOOG results are summarized in {\bf Table~8}.  The assumed model atmospheric parameters are given in column 2, where 
the uncertainties are $\sim\pm150$\,K for $T_{\rm eff}$, $\pm0.2$\,cm/s$^2$ for log\,$g$, and $\pm$0.5 km/s for $\xi_t$.   
Columns 3-4 give  log\,$\epsilon$ values (relative to hydrogen)
for the Fe~I and Fe~II ions (standard output from the `ABFIND' routine in MOOG).   Also given are 
the number of lines used to  derive the abundance.  Column~5 contains the average iron abundance, log\,$\epsilon$(Fe), 
weighted inversely by the Fe~I and Fe~II uncertainties,
and the last column gives [Fe/H] values, derived by adding $12.50\pm0.05$ to the column~5 abundances.  
Since only a limited number of MOOG runs were made the uncertainties are
larger than for the VWA results, the  largest discrepancies occurring for V782~Cyg and SW~And.  

\subsubsection{Parameter Estimation with SME}

Amospheric parameters for the program and standard RR~Lyr stars were also derived using the 
`Spectroscopy Made Easy' (SME) program of Valenti \& Piskunov (1996)\footnote{SME was written
in IDL and is freely available at  http://www.stsci.edu/$\sim$valenti/sme/. }.   
SME compares  observed spectra with synthetic spectra, where the latter are computed by solving the radiative
transfer equations.  Parameter estimates are calculated by minimizing (using Levenberg-Marquart) a $\chi^2$ statistic that
measures the lack of agreement between the observed and fitted intensities.   
Unlike VWA and MOOG, which are based on comparing calculated and measured EWs, SME 
matches sections of spectra.    SME does not actually calculate Fe~I and Fe~II abundances, but rather,
the abundances assumed for the Sun are scaled and used to calculate a mean metal abundance, [M/H], which
we assume to be comparable to [Fe/H].  

The spectra that were analyzed with SME were the co-added normalized Keck and CFHT spectra 
and the Kurucz-Hinckle solar spectrum.  The derived photospheric quantities 
were  $T_{\rm eff}$,   log\,$g$ (cgs units),  $\xi_t$,  $v_{\rm mac}$,  $v$sin$i$ and [M/H].  
For most of the SME analyses the synthetic spectra were calculated for the $\lambda$-interval
500-520~nm, for which VALD returned atomic parameters for 476 lines.  This interval is 
free of Balmer lines, around which the continua were usually poorly defined.  The
interval 480-520 nm was tried, but spurious results were found, presumably owing to the
presence of  H$\gamma$.  For several  stars the interval 440-460\,nm was also
analyzed, using 952 lines identified by VALD, and the results were averaged with those
from 500-520 nm.   

Our estimates of the atmospheric parameters were obtained iteratively by varying the initial values supplied to SME
until the `best' overall estimates were found. 
The usual reduction procedure was to take as starting values the parameter set calculated by VWA
(see Table~7) and then to solve simultaneously for two parameters at a time.  First $\xi_t$ and
$v_{\rm mac}$ were calculated,  then $v$\,sin$i$ and R (=$\lambda$/$\Delta\lambda$), then $T_{\rm
eff}$ and log$g$.  After each run the prior input values were replaced with the revised
values. Lastly, with the newer set of SME parameters, we solved for [M/H].  

{\bf Table 9} contains the final results of the SME analyses, sorted in order of increasing metal
abundance.  A total of 47 stars were analyzed.  The co-added spectrum of KIC\,9973633 was not
measured owing to its very low S/N.

\begin{deluxetable}{lclcrrc}
\tabletypesize{\scriptsize}
\tablewidth{0pt}
\tablecaption{Physical characteristics derived using SME  }
\label{tab:Table9}
\tablehead{
\colhead{Star}  & \colhead{\c{$T_{\rm eff}$} } &  \colhead{ log\,$g$} & \colhead{ $\xi_t$} & \colhead{ $v_{\rm mac}$ } & \colhead{ $v$\,sin$i$ }   & \colhead{ [M/H] }  \\ [0.1cm]
\colhead{(1)}  & \colhead{(2)}   & \colhead{(3)}           & \colhead{(4)}          &  \colhead{(5)}         & \colhead{(6)   } & \colhead{ (7) }  }  
\startdata
{\bf X\,Ari}{\dag}& 6200  & 1.85  & 4.8   &  9.5  &  2.7  & $-2.71$ \\ 
NR\,Lyr{\dag}     & 6363  & 2.56  & 4.4   &  7.9  &  4.8  & $-2.51$ \\ 
KIC\,9453414      & 6204  & 2.04  & 6.4   & 29.3  &  4.3  & $-2.15$ \\
V1510 Cyg         & 5706  & 1.76  & 4.9   & 10.8  &  4.9  & $-2.00$ \\
FN\,Lyr           & 6203  & 2.21  & 4.7   &  8.0  &  4.6  & $-1.91$ \\
NQ\,Lyr           & 5980  & 1.56  & 4.4   &  8.4  &  4.5  & $-1.87$ \\
V350\,Lyr         & 6131  & 2.29  & 5.1   & 16.7  &  3.0  & $-1.82$ \\
V445\,Lyr         & 6227  & 1.70  & 5.3   & 12.4  &  5.4  & $-1.78$ \\
{\bf VY\,Ser}     & 6167  & 2.10  & 4.4   & 11.3  &  3.2  & $-1.71$ \\
V2178\,Cyg        & 6301  & 2.35  & 4.2   &  9.2  &  4.3  & $-1.66$ \\
V894\,Cyg         & 6220  & 1.92  & 3.8   &  7.2  &  4.4  & $-1.65$ \\
{\bf ST\,Boo}{\dag}&6327  & 1.91  & 3.8   & 10.5  &  2.1  & $-1.61$ \\ 
V353\,Lyr         & 6093  & 1.84  & 4.8   &  4.8  &  4.1  & $-1.57$ \\
V450\,Lyr         & 6280  & 2.34  & 3.6   & 12.7  &  5.0  & $-1.51$ \\
KIC\,4064484      & 6473  & 2.39  & 4.0   & 18.4  &  4.9  & $-1.51$ \\
V360\,Lyr         & 6202  & 2.42  & 5.3   & 18.9  &  5.0  & $-1.48$ \\
V346\,Lyr         & 6108  & 2.34  & 4.2   &  9.6  &  5.0  & $-1.48$ \\
V354\,Lyr         & 6391  & 2.05  & 4.6   & 15.0  &  4.2  & $-1.43$ \\
V1107\,Cyg        & 6087  & 2.23  & 5.1   & 11.4  &  3.0  & $-1.40$ \\
{\bf VX\,Her}     & 6533  & 2.46  & 3.8   &  9.0  &  2.2  & $-1.38$ \\
KIC\,7030715      & 6221  & 1.99  & 4.2   & 13.5  & 10.0  & $-1.37$ \\
AW\,Dra           & 6461  & 2.02  & 4.4   &  5.9  &  3.6  & $-1.34$ \\
{\bf UU\,Cet}     & 6269  & 2.31  & 4.2   &  9.1  &  2.4  & $-1.33$ \\
{\bf RR\,Lyr}{\dag}&6445  & 3.13  & 5.1   & 11.7  &  5.6  & $-1.33$ \\ 
V368\,Lyr         & 6231  & 1.93  & 3.3   &  9.5  &  3.0  & $-1.29$ \\
KIC\,9658012      & 6314  & 2.15  & 4.3   & 15.0  & 12.9  & $-1.27$ \\
KIC\,9717032      & 6461  & 2.14  & 5.0   & 14.8  &  9.8  & $-1.24$ \\
{\bf RR\,Cet}     & 6459  & 2.40  & 4.0   &  8.8  &  3.8  & $-1.22$ \\
V1104\,Cyg        & 6352  & 2.08  & 2.0   &  0.4  &  2.0  & $-1.19$ \\
V355\,Lyr         & 6869  & 2.29  & 5.8   & 18.3  &  5.0  & $-1.15$ \\
V715\,Cyg         & 6265  & 2.19  & 3.7   & 11.1  &  4.6  & $-1.14$ \\
V808\,Cyg         & 6169  & 2.19  & 6.4   & 25.2  &  4.3  & $-1.14$ \\
V366\,Lyr         & 6263  & 2.26  & 5.1   & 19.8  &  5.3  & $-1.13$ \\
V783\,Cyg         & 6122  & 1.59  & 3.9   & 14.3  &  5.0  & $-1.10$ \\
KIC\,11125706     & 6145  & 1.67  & 2.3   & 14.7  &  6.6  & $-1.09$ \\
V838\,Cyg         & 6600  & 1.81  & 4.3   & 14.0  &  5.0  & $-1.01$ \\
V2470\,Cyg        & 6220  & 2.13  & 2.6   &  9.7  &  6.5  & $-0.59$ \\
V782\,Cyg         & 6446  & 3.17  & 3.7   & 12.0  &  4.6  & $-0.42$ \\
V839\,Cyg{\dag}   & 6982  & 2.46  & 3.5   & 18.4  &  5.4  & $-0.31$ \\ 
KIC\,3868420{\dag}& 7612  & 3.71  & 2.3   & 23.0  &  5.7  & $-0.28$ \\ 
KIC\,8832417      & 6999  & 3.30  & 3.5   & 15.3  &  4.8  & $-0.25$ \\
KIC\,5520878      & 7266  & 3.66  & 3.0   & 15.2  &  5.1  & $-0.18$ \\
KIC\,6100702      & 6471  & 2.83  & 3.1   & 10.2  &  4.7  & $-0.18$ \\
{\bf Sun}         & 5770  & 4.44  & 0.8   &  8.7  &  2.1  & $+0.00$ \\ 
V784\,Cyg         & 6305  & 2.84  & 3.6   & 13.4  &  5.7  & $+0.00$ \\
{\bf V445\,Oph}   & 6496  & 3.15  & 2.9   & 12.0  &  2.5  & $+0.14$ \\
{\bf SW\,And}     & 6996  & 3.70  & 4.2   & 16.2  &  2.9  & $+0.19$ 
\enddata
\tablecomments{
Most of the entries in this table are based on analysis of 476 lines in the 
wavelength interval 500-520\,nm.  The stars with a {\dag} following the name 
include additional measurements made using 952 lines in the interval 440-460\,nm. 
The names of the nine Keck standard stars (and the Sun) are printed in `boldface'.  
The V1104 Cyg results are based on analysis of the Keck spectrum (the CFHT spectra for
this star were too noisy to measure).    }
\end{deluxetable}

\subsubsection{Comparison of Atmospheric Parameters}

\begin{deluxetable*}{lcccccccccc}
\tabletypesize{\scriptsize}
\tablewidth{0pt}
\tablecaption{ Spectroscopic Iron-to-Hydrogen Abundances for the Keck Standard Stars } 
\label{tab:Table10}
\tablehead{
\colhead{Star} &       \multicolumn{6}{c}{[Fe/H] - previous papers}                                    & &   \multicolumn{3}{c}{[Fe/H] - this paper}  \\
\cline{2-7} \cline{9-11}     
\colhead{    } &\colhead{L94} & \colhead{SKK94}  &\colhead{C95}   & \colhead{L96}  &\colhead{Fer98} &\colhead{Wal10}& & \colhead{VWA}  & \colhead{MOOG}  & \colhead{SME}  \\
\colhead{(1)}  &\colhead{(2)}   &\colhead{(3)}     & \colhead{(4)}  &\colhead{(5)}     &\colhead{(6)}   &\colhead{(7)  }& & \colhead{(8)}  & \colhead{(9)}   & \colhead{(10)}     } 
\startdata
X~Ari          & $-2.40$ & $-2.16$  & $-2.44$ &  $-2.53$         & $-2.43$        & $-2.68$       & &  $-2.74\pm0.09$ &  $-2.59\pm0.12$  &  $-2.71\pm0.10$   \\ 
VY Ser         & $-1.82$ & $-1.83$  & $-1.65$ &  $-1.75$         & $-1.79$        & \nodata       & &  $-1.71\pm0.07$ &  $-1.58\pm0.11$  &  $-1.71\pm0.10$   \\ 
ST~Boo         & $-1.86$ & \nodata  & $-1.62$ &  \nodata         & $-1.76$        & $-1.77$       & &  $-1.62\pm0.06$ &  $-1.57\pm0.11$  &  $-1.61\pm0.10$   \\ 
RR~Cet         & $-1.52$ & $-1.55$  & $-1.32$ &  \nodata         & $-1.45$        & \nodata       & &  $-1.27\pm0.12$ &   \nodata        &  $-1.22\pm0.10$   \\ 
UU~Cet         & \nodata & $-1.88$  & $-1.32$ &  \nodata         & $-1.28$        & \nodata       & &  $-1.33\pm0.08$ &   \nodata        &  $-1.33\pm0.10$   \\ 
RR~Lyr         & $-1.37$ & $-1.37$  & $-1.32$ &  $-1.47$         & $-1.39$        & \nodata       & &  $-1.27\pm0.12$ &  $-1.36\pm0.10$  &  $-1.33\pm0.10$   \\ 
VX~Her         & \nodata & $-1.37$  & $-1.52$ &  \nodata         & $-1.58$        & $-1.48$       & &  $-1.23\pm0.12$ &  $-1.21\pm0.11$  &  $-1.38\pm0.10$   \\ 
SW~And         & $-0.38$ & $-0.41$  & $-0.00$ &  $-0.23$         & $-0.24$        & $-0.16$       & &  $+0.20\pm0.08$ &  $-0.12\pm0.13$  &  $+0.19\pm0.10$   \\ 
V445~Oph       & $-0.23$ & $-0.30$  & $+0.23$ &  \nodata         & $-0.19$        & $+0.24$       & &  $+0.13\pm0.10$ &  $+0.02\pm0.11$  &  $+0.14\pm0.10$      
\enddata
\tablecomments{The columns contain: (1) the star name;  
(2) [Fe/H] from Table~9 of L94, where [Fe/H] is based on the Ca\,II K-line;
(3) [Fe/H] from Table~3 of Suntzeff, Kraft \& Kinman (1994), based on $\Delta S$ measurements;
(4) [Fe/H] from Table~12 of C95,  made more metal rich by 0.06 since Clementini {\it et al.} (1995) assumed for the Sun log\,$\epsilon$(Fe)=7.56, whereas we are assuming 7.50;
(5) [Fe/H] from Table~5 of Lambert {\it et al.} (1996) - the log\,$\epsilon$ given, which are the Fe\,II values, have had the assumed solar Fe abundance (7.50) subtracted;   
(6) [Fe/H] from Table~1 of Fernley {\it et al.} (1998) -- these same values were adopted by Feast {\it et al.} (2008);
(7) [Fe/H] from Table~2 of Wallerstein \& Huang (2010), based on analysis of Apache Point Observatory spectra;    
(8-10) spectroscopic [Fe/H] values derived here using VWA, MOOG and SME.  }
\end{deluxetable*}

In {\bf Table\,10} the VWA, MOOG and SME iron abundances  are
compared.   Also included for comparison with our values are the [Fe/H] values derived by L94,
Suntzeff, Kraft \& Kinman (1994),  C95,  Lambert {\it et al.}
(1996, hereafter L96), Feast {\it et al.} (2008), and Wallerstein \& Huang
(2010).   Because C95 assumed log\,$\epsilon$(Fe)=7.56, whereas we adopted the value  
log\,$\epsilon$(Fe)=7.50, the C95 metallicity values (from their Table 12) have been increased by 0.06 dex.  
L96 report log\,$\epsilon$(FeII) values, to which we have added 7.50 to give the
abundances in column~5.  The L94 metallicities are
based on measurements of the Ca\,II K-line.

X\,Ari, with [Fe/H]$\sim$$-2.65$\,dex, is probably more 
metal deficient than NR~Lyr, the most metal-poor RR~Lyr star in the {\it Kepler} field, which we 
estimated to have   [Fe/H]=$-2.54\pm0.11$\,dex (Table 7).      
Recently Haschke {\it et al.} (2012) derived a spectroscopic [Fe/H]=$-$2.61\,dex for X~Ari, a value 
consistent with our estimate  and with the value derived by Wallerstein $\&$ Huang (2010).
It is unclear whether the metal-rich end  is defined by the standard star V445~Oph or by SW~And.  
Our VWA and SME analyses suggest that  the [Fe/H] values for both stars are greater 
than that of the Sun, while the MOOG analyses suggest that SW~And may be more metal-poor and  V445~Oph more metal rich than the Sun.  
Most previous analyses lean toward V445~Oph being the more
metal rich, as does our photometric [Fe/H] (see Figs.\,4 and 12).  
In any case, both have metallicities similar to that of the most
metal-rich star in the {\it Kepler} field, V784~Cyg.

For VY~Ser a metal abundance [Fe/H]=$-1.77\pm0.10$\,dex was derived by Carney \& Jones (1983).  The
good agreement between this value and our  VWA, MOOG and SME estimates is  undoubtedly due to  their spectra and ours having 
been taken at similar phases (0.61 versus our 0.48).  Furthermore, 
their log\,$g$=2.3$\pm$0.3  and  $\xi_t$=4.2$\pm$0.5 km/s agree with our SME values of 2.1 (cgs units) and 4.4 km/s; 
and their $T_{\rm eff}$=6000$\pm$150\,K is in
accord with our SME estimate of 6167\,K.



In {\bf Fig.\,8} the SME estimates of the atmospheric parameters $T_{\rm eff}$, log\,$g$, $\xi_t$ and [M/H]
are compared (solid squares) with the values derived using VWA.  While the results are 
similar, the temperatures from SME tend to be smaller than the VWA values (upper left panel), the
mean difference being $\sim$100\,K, with an RMS scatter $\sim$100\,K.  The surface gravities from
SME  are also  smaller on average than the VWA values (lower left panel) --  mean $\Delta$log\,$g$$\sim-0.5$ 
with an RMS scatter of 0.3.   The trend seen in  $\xi_t$
(upper right panel) implies that the SME-derived values are systematically larger than the VWA
values in most cses, where the difference increases  as $\xi_t$ increases (not shown in the
diagram is the extreme outlier V360~Lyr, where  $\Delta\xi_t=-4.7$ km/s is due primarily to  the
large $\xi_t$ (=10 km/s) adopted for the VWA analysis).   Despite these differences the lower right
panel shows that there is reasonably good agreement of the [M/H] abundances derived with SME and the
[Fe/H] values derived using VWA (and MOOG) -- the outliers at  the bottom of the panel are
V1107~Cyg,  V839~Cyg, and VX~Her (a standard star). 
 For the VWA analyses, the $v_{\rm mac}$
values were generally set to 2-3 km/s and  $v$\,sin$i$ values of 2 km/s  and 7-10 km/s were adopted for the Keck
and  CFHT spectra, respectively.  These values differ considerably from  the $v_{\rm mac}$ values derived by
SME, which  have a mean of 13.0 km/s (standard deviation $\sigma$=5.4 km/s, N=43 stars), and 
from  the $v$\,sin$i$ values (see bottom panel of Fig.9) derived by SME, which have a mean of 4.3\,km/s ($\sigma=1.2$\,km/s, N=40, excluding the three stars
with  $v$\,sin$i$\,$>9$\,km/s). 

\begin{figure} \figurenum{8} \epsscale{1.2} \plotone{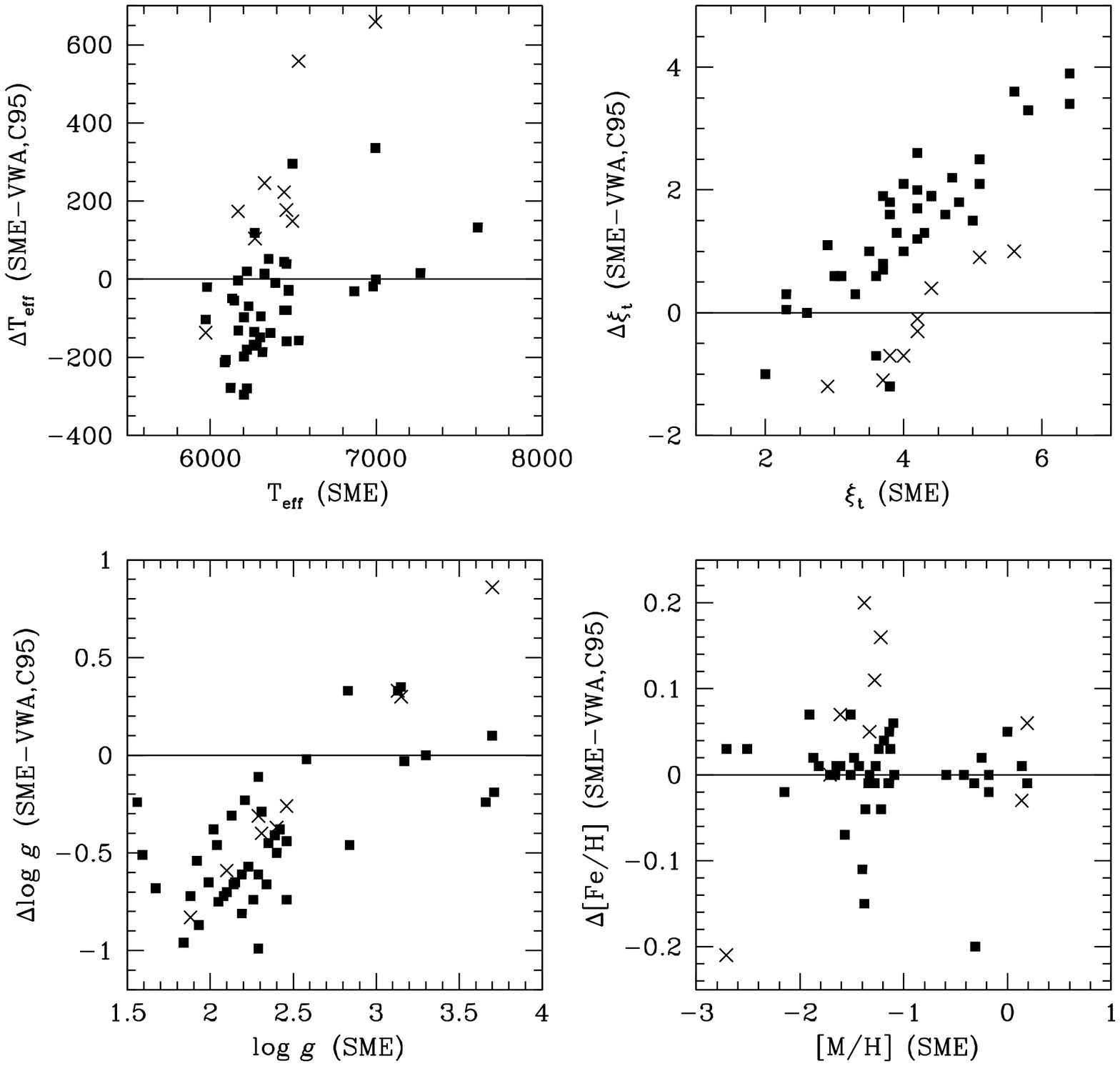} \caption{Comparison of  physical
parameters derived using SME with those derived using VWA (or assumed for the VWA analyses).  Also
shown (as crosses) are the parameter differences for the nine Keck standards, as derived using SME
compared with the values derived independently by C95. }
\label{fig:smeVSvwa} \end{figure}

Also plotted in Fig.\,8  (as crosses) are the differences between our SME-derived parameters for the
nine Keck RR~Lyrae standard stars and the corresponding values derived by C95.  Since the
pulsation phases  for our Keck spectra differ from those of  the Palomar 1.5-m spectra analyzed by C95
the effective temperatures are expected to differ.  On the other hand, because
the RRab stars lie on the red side of the instability strip and both studies used spectra taken away
from maximum light the $T_{\rm eff}$ should be similar;  in fact, $\Delta T_{\rm
eff}$(SME$-$C95) (upper left panel) does not exceed 700~K and is  typically  $\sim$200\,K.  
The largest differences occur for
VX~Her (558\,K) and for SW~And (659\,K).  The difference for VX~Her is readily explained by the
different $\phi_{\rm puls}$ values: the average pulsation phase for the five C95 spectra is
$\phi_{\rm puls}$=0.61 and therefore they measured $T_{\rm eff}$=5978\,K;  on 
the other hand the Keck spectrum was taken at $\phi_{\rm puls}$=0.32 and the average $T_{\rm eff}=6660\pm70$\,K 
from the VWA, MOOG and SME analyses is hotter.   
The large temperature difference for SW\,And is problematic. The average
$\phi_{\rm puls}$ for the three C95 spectra is 0.82, which is to be compared with $\phi_{\rm
puls}$=0.56 for our Keck spectrum.  According to For, Sneden \& Preston (2011, Figs.8-10) both
phases are near minimum temperature and therefore  $T_{\rm eff}\sim$6200\,K is expected in both cases.   
The $T_{\rm eff}=6337$\,K derived by C95 is consistent with this
prediction, as is the $T_{\rm eff}$ from our MOOG analysis;  however, both the VWA and SME analyses
suggest a higher temperature, and as a consequence 
a slightly  higher [Fe/H] value.

The comparison of surface gravities (Fig.\,8, lower left panel) is intriguing, with the differences for the
nine standard stars lying along the line $\Delta$log$g$\,$=0.89\,$log\,$g$(SME)$- 2.46$, {\it i.e.}, the 
gravities derived by SME tend to be smaller than
the C95 values when log\,$g$ is less than 2.76 and larger when log\,$g$ is greater than 2.76.
A linear trend is also seen when our VWA  $\xi_t$ estimates for the standard stars (crosses in the upper
right panel) are compared with  the  C95 microturbulent parameter, where the former values are offset by 
$\sim$2 km/s. 
Despite these atmospheric parameter differences, the metallicities derived here and by 
C95 are all within $\sim$0.2~dex of each other (bottom right panel), with a difference  $<$\,0.1 dex in most cases.

\subsubsection{Correlations of Atmospheric Parameters}

\begin{figure} \figurenum{9} \epsscale{1.15} \plotone{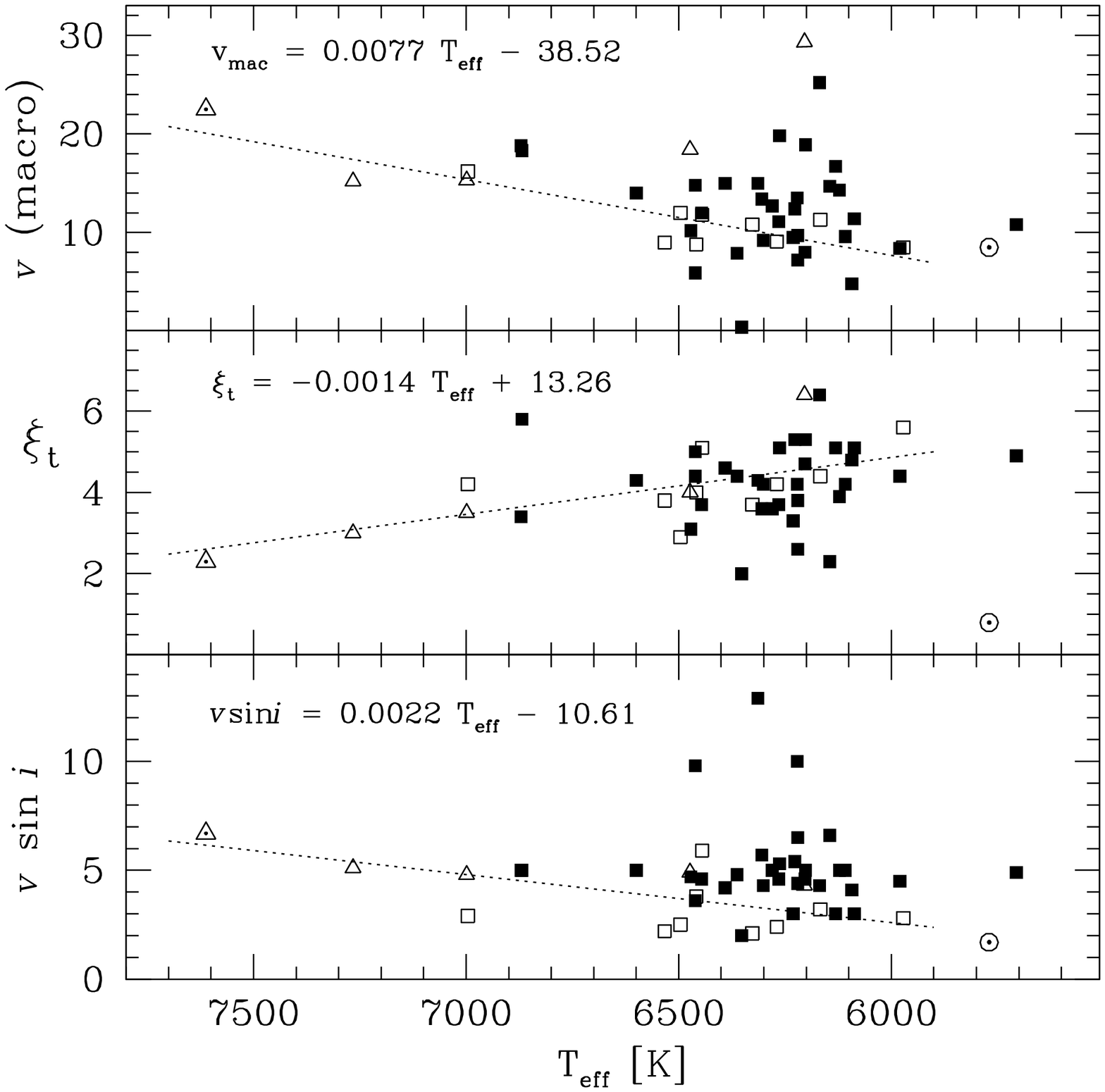} \caption{ Macroturbulent velocity
(top panel), microturbulent velocity (middle panel), and projected rotation velocity (bottom panel) for
the program and standard stars, all derived with SME, and all with units of [km/s].  The equations of the 
fitted lines (based on the high-S/N standard star data only) are as indicated on the graphs.  
Included in the diagrams are the program star RRab stars (solid black squares), 
the RRc stars (open triangles),  the  bright Keck RRab standard stars (open squares), the Sun (circled dot) and
the candidate HADS star KIC\,3868420 (dotted triangle).  } \label{fig:loggVSfeh}
\end{figure}

{\bf Fig.\,9} shows the SME-derived atmospheric parameters $v_{\rm mac}$,  $\xi_t$,  and $v$\,sin$i$, 
plotted as a function of $T_{\rm eff}$.   Each panel includes the values derived for the program RR~Lyr stars, for the
Sun, for the candidate HADS star KIC\,3868420, and for the bright
Keck RR~Lyrae standard stars.
As  expected, the RRc stars, KIC\,5520878 and KIC\,8832417 (both of which
are metal-rich) are the hottest among the RR~Lyr stars.  
The  candidate HADS star KIC\,3868420 (also metal-rich) is even hotter. 
Differences  between  KIC\,3868420 and the {\it bona fide} hot RRc stars appear to be comparatively small.
While this observation lends support to the possibility that it may be an unusual metal-rich short-period RRc star,
possibly an RRe (second overtone?) pulsator (Walker \& Nemec 1996; Kov\'acs 1998),
identification as a hot HADS star of 1.6 to 2 solar masses seems more likely.  
Balona \& Nemec (2012) found  that $\delta$~Sct stars hotter than log\,$T_{\rm eff}=3.88$ are quite common and
suggested that KIC\,3868420, based on its  photometric and kinematic properties, 
is not an SX~Phe star ({\it i.e.}, a halo population $\delta$~Sct star such as are found among the blue
stragglers in globular clusters).
The metal-poor RRc star KIC\,9453414 has the highest derived macroturbulent velocity of
all the program stars, {\it i.e.}, $v_{\rm mac}=29$ km/s.  The three stars with the greatest projected
rotation relocities are the non-Blazhko RRab stars KIC\,9658012, KIC\,9717032 and KIC\,K7030715 -- all
the other stars have $v$\,sin$i$ values smaller than 7 km/s.   The RRab star with the coolest
$T_{\rm eff}$ is V1510~Cyg, a non-Blazhko star for which we were unable to perform a successful VWA analysis of the 
CFHT spectra.    V808~Cyg (a Blazhko star) also has a high $v_{\rm mac}$ and
V1104~Cyg (another Blazhko star) has the extremely low value of $v_{\rm mac}=0.4$\,km/s (see top panel of Fig.\,9) 
For the standard stars (all with high S/N spectra) and the
RRc stars, in Fig.\,9 three linear trends are apparent:
the hottest stars tend to have higher $v_{\rm mac}$, lower $\xi_t$, and higher
$v$\,sin\,$i$ velocities than the cooler stars.  Equations of the fitted trend lines 
are given on the graphs.   It is  interesting to note that the $v_{\rm mac}$ and $v$\,sin\,$i$  trends are similar to  those
for the 23 bright solar-type stars studied by Bruntt {\it et al.} (2010b) and
summarized in their Fig.11, although at a given $T_{\rm
eff}$ the RR\,Lyr stars appear to have higher $v_{\rm mac}$ values than the
solar-type stars.

\begin{figure*} \figurenum{10} \epsscale{1.1}
\plottwo{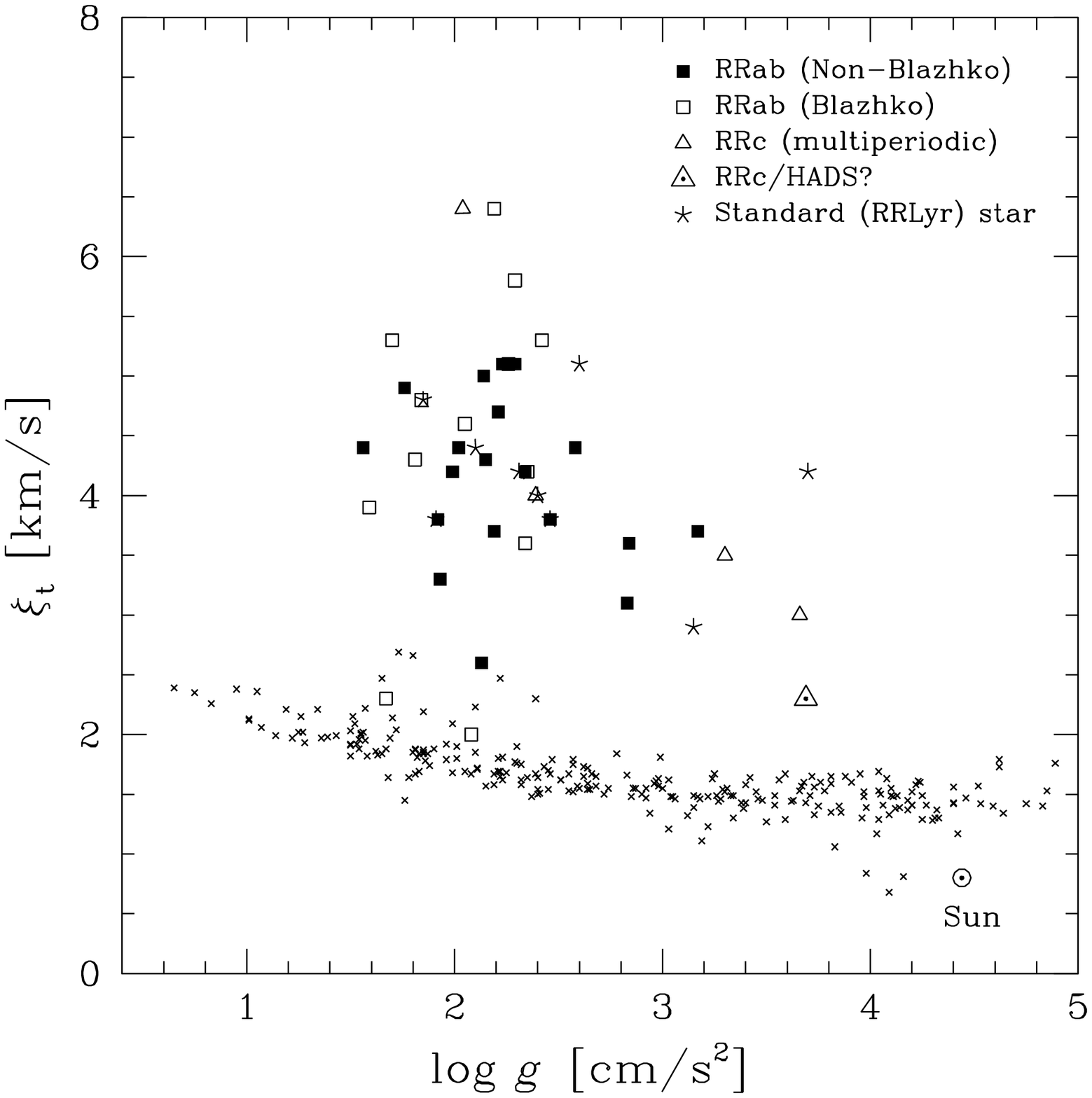}{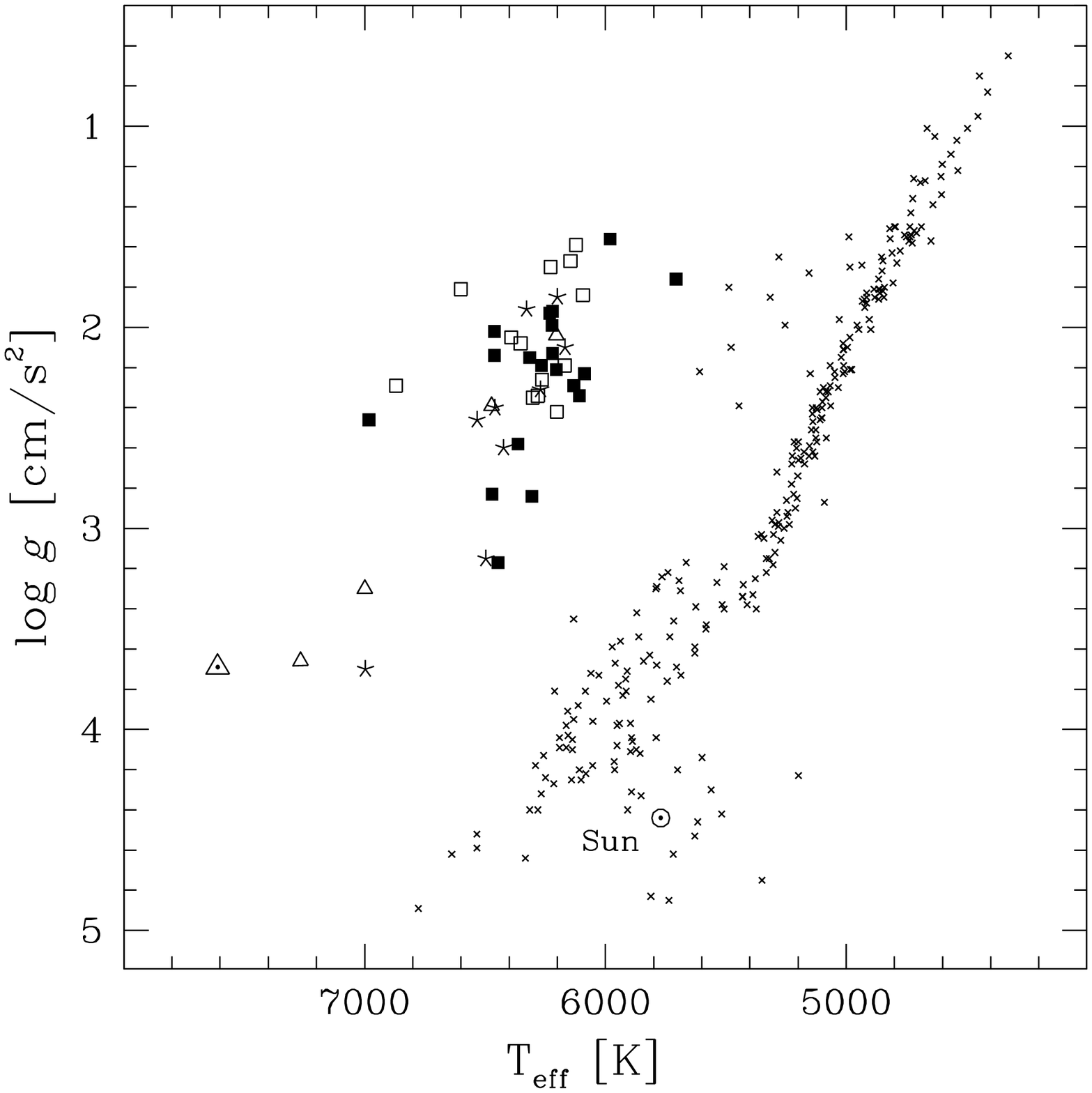}
\plottwo{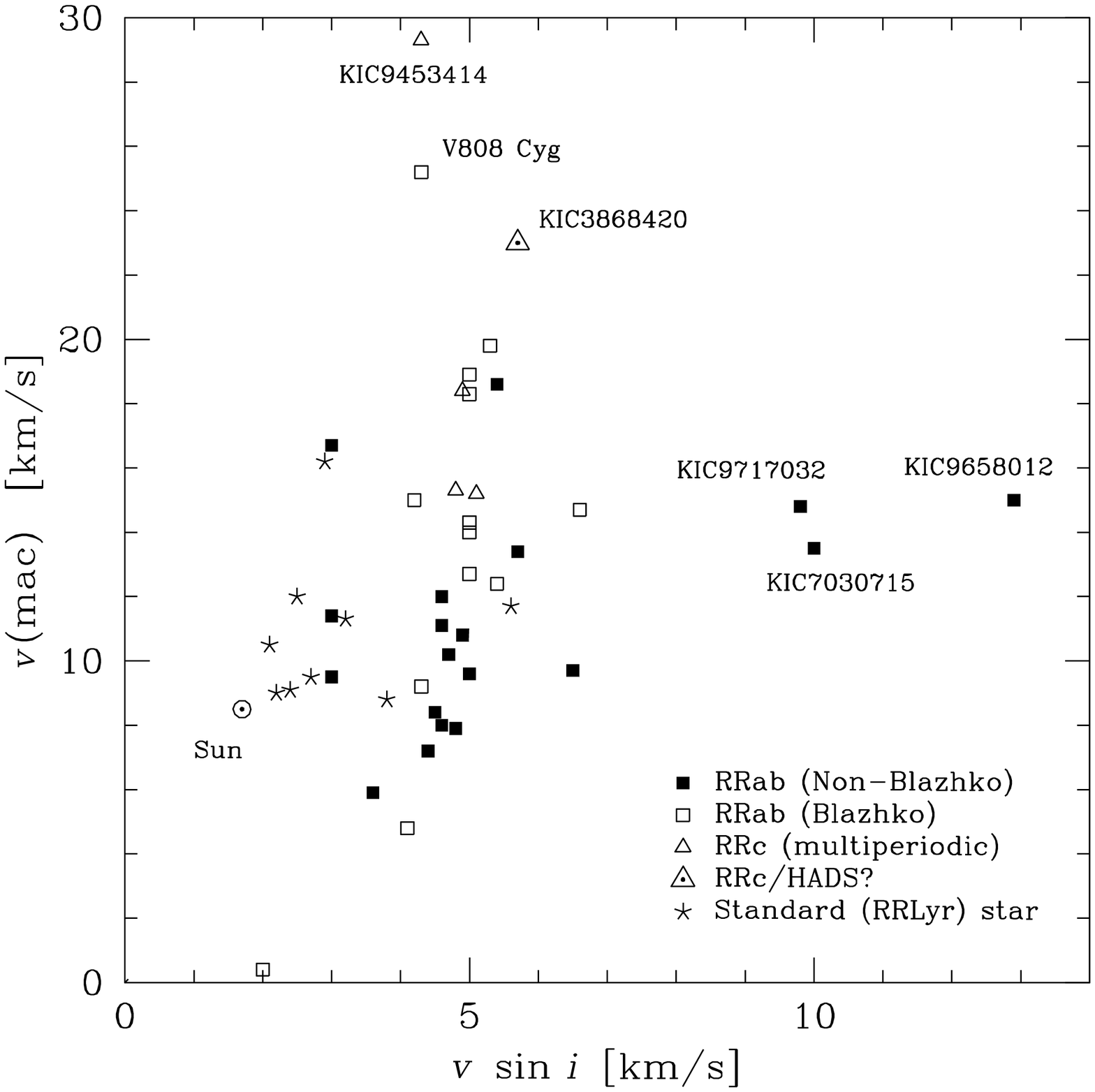}{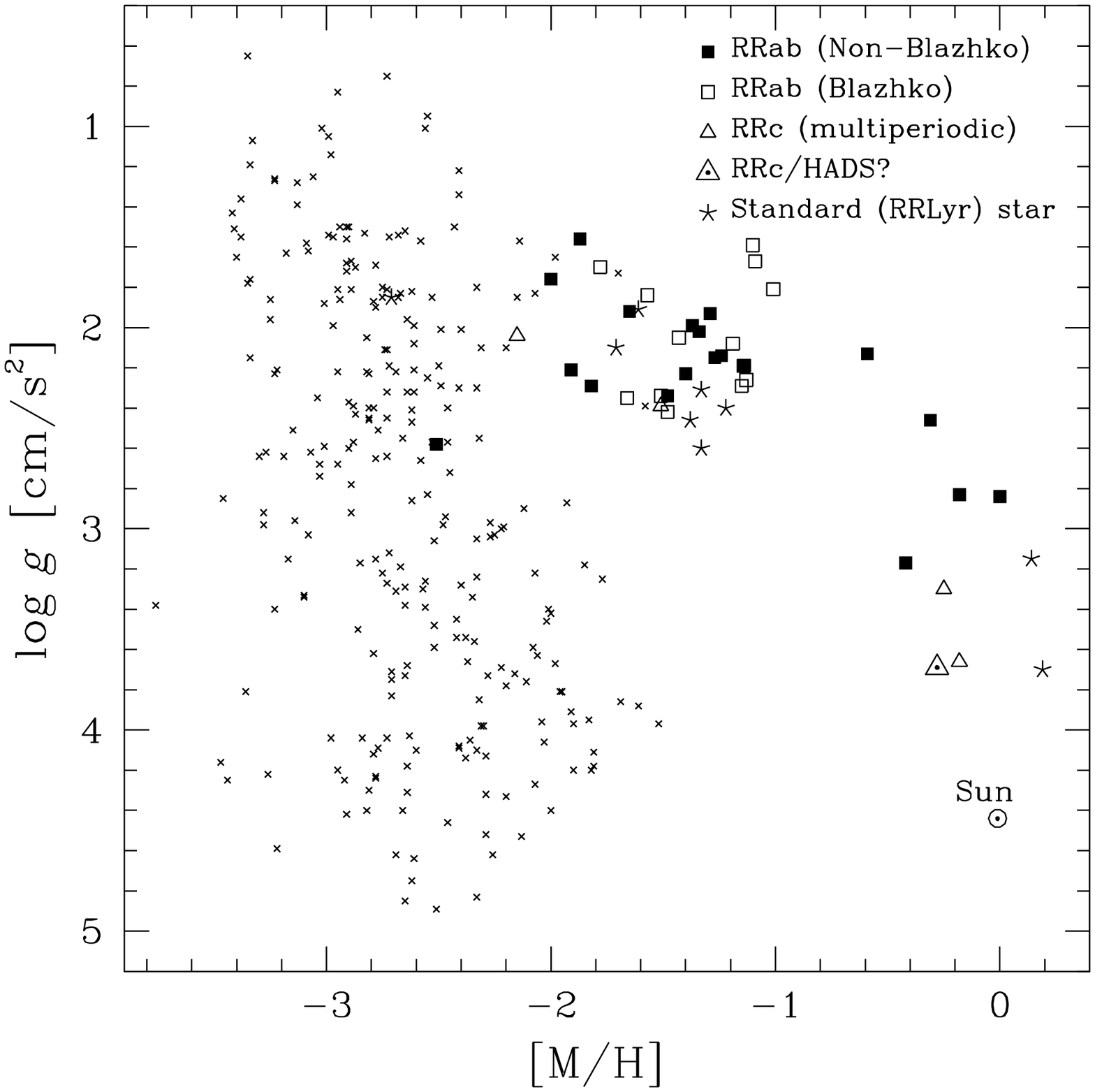}
\caption{Four diagrams comparing the atmospheric parameters for the {\it Kepler}-field RR~Lyrae
stars, the nine RR~Lyrae standards, and the Sun (all derived with SME), with the same parameters
(also derived with SME) for the HERES metal-poor halo red giants and red horizontal branch stars
(from Barklem {\it et al.} 2005).   {\bf [Upper Left]}  Microturbulent velocity, $\xi_t$ [km/s], versus
surface gravity, log\,$g$ [cm/s$^2$] - the HERES stars are plotted as small crosses; {\bf [Upper
Right]} Surface gravity, log\,$g$ [cm/s$^2$], versus effective temperature, $T_{\rm eff}$ [K] - here
the HERES stars clearly are seen to be mainly red giants, subgiants and $\sim$10 red horizontal
branch stars;  {\bf [Lower Left]} Macroturbulent velocity, $v_{\rm mac}$, versus projected rotation
velocity, $v$\,sin$i$ [km/s], with labels identifying the most extreme stars ($v_{\rm mac}$ and
$v$\,sin$i$ were not given for the HERES stars); {\bf [Lower Right]} Surface gravity, log\,$g$
[cm/s$^2$],  versus metal abundance, [M/H].  } \label{fig:4panelsme} \end{figure*}

In {\bf Fig.10} the atmospheric parameters derived with SME for the RR~Lyr stars are compared with
the same quantities (also derived using SME) for the 253 halo metal-poor red giants and red
horizontal branch (RHB) stars investigated by Barklem {\it et al.} (2005) as part of the HERES
survey.  In all four panels KIC\,3868420 (dotted triangle)  appears to be similar
to the two metal-rich RRc stars KIC\,5520878 and KIC\,8832417, suggesting again  that
it might be a short period RRc star and not a HADS star.
The $\xi_t$ {\it vs.} log\,$g$  diagram (upper left panel) shows that  the RR~Lyrae stars  have
systematically higher microturbulent velocities than 
the HERES metal-poor red giants and RHB stars with the same log\,$g$.  
The log\,$g$ {\it vs.} $T_{\rm eff}$ diagram (upper right panel),  where 
the log\,$g$ scale is reversed,  closely resembles an H-R diagram.  At a
given surface gravity the RR~Lyrae stars are hotter than the RHB stars, which
in turn are hotter than the metal-poor red giants.   As before the hottest stars are the
RRc stars (and KIC\,3868420), and the two metal-rich RRc stars are seen to have higher
surface gravities and temperatures  than the two metal-poor RRc stars.  
The $v_{\rm mac}$ versus projected rotational velocity diagram (lower left panel) identifies those
stars with the most extreme values of $v_{\rm mac}$  and  $v$\,sin$i$ (see earlier discussion).
The mean $v_{\rm mac}$, excluding the three outliers with $v_{\rm mac}$ greater than 20 km/s, equals
12.1 km/s ($\sigma$=4.2\,km/s, 40 RR\,Lyr stars), and the corresponding mean $v$\,sin$i$ (after eliminating 
the three $v$\,sin\,$i$ outliers) is 4.3 km/s ($\sigma$=1.2\,km/s, 40 RR\,Lyr stars).  
The surface gravity versus metallicity diagram (lower right panel) reveals a clear trend between
log\,$g$ and [M/H]:  the metal-rich RR~Lyrae stars have larger surface gravities (hence smaller
luminosities) than the more metal-poor RR~Lyrae stars, which was explained  by Sandage's (1958) 
`stacked horizontal-branch luminosity levels' model (see Bono {\it et al.} 1997, and discussion in N11).
No significant differences between the Blazhko and non-Blazhko 
stars are evident in any of the diagrams of Fig.\,10.

\section{PHOTOMETRIC METAL ABUNDANCES FROM  \\ $P$-$\phi_{\rm 31}$-Fe/H] RELATIONSHIPS}

Derivation of metal abundances from
correlations relating pulsation period, spectroscopic [Fe/H], and  
one or more parameters that characterize the shape of the photometric light curve (e.g., $\phi_{\rm 31}^s$, $A_{\rm tot}$, risetime, etc.) was discussed in $\S3$.
In this section we use the accurate pulsation periods and Fourier light curve parameters presented in $\S$3,
and the spectroscopic metal abundances derived in $\S4$, to derive 
new $P$-$\phi_{\rm 31}$-[Fe/H] regressions for the {\it Kepler}-field 
RRab and RRc stars.   Using our fitted models we estimate  [Fe/H]$_{\rm phot}$ for the program stars
and investigate the applicability of the models to Blazhko and  non-Blazhko stars.
Because the spectroscopic [Fe/H] values are derived from high dispersion spectra analyzed using standard reduction procedures, 
the derived metallicities are on the scale of the high dispersion spectroscopy ({\it i.e.}, the Carretta {\it et al.} 2009, hereafter C9 scale).  

\begin{figure} \figurenum{11} 
\epsscale{1.17} \plotone{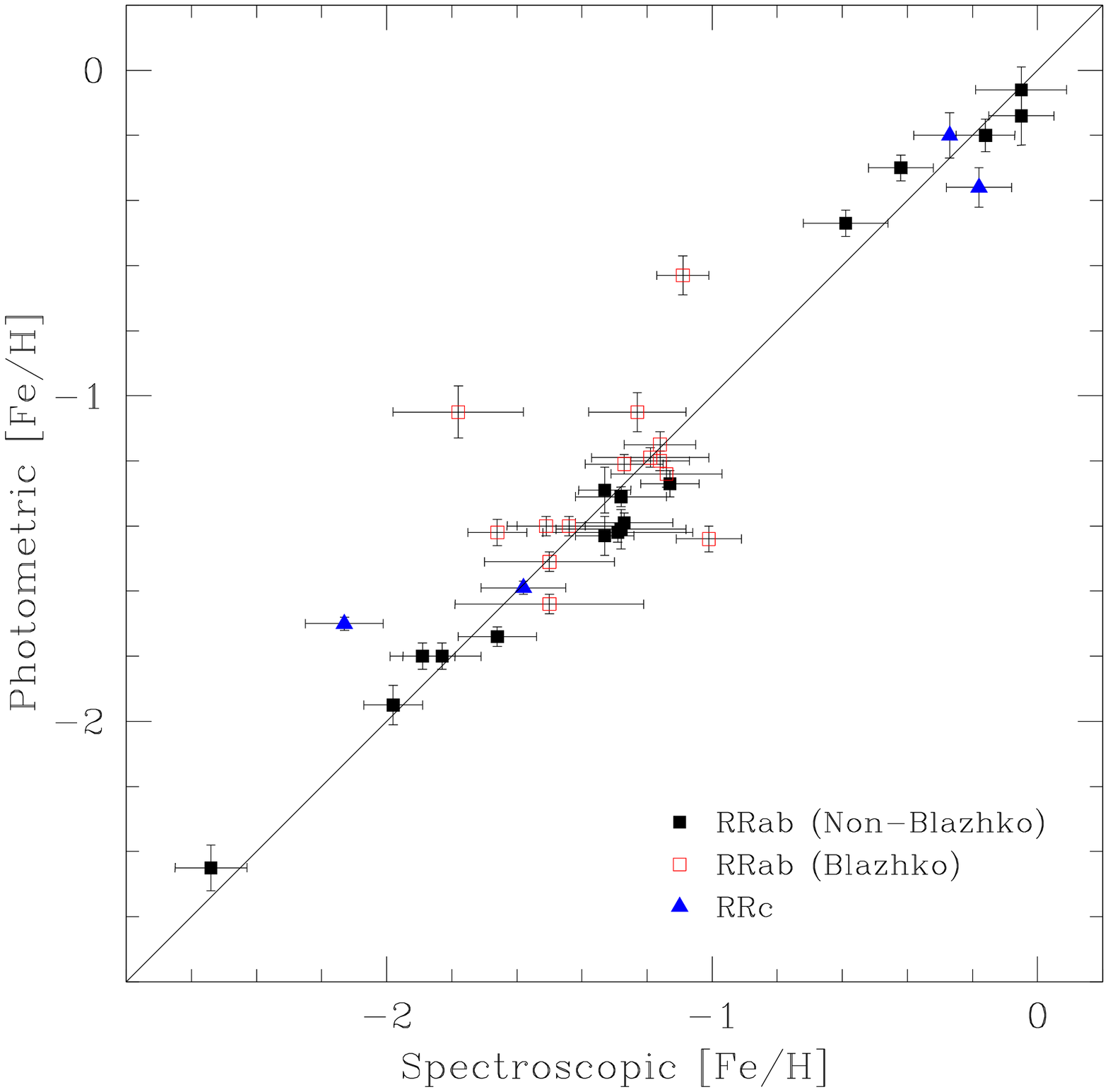}
\epsscale{1.1} \plotone{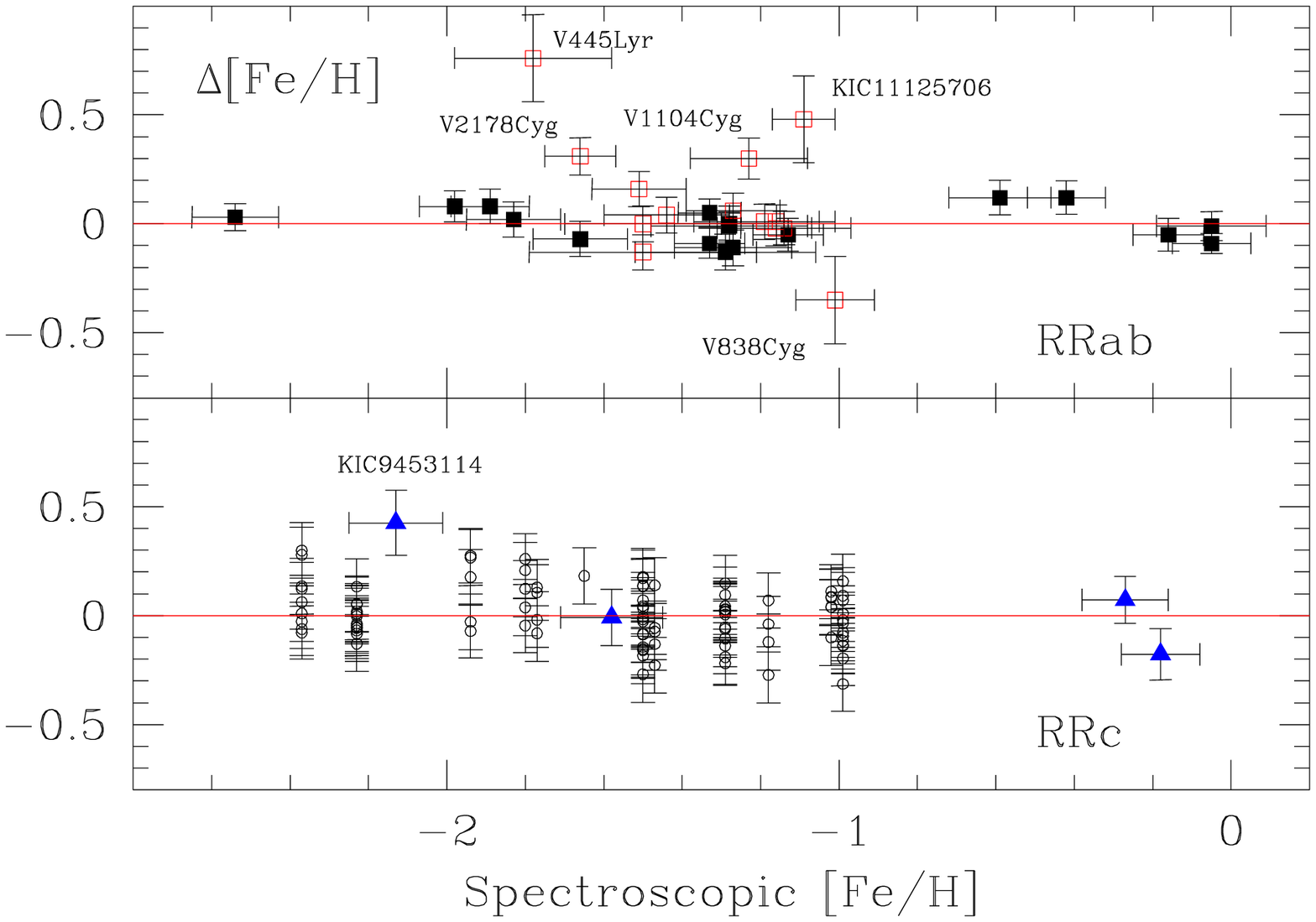}
\caption{Comparison of the spectroscopic and photometric metallicities for the {\it Kepler}-field RR~Lyrae
stars.   The diagonal line in the top panel is the 1:1 relation.
The bottom panel contains the [Fe/H] differences (photometric minus spectroscopic)
for the RRab (upper) and RRc (lower) stars, where the vertical error bars are the standard errors of the differences. 
Included in the lowest graph are the residuals for the 
M07 globular cluster RRc calibration data set (small open circles).
A color version of this figure is given in the on-line version of the paper.    } 
\label{fig:PHvsSP} 
\end{figure}

\subsection{RRab Stars }

Several models were evaluated  for the {\it Kepler}-field RRab stars and the optimum fit was achieved with the following nonlinear model:
\begin{equation}
 [{\rm Fe/H}] = b_0 + b_1 P +  b_2 \phi_{31}^s + b_3 \phi_{31}^s P + b_4 (\phi_{31}^s)^2,
\end{equation}
\noindent where   $\phi_{\rm 31}^s$ is the mean $\phi_{\rm 31}^s$({\it Kp}) value and  [Fe/H] is the VWA spectroscopic value (see Table~7).
Excluded from the final calibration were five Blazhko stars with large residuals: 
V445~Lyr and V2178~Cyg, both of which exhibit extremely large amplitude and frequency modulations (see Table~2, and 
right side of Fig.\,3),
and KIC\,11125706, V838\,Cyg and V1104\,Cyg,  which have small amplitude and frequency modulations (see Table~2, Figs.\,2-3 and $\S3.2.1$). 
Also excluded from the calibration were two non-Blazhko stars with unreliable metallicities, V346\,Lyr and V1510\,Cyg.   
The resulting  estimated  coefficients and their standard errors are:  $b_0= -8.65 \pm 4.64$, $b_1 = -40.12\pm 5.18$, $b_2 = 5.96 \pm 1.72$, $b_3= 6.27\pm 0.96$ and $b_4= -0.72\pm 0.17$.
The RMS error of the fit was 0.084\,dex, with adjusted $R^2$=0.98.  Seventeen of the 26 calibration stars were non-Blazhko variables and nine were Blazhko stars. 
The  photometric [Fe/H] values derived from the fitted model are given in 
column 9 of Table~1, where the estimated standard errors were calculated using the usual
Gaussian theory formula for nonlinear least squares regression (analogous to Eqns.4-5 of JK96, except that we  have ignored the 
errors in $\phi_{\rm 31}^s$ and $P$ owing to the high accuracy of the {\it Kepler} photometry).

In {\bf Fig.~11} the fitted photometric metallicities are plotted against the spectroscopic metallicities (top panel).
Also shown (bottom panel) are the differences,  $\Delta$[Fe/H] = [Fe/H]$_{\rm phot} - $ [Fe/H]$_{\rm spec}$, plotted against [Fe/H]$_{\rm spec}$.  
Included in the graphs are the five Blazhko stars that were excluded from the final calibration (labelled in the
bottom panel).  All of the calibrating stars lie close to the 1:1 line, with $\mid$$\Delta$[Fe/H]$\mid < 0.13$\,dex for all 17 non-Blazhko stars.    

The largest metallicity differences correspond to the five Blazhko stars excluded from the calibration.
It is perhaps not surprising that the two most extreme Blazhko stars (V445~Lyr, V2178~Cyg) showed the greatest disagreement. 
Since both stars exhibit considerable variation in $\phi_{\rm 31}^s$ over a Blazhko cycle, and since the calibration model (Eqn.\,2) is nonlinear,
it is questionable (Jensen's inequality) whether substitution of mean $\phi_{\rm 31}$ rather than averaging the `instantaneous' [Fe/H]
values is the appropriate way of calculating [Fe/H]$_{\rm phot}$.
The [Fe/H]$_{\rm phot}$ given in Table~1 were calculated using the first ({\it i.e.}, substitution of mean-$\phi_{\rm 31}^s$) method.
To investigate the second method we computed [Fe/H]$_{\rm phot}$ for the five Blazhko stars by averaging (over complete Blazhko cycles) 
the metal abundances evaluated (Eqn.\,2) at each `instant' along the $\phi_{\rm 31}^s$ time series. 
As expected,  the two methods gave similar results for the three low amplitude/frequency modulators.  
However, for V445~Lyr and V2178~Cyg  averaging over the instantaneous abundances gave [Fe/H]$_{\rm phot}$ =  
$-1.69$ and $-1.65$\,dex, respectively, both of which are much closer to the spectroscopic abundances:
$-1.78$\,dex for V445~Lyr (which is from the SME analysis since the VWA analysis was inconclusive), and $-1.66\pm0.13$\,dex for V2178~Cyg.
 
Of course the spectroscopic metallicities are also uncertain and contribute to the differences.
Four of the five Blazhko stars seem to have reliable (and consistent) metal abundances from the VWA and SME analyses. 
The largest difference, that for V445~Lyr (Fig.11), was calculated assuming the SME abundance, $-1.78$\,dex (which appears to be consistent with the 
partial spectrum shown in Fig.7a).  Consideration of the pulsation phases at the times of the spectroscopic observations, and of the
Blazhko properties of the stars, reveal no obvious patterns: the pulsation phases range from 0.31 to 0.43;  
four of the five stars have Blazhko periods $\sim$50 days, while V2178~Cyg has a considerably longer $P_{\rm BL}$=234$\pm$10\,d;
and the Blazhko phases range from 0.32 to 0.82.  

We conclude that since the large metallicity differences found for extreme Blazhko stars are reduced when  `instantaneous' [Fe/H]$_{\rm phot}$ values are averaged
instead of calculated using the mean-$\phi_{\rm 31}$ method, and that the two methods give the same result for the non-Blazhko and
less extreme Blazhko stars, that this method might be preferrable for future [Fe/H]$_{\rm phot}$ calculations.

In {\bf Fig.~12} two sets of isometallicity curves have been added to the $P$-$\phi_{\rm 31}$ diagram (see Fig.\,4). 
The solid curves (black) were calculated by solving our nonlinear relation (equation 2) for five [Fe/H] values (ranging from $-2.0$ to $-0.1$\,dex), and 
the dashed lines (blue) were calculated using the well-known JK96 formula (their equation 3) and solving for [Fe/H]=$-2.0$, $-1.0$ and $0.0$ dex. 
Since our photometry is on the {\it Kp} system it was necessary to transform the JK96 formula to the {\it Kp} system, which was done using
equation 2 of N11, {\it i.e.},  $\phi_{\rm 31}^s$($V$) = $\phi_{\rm 31}^s$({\it Kp})$ - 0.151(\pm0.026)$, resulting in  
\begin{equation}
{\rm [Fe/H]} = -5.241 - 5.394 P + 1.345 \thinspace \thinspace \phi_{31}^s({\it Kp}).
\end{equation}
\noindent  The original version of this relation was established using a calibration data set consisting of 81 galactic-field RR~Lyr stars with reliable $V$ light curves and
high dispersion spectroscopic (HDS) abundances.  It was considered optimal after various linear and nonlinear models were investigated, and 
gives photometic abundances on the HDS metallicity scale established by Jurcsik (1995).
It is applicable for stars meeting the  {\it compatibility condition}, $D_m$$<$3, a criterion that 
fails to hold for stars with peculiar light curves, such as certain Blazhko variables and RR~Lyr stars in advanced evolutionary stages. 

In general our nonlinear model and the JK96 model agree for [Fe/H]$<$$-1.0$\,dex, with progressively larger discrepancies occurring as [Fe/H] decreases.  
It is well-known that for the lowest [Fe/H] stars the JK96 formula gives metallicities too metal-rich by $\sim$0.3 dex, the problem having been identified by
JK96 themselves and discussed further by Nemec (2004) and Smolec (2005). 
Our model appears to correct this problem (compare the [Fe/H]$=-2.0$\,dex curves in Fig.12), in part due to the wider inclusion in our sample
of stars more metal-poor than $-2.0$\,dex, as well as our use of a nonlinear rather than linear model.  It should  also be noted that JK96 adopted 
a spectroscopic [Fe/H] of $-2.10$ dex for X~Ari, whereas the spectroscopic metallicities in Table~10 suggest a much lower value,  [Fe/H]$\sim$$-2.65$\,dex. 

The {\it Kepler}-field RR Lyr stars divide into two groups, those with  $\phi_{\rm 31}^s$({\it Kp})$\sim$5 having low metallicities, [Fe/H]$<$$-1$ dex,
and a high metallicity group with $\phi_{\rm 31}^s$({\it Kp})$\sim$5.7.  Four of the five {\it Kepler} stars comprising the latter group were 
identified as metal-rich by N11 (V782~Cyg, V784~Cyg, KIC\,6100702 and V2470\,Cyg);  to this group we
can now add V839~Cyg (and the Keck standard stars V445~Oph and SW~And).

\begin{figure}  \figurenum{12} \epsscale{1.15}
\plotone{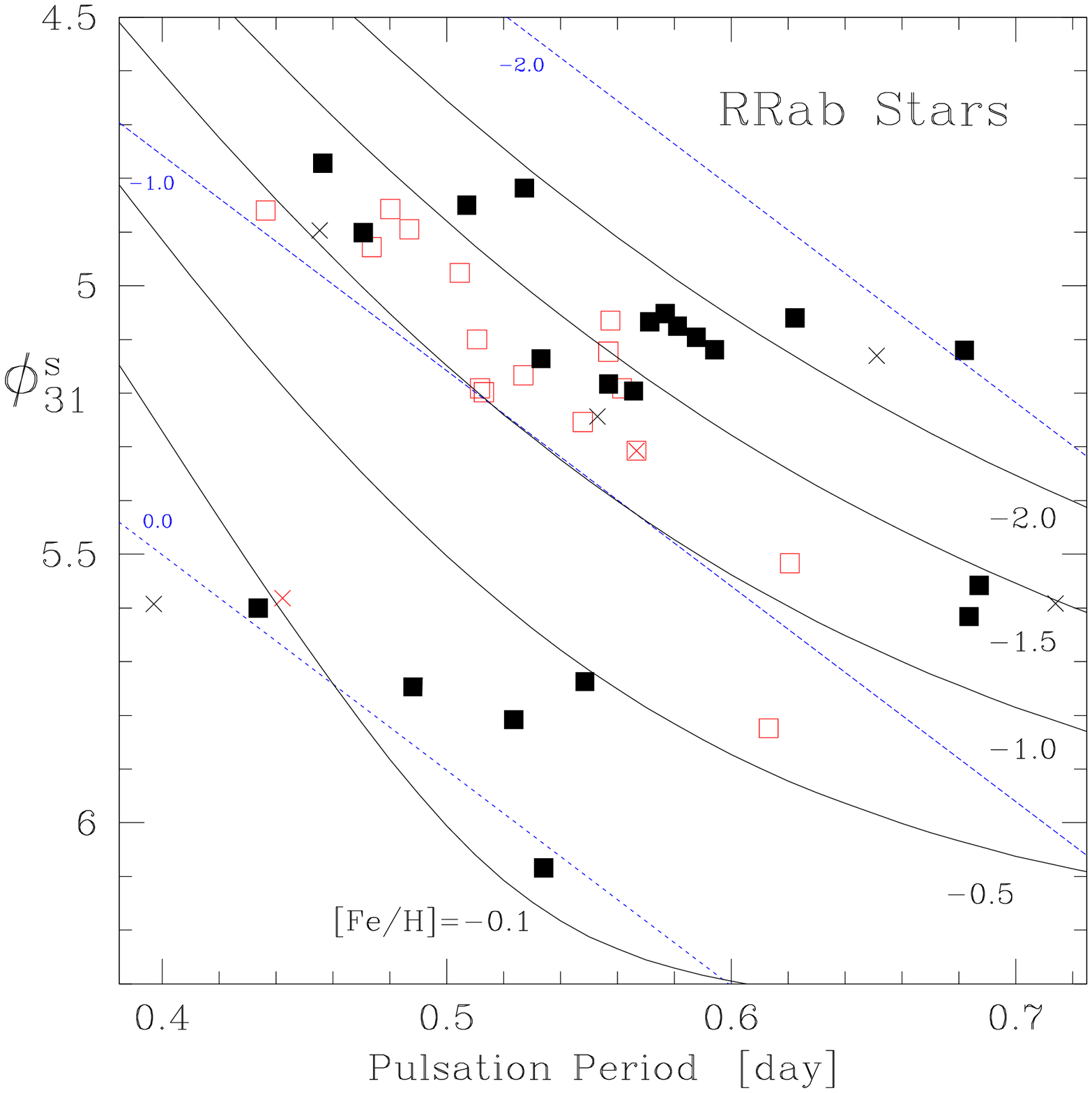}
\caption{Period-$\phi_{\rm 31}$ diagram for the {\it Kepler}-field RRab stars.   Also show are five isometallicity curves (black solid curves) derived from
our nonlinear $P$-$\phi_{31}^s$-[Fe/H] relation (eqn.~2), and three isometallicity curves (blue dotted lines) derived from the JK96 linear relation.     
The {\it Kepler}-field non-Blazhko RRab stars (black filled squares) and Blazhko RRab stars (red open squares) are also plotted, along with six Keck 
spectroscopic standard stars (crosses).  The $\phi_{\rm 31}^s$ values were derived from sine-series Fourier decomposition of the {\it Kp}-passband 
observations.  A color version of this figure is given in the on-line version of the paper.  } 
\label{fig:P-phi31-RRab} 
\end{figure}

\begin{figure} \figurenum{13} \epsscale{1.15} \plotone{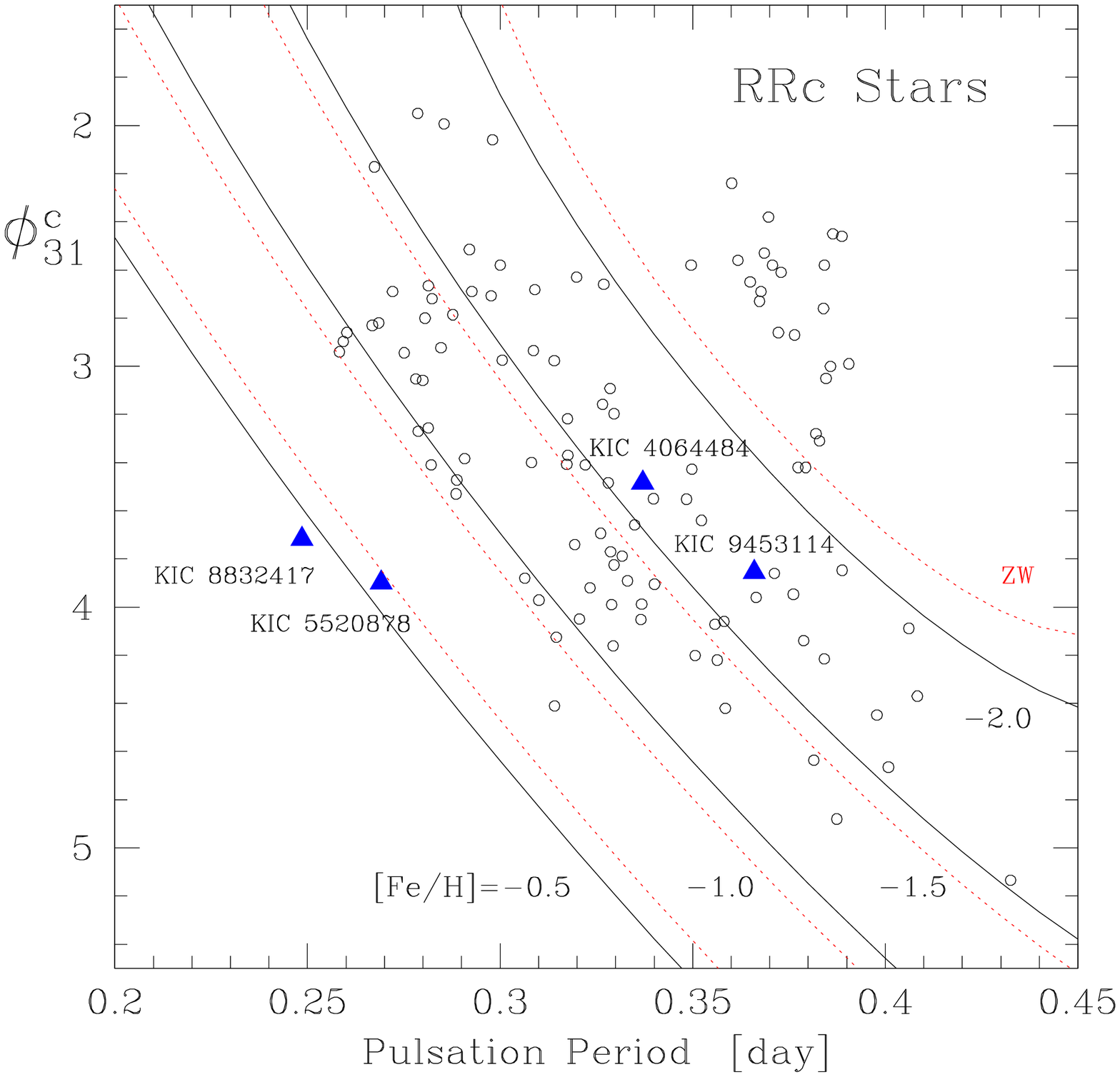}
\caption{Period-$\phi_{\rm 31}$ diagram for RRc stars.  The {\it Kepler}-field RRc stars (blue triangles) have been labelled with KIC number.
Also plotted are the 106 Galactic globular cluster RRc stars that were used by M07 in the derivation of 
their $P$-$\phi_{\rm 31}^c$-[Fe/H] relations (small open circles), and 
two sets of isometallicity curves (the black curves are for the C9 scale, and the red dotted curve for the ZW84 scale).
The $\phi_{\rm 31}^c$ values were derived from cosine Fourier decomposition of $V$-passband observations.
A color version of this figure is given in the on-line version of the paper.   } 
\label{fig:P-phi31-RRc} 
\end{figure}

\subsection{RRc stars}

The $P$-$\phi_{\rm 31}$-[Fe/H] relation for RRc stars was investigated previously by M07.
Two nonlinear formulas were derived, one for metallicities on the ZW84 scale (Zinn \& West 1984) and the other 
for the CG97 scale (Carretta \& Gratton 1997). 
For calibration purposes 106 stars in 12 galactic globular clusters were used, with cluster mean metallicities  
ranging from $-2.2$ to $-1.0$ dex.       

With only four RRc stars in our sample it was not possible to derive independently a similar relation for the {\it Kepler}-field RRc stars.  
We therefore added the  {\it Kepler} stars to the M07 calibration data set (kindly sent to us by Dr.\,Morgan)
and derived a new relation.   The inclusion of the {\it Kepler} stars extends the [Fe/H] range to 
considerably higher metallicities. 

Since most of the data were from the M07 sample we chose to work with $\phi_{\rm 31}^c$($V$) Fourier phase parameters.  
The $\phi_{\rm 31}^s$({\it Kp}) values for the  {\it Kepler} stars were transformed to $\phi_{\rm 31}^c$(V) values
using equation 2 of N11, which, while probably not optimal for the 
hotter RRc stars (since it was derived for non-Blazhko RRab stars) is all that is presently available; 
the sine-to-cosine transformation was made using  $\phi_{\rm 31}^c = \phi_{\rm 31}^s - \pi$.  

The metallicities assumed for the {\it Kepler} stars were those given in Table\,7, 
and the [Fe/H] values assumed for the globular cluster stars were the cluster means given in Table A.1 of C9.
The latter are very similar to the metal abundances adopted by W. Harris in the latest edition (2010) of his on-line 
catalog of globular cluster parameters (Harris 1996).

Using the revised data set and the $\phi_{\rm 31}^c$($V$) phase parameters we refit the M07 model,  
\begin{equation}
 [{\rm Fe/H}] = b_0 + b_1 P +  b_2 \phi_{31}^c +  b_3 \phi_{31}^c P + b_4 P^2 + b_5 (\phi_{31}^c)^2,
\end{equation}
\noindent where the metallicites are now on the C9 scale.
After removing nine outliers (V17 in M9; V26 in NGC\,6934; V27 and V19 in M2;  V50 in M15; V3, V11 and V14 in NGC\,4147; and KIC\,9453114)
the estimated  model coefficients and their standard errors were found to be: 
$b_0= 1.70 \pm0.82$, $b_1=-15.67\pm5.38$, $b_2=0.20\pm0.21$, $b_3= -2.41\pm0.62$, $b_4=18.00\pm8.70$, and $b_5=  0.17\pm0.04$.
The root MSE of the fit was 0.13 dex, the adjusted $R^2=0.94$, and the calibration sample size  $N$=101 stars.    
The fitted [Fe/H]$_{\rm phot}$ (and their standard errors) for the {\it Kepler}-field RRc stars are given in column (9) of Table~1.

The metallicity differences for the four {\it Kepler}-field RR Lyr stars (blue triangles) and for the
globular cluster stars (small open circles) are plotted in the bottom panel of Fig.\,11.  
For the three {\it Kepler} stars that were included in the calibration the differences are small, all under 0.18\,dex;  
for KIC\,9453114 the derived [Fe/H]$_{\rm phot}$ is 0.43\,dex more metal rich than the [Fe/H]$_{\rm spec}$, $-2.13\pm0.12$\,dex.  
For the cluster stars the groups correspond to the individual clusters, with the vertical variation
reflecting the range of the $P$ and $\phi_{\rm 31}$ values within each cluster (for clarity the uncertainties in the
[Fe/H]$_{\rm spec}$ have been omitted).

In {\bf Fig.~13} the $P$-$\phi_{\rm 31}^c$ diagram is plotted for the RRc stars in the {\it Kepler}-field and in the M07 globular clusters 
(the symbols are the same as in Fig.11). 
Also shown are two sets of isometallicity curves for [Fe/H] ranging from  $-2.0$ dex to $-0.5$ dex.
The solid curves (black) were calculated using the new $P$-$\phi_{\rm 31}^c(V)$-[Fe/H] relation (equation 4)
and solving for the four different [Fe/H] values; and the dotted curves (red) are the same curves
transformed (using the transformation equation given in C9) to the ZW84 scale.

\section{SUMMARY}

Metal abundances, RVs and atmospheric properties have been derived for 41 RR\,Lyr stars located in
the field of view of the {\it Kepler} space telescope.  The spectroscopic  [Fe/H] values range from
$-2.54\pm0.13$ (NR\,Lyr) to $-0.05\pm0.13$ dex (V784\,Cyg).   Four stars that were suspected by N11 from Q0-Q5 
{\it Kepler} photometry of being metal-rich (KIC\,6100702, V2470\,Cyg, V782\,Cyg and V784~Cyg) are here confirmed as
such, and three more metal-rich RR\,Lyr stars are identified: V839\,Cyg, KIC\,5520878 and KIC\,8832417
(the last two being RRc stars).

For all but five Blazhko RRab stars and one RRc star the [Fe/H]$_{\rm spec}$ are in good
agreement with newly calculated [Fe/H]$_{\rm phot}$ values derived from detailed analyses of
$\sim$970 days of quasi-continuous high-precison Q0-Q11 long- and short-cadence {\it Kepler}
photometry.  Revised pulsation periods and times of maximum light are given for all the stars (these
were needed for calculating the pulsation phases at the times of the spectroscopic observations and
the  [Fe/H]$_{\rm phot}$ values), and updated Blazhko periods and times of maximum amplitude are
given for the amplitude modulated variables (needed for calculating Blazhko phases).  

We conclude that empirical Fourier-based $P$-$\phi_{\rm 31}$-[Fe/H] relations {\it can} be used to
derive [Fe/H]$_{\rm phot}$ for non-Blazhko and most Blazhko RRab stars (provided that the
modulation is not too extreme and that sufficient Blazhko cycles are used when calculating the
average $\phi_{\rm 31}$);  similar conclusions previously were reached by Smolec (2005) for Blazhko
and non-Blazhko stars in the LMC and Galactic bulge (OGLE survey), and by Jurcsik {\it et al.}
(2012) for Galactic field stars observed by the Konkoly Blazhko Survey.  For three of the four RRc stars the
[Fe/H]$_{\rm phot}$ calculated from empirical Fourier relations are in good agreement with the 
measured [Fe/H]$_{\rm spec}$ values.   We also suggest that the use of `instantaneous' [Fe/H]$_{\rm phot}$ values 
might be preferrable to the usual method of calculating [Fe/H]$_{\rm phot}$ using mean-$\phi_{\rm 31}$ values.   

All the Blazhko stars are found to exhibit both amplitude and frequency modulations, and 
five of the 21 non-Blazhko RRab stars are more metal-rich than [Fe/H]$\sim -1.0$
dex while none of the 16 Blazhko stars is more metal-rich than this.  
Several stars are found to have special photometric characteristics:

(1) V838~Cyg, classified originally as a non-Blazhko star (B10, N11), has been discovered here to be a Blazhko star with
a complex power spectrum and the smallest amplitude modulation yet detected in such a star, $\Delta A_1 = 0.0024$ mag -- it 
joins KIC\,11125706 as the current record holders;  

(2) V2178~Cyg is an extreme Blazhko star with a unique up-down pattern seen in 
the time variation of its Fourier $\phi_{\rm 31}$ parameter (Fig.~4);  

(3)  V354~Lyr has a Blazhko period of just under two years, the longest $P_{\rm BL}$ of any {\it Kepler} Blazhko star;  

(4)  the candidate RRc star KIC\,3868420 is shown to have a high metallicity, a high surface gravity, a high $v_{\rm mac}$ and a low $RV$;  while
its atmospheric characteristics are not unlike those of the two metal-rich RRc stars in the {\it Kepler} field, 
we conclude based on the shape of its light curve and its rather short period that it more probably is a HADS star than an RRc star;     

(5)  V349~Lyr (KIC\,7176080) has been classified previously as a Blazhko star with
an amplitude modulation period greater than 127 days, and as a non-Blazhko star;  
analysis of the currently available data (Q1-Q13) supports the latter classification. 

Finally, atmospheric parameters are derived for the stars, from which we conclude that the RR~Lyrae stars have
higher microturbulent velocities (and are hotter) than the red horizontal branch and metal-poor red giants of the same surface gravity.
The observations also directly confirm that the more metal-rich RR Lyrae stars have higher surface gravities than the more metal
poor RR Lyrae stars.

The CFHT and Keck spectra contain much more chemical information than the [Fe/H] values and atmospheric parameters presented here.  
Of particular interest are relative abundance ratios [X/Fe], especially [$\alpha$/Fe] ratios which are related to the
supernovae history of the Galaxy.  The expectation is that the $\alpha$-elements (O, Ne, Mg, Si, S, Ca, Ti)
produced mainly by Type II supernovae, will be enhanced with respect to Fe (produced mainly by Type Ia supernovae) 
by $\sim$0.4 for the stars more metal poor than [Fe/H]$=-0.8$ dex (see C95).  Such measurements 
will be presented elsewhere. 



\section{Acknowledgments}

Funding for the {\it Kepler} Discovery mission is provided by NASA's Science Mission Directorate.  The authors thank the entire 
{\it Kepler} team without whom these results would not be possible. 
We thank Nadine Mansett and her team of service observers at the Canada-France-Hawaii 3.6-m telescope for successfully 
acquiring the CFHT spectra.  JN would like to thank Jozsef Benk\H o, Johanna Jurcsik, L\'aszlo Moln\'ar 
and R\'obert Szab\'o for discussions on Blazhko stars and for their hospitality at KASC5 and in Budapest.  He also 
greatly appreciates discussions with Katrien Kolenberg regarding her RR~Lyrae work, with Robert Stellingwerf 
concerning his PDM2 program, with BiQing~For concerning the results of 
her Ph.D. thesis, with Andrzej Pigulski for discussions about the 
ASAS-North survey.  John Feldmeier sent information on the two BOKS survey RR~Lyr stars, G\'eza Kov\'acs kindly provided JN with his
Fourier decomposition software, and Siobahn Morgan sent 
her file of Fourier parameters for RRc stars in galactic globular clusters.
Finally, special thanks go to Amanda~Linnell~Nemec, Young-Beom~Jeon and Robert Szab\'o for their helpful comments on the manuscript,
and to George Preston for his useful referee report. 
JN acknowledges support from the Camosun College Faculty Association; JC and BS acknowledge NSF grant AST-0908139; and 
PM is supported by Polish NCN grant DEC-2012/05/B/ST9/03932.  
This project has been supported by the Hungarian OTKA Grants K76816, K83790 an MB08C 81013 and the "Lend\"ulet-2009" Young
Researchers Program of the Hungarian Academy of Sciences.  AD was supported by the Hungarian E\"otv\"os fellowship and
by the J\'anos Bolyai Research Scholarship of the Hungarian Academy of Sciences.
Finally, all the authors wish to recognize and acknowledge the very significant cultural role and reverence that the 
summit of Mauna Kea has always had within the indigenous Hawaiian community;  we are most fortunate to have had the opportunity 
to conduct observations from this mountain.


\begin{thebibliography}{}
\bibitem[Alcock {\it et al.} (1998)]{alc98} Alcock, C., Allsman, R., Alves, D.R. {\it et al.} 1998, \apj, 492, 190 
\bibitem[Alcock {\it et al.} (2000)]{alc00} Alcock, C., Allsman, R., Alves, D.R. {\it et al.} 2000, \apj, 542, 257  
\bibitem[Alcock {\it et al.} (2003)]{alc03} Alcock, C., Alves, D.R., Becker, A. {\it et al.} 2003, \apj, 598, 597  
\bibitem[Alcock {\it et al.} (2004)]{alc04} Alcock, C., Alves, D.R., Axelrod, T.S. {\it et al.} 2004, \aj, 127, 334  
\bibitem[Arp (1955)]{arp55} Arp, H.C. 1955, \aj, 60, 317
\bibitem[Asplund {\it et al.} (2009)]{asp09} Asplund, M., Grevesse, N. {\it et al.} 2009, \araa, 2009, 47, 481
\bibitem[Balona \& Nemec (2012)]{bn12} Balona, L.A. \& Nemec, J.M. 2012, \mnras, 426, 2413
\bibitem[Barklem {\it et al.} (2005)]{bar05} Barklem, P.S., Christlieb, N. {\it et al.} 2005, \aap, 439, 129
\bibitem[Benedict {\it et al.} (2011)]{fben11} Benedict, G.F., McArthur, B.E. {\it et al.} 2011, \aj, 142:187
\bibitem[Benk\H o {\it et al.} (2010)]{ben10} Benk\H o, J.M., Kolenberg, K. {\it et al.} 2010, \mnras, 409, 1585 (B10) 
\bibitem[Blazhko (1907)]{bla07} Blazhko, S. 1907, Astron. Nachr., 175, 325
\bibitem[Bono {\it et al.} (1997)]{bon97} Bono, G., Caputo, F., Cassisi, S. {\it et al.} 1997, \apj, 483, 811    
\bibitem[Bono, Caputo \& DiCriscienzo (2007)]{bcc07} Bono, G., Caputo, F. \& DiCriscienzo, M. 2007, \aap, 476, 779
\bibitem[Brown {\it et al.} (2011)]{bro11} Brown, T.M., Latham, D.W {\it et al.} 2011,  \aj, 142, 112
\bibitem[Bruntt {\it et al.} (2002)]{bru02} Bruntt, H., Catala, C. {\it et al.} 2002, \aap, 389, 345 
\bibitem[Bruntt {\it et al.} (2008)]{bru08} Bruntt, H., De Cat, P. \& Aerts, C. 2008, \aap, 478, 487   
\bibitem[Bruntt {\it et al.} (2010a)]{bru10a} Bruntt, H., Deleuil, M. {\it et al.} 2010a, \aap, 519, A51  
\bibitem[Bruntt {\it et al.} (2010b)]{bru10b} Bruntt, H., Bedding, T. {\it et al.} 2010b, \mnras, 405, 1907   
\bibitem[Buchler \& Moskalik (1992)]{buc92} Buchler, J.R. \& Moskalik, P. 1992, \apj, 391, 736  
\bibitem[Buchler \& Koll\'ath (2011)]{buc11} Buchler, J.R. \& Koll\'ath, Z. 2011, \apj, 731:24
\bibitem[Cacciari, Corwin \& Carney (2005)]{ccc05} Cacciari, C., Corwin, T.M. \& Carney, B. 2005, \aj, 129, 267
\bibitem[Caputo {\it et al.} 2000]{cap00} Caputo, F., Castellani, V. {\it et al.} 2000, \mnras, 316, 819
\bibitem[Carney \& Jones (1983)]{cj83} Carney, B.W. \& Jones, R. 1983, \pasp, 95, 246 
\bibitem[Carretta {\it et al.} (1997)]{car97} Carretta, E. \& Gratton, R.G.  1997, \aaps, 121 95 (CG97)
\bibitem[Carretta {\it et al.} (2009)]{c9} Carretta, E., Bragaglia, A.  {\it et al.} 2009, \aap, 508, 695 (C9)
\bibitem[Cassisi {\it et al.} (1999)]{cas99} Cassisi, S., Castellani, V. {\it et al.} 1999, \aaps, 134, 103 
\bibitem[Clement {\it et al.} (2001)]{cle01} Clement, C., Muzzin, A. {\it et al.} 2001, \aj, 122, 2587
\bibitem[Clementini et al. (1995)]{cle95} Clementini, G.,  Carretta, E. {\it et al.} 1995, \aj, 110, 2319 (C95)
\bibitem[Cohen (2011)]{Coh11} Cohen, J.G. 2011, \apjl, 7430, L38
\bibitem[Cohen \& Melendez (2005a)]{cm05a} Cohen, J.G. \& Melendez, J. 2005a, \aj, 129, 303
\bibitem[Cohen \& Melendez (2005b)]{cm05b} Cohen, J.G. \& Melendez, J. 2005b, \aj, 129, 1607
\bibitem[Cohen \& Huang (2009)]{ch09} Cohen, J.G. \& Huang, W.  2009, \apj, 701, 1053
\bibitem[Collinge et al. (2006)]{col06} Collinge, M.J., Sumi, T. \& Fabrycky, E. 2006, \apj, 651, 197
\bibitem[Donati {\it et al.} (1997)]{don97} Donati, J.-F. {\it et al.} 1997, \mnras, 291, 658
\bibitem[Drukier {\it et al.} (2007)]{dru07} Drukier, G.A. {\it et al.} 2007, \aj, 133, 1041
\bibitem[Feast {\it et al.} (2008)]{fea08} Feast, M.W., Laney, C.D.  {\it et al.}  2008 \mnras, 386, 2115
\bibitem[Feldmeier {\it et al.} (2011)]{fel11} Feldmeier, J. {\it et al.} 2011, \aj, 142:2
\bibitem[Fernley {\it et al.} (1998)]{fer98} Fernley, J., Barnes, T.G. {\it et al.} 1998, \aap, 330, 515
\bibitem[For, Sneden \& Preston (2011)]{for11} For, B.-Q., Sneden, C. \& Preston, G.W. 2011, \apjs, 197, 29
\bibitem[Garofalo {\it et al.} (2013)]{gar13} Garofalo, A., Cusano, F., Clementini, G. {\it et al.} 2013, \apj, 767, 62
\bibitem[Gilliland {\it et al.} (2010)]{gil10} Gilliland, R.L., Jenkins, J.M.  {\it et al.} 2010, \apj, 713, L160
\bibitem[Gratton {\it et al.} (2004)]{gra04} Gratton, R., Bragaglia, A. {\it et al.} 2004, \aap, 421, 937 
\bibitem[Grevesse, Asplund \& Sauval (2007)]{gre07} Grevesse, N., Asplund, M. {\it et al.}  2007, Space Sci. Rev. 130, 105
\bibitem[Guggenberger {\it et al.} (2012)]{gug12} Guggenberger, E.  {\it et al.} 2012, \mnras, 424, 649 (G12) 
\bibitem[Gustafsson {\it et al.} (1975)]{gus75} Gustafsson, B., Bell, R.A. {\it et al.} 1975, \aap, 42, 407
\bibitem[Gustafsson {\it et al.} (2008)]{gus08} Gustafsson, B., Edvardsson, B. {\it et al.} 2008, \aap, 486, 951
\bibitem[Harris (1996)]{har96} Harris, W.E. 1996, \aj, 112, 1487
\bibitem[Haschke {\it et al.} (2012)]{has12} Haschke, R., Grebel, E.K. {\it et al.} 2012, \aj, 144:88
\bibitem[Heiter {\it et al.} (2002)]{hei02} Heiter, U. {\it et al.}  2002, \aap, 392, 619
\bibitem[Heiter \& Eriksson (2006)]{hei06} Heiter, U. \& Eriksson, K. 2006, \aap, 452, 1039
\bibitem[Hinkle {\it et al.} (2000)]{hin00} Hinkle, K. {\it et al.} 2000, Visible and Near-IR Atlas of the Arcturus Spectrum 3727-9300 \AA (San
Francisco: ASP) 
\bibitem[Holweger (2001)]{hol01} Holweger, H.  2001, AIP Conf. Proc. 598, 23
\bibitem[Jenkins {\it et al.} (2010)]{jen10} Jenkins, J.M., {\it et al.} 2010, \apj, 713, L87
\bibitem[Jurcsik (1995)]{jur95} Jurcsik, J.  1995, Acta Astron., 45, 653   
\bibitem[Jurcsik \& Kov\'acs (1996)]{JK96} Jurcsik, J. \& Kov\'acs, G. 1996, \aap, 312, 133 (JK96)
\bibitem[Jurcsik {\it et al.} (2005)]{Jur05} Jurcsik, J. {\it et al.} 2005, \aap, 430, 1049      
\bibitem[Jurcsik {\it et al.} (2006)]{Jur06} Jurcsik, J. {\it et al.} 2006, \aj, 132, 61            
\bibitem[Jurcsik {\it et al.} (2009a)]{Jur09a} Jurcsik, J. {\it et al.} 2009a, \mnras, 393, 1553    
\bibitem[Jurcsik {\it et al.} (2009b)]{Jur09b} Jurcsik, J. {\it et al.} 2009b, \mnras, 397, 350     
\bibitem[Jurcsik {\it et al.} (2009c)]{Jur09c} Jurcsik, J. {\it et al.} 2009c, \mnras, 400, 1006  
\bibitem[Jurcsik {\it et al.} (2012)]{Jur12} Jurcsik, J. {\it et al.} 2012, \mnras, 423, 993  
\bibitem[Koch {\it et al.} (2010)]{koc10} Koch, D.G., Borucki, W.J., Basri, G., {\it et al.} 2010, \apj, 713, L79
\bibitem[Kolenberg (2002)]{kol02} Kolenberg, K. 2002, Ph.D. thesis, Katholieke Universiteit Leuven.
\bibitem[Kolenberg {\it et al.} (2010)]{kol10} Kolenberg, K., Fossati, L. {\it et al.} 2010, \apjl, 713, L198 (K10)
\bibitem[Kolenberg {\it et al.} (2011)]{kol11} Kolenberg, K., Bryson, S. {\it et al.} 2011, \mnras, 411, 878 (K11)
\bibitem[Koll\'ath, Moln\'ar \& Szab\'o 2011]{kms11} Koll\'ath, Z., Moln\'ar, L. \& Szab\'o, R. 2011, \mnras, 414, 1111
\bibitem[Kov\'acs (1998)]{kov98} Kov\'acs, G. 1998, ASP Conf. Ser., 135, 52 
\bibitem[Kov\'acs (2005)]{kov05} Kov\'acs, G. 2005, \aap, 438, 227    
\bibitem[Kov\'acs (2009)]{kov09} Kov\'acs, G. 2009, AIP Conf.Proc.1170, {\it Stellar Pulsation: Challenges for Theory and Observation}, eds. J.A.Guzik
\& P.A.Bradley, p.261   
\bibitem[Kov\'acs \& Jurcsik (1996)]{kj96} Kov\'acs, G. \& Jurcsik, J.  1996, \apj, 466, L17 
\bibitem[Kov\'acs \& Walker (2001)]{KW01} Kov\'acs, G. \& Walker, A.R. 2001, \aap, 371, 579
\bibitem[Kov\'acs \& Zsoldos, E. (1995)]{KZ95} Kov\'acs, G. \& Zsoldos, E. 1995,  \aap, 293, L57
\bibitem[Kunder \& Chaboyer (2008)]{KC08} Kunder, A. \& Chaboyer, B. 2008, \aj, 136, 2441  
\bibitem[Kunder \& Chaboyer (2009)]{KC09} Kunder, A. \& Chaboyer, B. 2009, \aj, 138, 1284  
\bibitem[Kupka {\it et al.} (1999)]{Kup99} Kupka, F., Piskunov, N.E. {\it et al.} 1999, \aaps, 138, 119    
\bibitem[Kurucz {\it et al.} (1984)]{Kur84} Kurucz R. L., Furenlid I., Brault J., Testerman L., 1984, Solar Flux Atlas from 296 to 1300 nm. National Solar Observatory, Sunspot, New Mexico , USA
\bibitem[Lambert {\it et al.} (1996)]{lam96} Lambert, D.L., Heath, J.E. {\it et al.}  1996, \apjs, 103, 183 (L96)
\bibitem[Layden (1994)]{lay94} Layden, A.C. 1994, \aj, 108, 1016 (L94)
\bibitem[Le Borgne {\it et al.} (2012)]{leb12} Le Borgne, J.-F., Klotz, A., Poretti, E., {\it et al.} 2012, \aj, 144:39 
\bibitem[Lenz \& Breger (2005)]{len05} Lenz, P. \& Breger, M.  2005, CoAst, 146, 53
\bibitem[Liu \& Janes (1989)]{lj89} Liu, T. \& Janes, K.A. 1989, \apjs, 69, 593
\bibitem[Liu \& Janes (1990a)]{lj90a} Liu, T. \& Janes, K.A. 1990a, \apj, 354, 273
\bibitem[Liu \& Janes (1990b)]{lj90b} Liu, T. \& Janes, K.A. 1990b, \apj, 360, 561
\bibitem[Moln\'ar {\it et al.} (2012)]{mol12} Moln\'ar, L., Koll\'ath, Z., Szab\'o, R. {\it et al.} 2012, \apj, 757, L13
\bibitem[Morgan, Wahl \& Wieckhorst (2007)]{m07} Morgan, S.M., Wahl, J.N. {\it et al.}  2007, \mnras, 374, 1421 (M07)
\bibitem[Moskalik \& Buchler (1991)]{mos91} Moskalik, P.  \& Buchler, J.R. 1991, \apj, 366, 300  
\bibitem[Moskalik {\it et al.} (2012)]{mos12} Moskalik, P., Smolec, R. {\it et al.} 2012, arXiv:1208.4251v1
\bibitem[Moskalik \& Ko\l{}aczkowski (2009)]{mos09} Moskalik, P. \& Ko\l{}aczkowski, Z. 2009, \mnras, 394, 1649
\bibitem[Moskalik \& Poretti (2003)]{MP03} Moskalik, P.  \& Poretti, E. 2003, \aap, 398, 213
\bibitem[Nardetto {\it et al.} (2004)]{nar04} Nardetto, N. {\it et al.}  2004, \aap, 428, 131
\bibitem[Nardetto {\it et al.} (2009)]{nar09} Nardetto, N. {\it et al.}  2009, \aap, 502, 951  
\bibitem[Nemec (2004)]{nem04} Nemec, J.M. 2004, \aj, 127, 2185   
\bibitem[Nemec, Walker \& Jeon (2009)]{nem09} Nemec, J.M., Walker, A. \& Jeon, Y.-B. 2009, \aj, 138, 1310  
\bibitem[Nemec {\it et al.} (2011)]{nem11} Nemec, J.M., Smolec, R.  {\it et al.} 2011, \mnras, 417, 1022 (N11)  
\bibitem[Oke, J.B. (1966)]{oke66} Oke, J.B. 1966, \apj, 145, 468
\bibitem[Oosterhoff (1939)]{oos39} Oosterhoff, P. Th. 1939, The Observatory, 62, 104
\bibitem[Oosterhoff (1944)]{oos44} Oosterhoff, P.Th. 1944, Bull. of the Astron. Inst. Neth., 10, 55
\bibitem[Pietrukowicz et al. (2012)]{pie12} Pietrukowicz, P., Udalski, A. {\it et al.} 2012, \apj, 750:169
\bibitem[Piskunov (1992)]{Pis92} Piskunov, N.E. 1992, in `Stellar Magnetism', eds., Yu.V. Glagolevskij \& I.I. Romanyuk (St. Petersburg, Nauka), p.92
\bibitem[Preston (1959)]{pre59} Preston, G.W. 1959, \apj, 130, 507
\bibitem[Preston (2009)]{pre09} Preston, G.W. 2009, ``RR Lyrae Atmospherics: Wrinkles Old and New", Henry Norris Russell Lecture,
          ftp://ftp.obs.carnegiescience.edu/pub/gwp/HNRLecture
\bibitem[Preston (2011)]{pre11} Preston, G.W. 2011, \aj, 141:6
\bibitem[Rentzsch-Holm (1996)]{ren96} Rentzsch-Holm, I. 1996, \aap, 312, 966
\bibitem[Rey {\it et al.} (2000)]{rey00} Rey, S.-C., Lee, Y.-W. {\it et al.} 2000, \aj, 119, 1824
\bibitem[Roederer \& Sneden (2011)]{rs11} Roederer, I.U. \& Sneden, C. 2011, \aj, 142:22
\bibitem[Sandage (1958)]{san58} Sandage, A.R. 1958, in {\it Stellar Populations}, ed. D.O'Connell ({\it Ricerche Astr. Specola Vaticana}, Vol.5), p.41
\bibitem[Sandage (1981)]{san81} Sandage, A.R. 1981, \apj, 248, 161
\bibitem[Sandage (1990)]{san90} Sandage, A.R. 1990, \apj, 350, 603
\bibitem[Sandage (2004)]{san04} Sandage, A.R. 2004, \aj, 128, 858
\bibitem[Sandage (2010)]{san10} Sandage, A.R. 2010, \apj, 722, 79
\bibitem[Sch\"orck {\it et al.} (2009)]{sch09} Sch\"orck, T., Christlief, N., Cohen, J. {\it et al.} 2009, \aap, 507, 817
\bibitem[Sesar (2012)]{ses12} Sesar, B. 2012, \aj, 144, 114
\bibitem[Sesar et al. (2010)]{ses10} Sesar, B., Ivezic, Z. {\it et al.} 2010, \apj, 708, 717
\bibitem[Simon (1985)]{sim85} Simon, N.R. 1985, \apj, 299, 723  
\bibitem[Simon (1988)]{sim88} Simon, N.R. 1988, \apj, 328, 747  
\bibitem[Simon \& Lee (1981)]{sl81} Simon, N.R. \& Lee, A.S. 1981, \apj, 248, 291
\bibitem[Simon \& Teays (1982)]{st82} Simon, N.R. \& Teays, T.J. 1982, \apj, 261, 586
\bibitem[Simon \& Clement (1993)]{sc93} Simon, N.R. \& Clement, C. 1993, \apj, 410, 526
\bibitem[Smolec (2005)]{smo05} Smolec, R. 2005, Acta Astron. 55, 59  
\bibitem[Smolec {\it et al.} (2012)]{smo12} Smolec, Soszynski, Moskalik {\it et al.} 2012, \mnras, 419, 2407
\bibitem[Sneden (1973)]{sne73} Sneden, C. 1973, Ph.D. thesis, Univ. Texas
\bibitem[Sneden (2002)]{sne02} Sneden, C. 2002, Manual (MOOG: An LTE Stellar Line Analysis Program)
\bibitem[Sneden, For \& Preston (2011)]{sne11} Sneden, C., For, B.-Q. and Preston, G.W. 2011, Carnegie Obs. Astrophys. Ser., Vol.5: RR~Lyrae Stars,
Metal-Poor Stars, and the Galaxy, ed. A.McWilliam (Pasadena: Carnegie Observatories)
\bibitem[Soszynski {\it et al.} (2009)]{sos09} Soszynski, I., Udalski, A. {\it et al.} 2009, Acta Astron. 59, 1
\bibitem[Soszynski {\it et al.} (2010)]{sos10} Soszynski, I., Udalski, A. {\it et al.} 2010, Acta Astron. 60, 165-178  
\bibitem[Soszynski {\it et al.} (2011)]{sos11} Soszynski, I., Dziembowski, W.A.  {\it et al.} 2011, Acta Astron. 61, 1  
\bibitem[Sousa {\it et al.} (2007)]{sou07} Sousa, S.G., Santos, N.C. {\it et al.} 2007, \aap, 469, 783
\bibitem[Sousa {\it et al.} (2008)]{sou08} Sousa, S.G., Santos, N.C., Mayor, M. {\it et al.} 2008, \aap, 487, 373
\bibitem[Stellingwerf, R. (1978)]{ste78}  Stellingwerf, R. 1978, \apj, 224, 953
\bibitem[Suntzeff {\it et al.} (1994)]{SKK94} Suntzeff, N.B., Kraft, R.P \& Kinman, T.D. 1994, \apjs, 93, 271
\bibitem[Szab\'o {\it et al.} (2010)]{sza10} Szab\'o, R., Koll\'ath, Z., {\it et al.} 2010, \mnras, 409, 1244
\bibitem[Szeidl (1988)]{sze88} Szeidl, B. 1988, in Multimode Stellar Pulsations, eds. Kov\'acs, Szabados \& Szeidl (Budapest, Kultura), 45 
\bibitem[Szeidl {\it et al.} (2012)]{sze12} Szeidl, B., Jurcsik, J. {\it et al.} 2012, \mnras, 424, 3094
\bibitem[Udalski {\it et al.} (1997)]{uda97} Udalski, A., Olech, A. \& Szymanski, M. 1997, Acta Astron. 47, 1
\bibitem[Valenti \& Piskunov (1996)]{vp96} Valenti, J.A. \& Piskunov, N.  1996, \aaps, 118, 595 
\bibitem[Vogt {\it et al.} (1994)]{vog94} Vogt, S.S. {\it et al.}, 1994, Proc. SPIE, 2198, 362
\bibitem[Walker \& Nemec (1996)]{wn96} Walker, A.R. \& Nemec, J.M. 1996. \aj, 112, 2026
\bibitem[Wallerstein \& Huang (2010)]{wal10} Wallerstein, G. \& Huang, W. 2010, Mem.S.A.It, 81, 952
\bibitem[Wils {\it et al.} (2006)]{wil06} Wils, P., Lloyd, C. \& Bernhard, K. 2006, \mnras, 368, 1757
\bibitem[Zinn \& West (1984)]{zw84} Zinn, R. \& West, R.J. 1984, \apjs, 55, 45 (ZW84)
\end{thebibliography}
\end{document}